%% file: NEDMI.tex
\documentclass[11pt]{article}
\usepackage{verbatim,amsmath,amssymb}
\usepackage{epsfig,float,color}
\usepackage[font=small,labelfont=bf]{caption}
\usepackage{geometry}
\usepackage{setspace}
\usepackage[table]{xcolor}
\usepackage{enumerate}
\usepackage{enumitem}
\usepackage{natbib}
\usepackage{amsthm,mathrsfs,amsfonts,dsfont} 
\usepackage{bm}
\usepackage{hyperref}
\usepackage[labelfont=bf]{caption}
\usepackage{lipsum} 											
\usepackage{booktabs}											
\usepackage{siunitx}

\definecolor{darkblue}{rgb}{0,0,1}
\definecolor{col1}{rgb}{1,0,1} 			
\definecolor{col2}{rgb}{0,0.5,0}		
\definecolor{col3}{rgb}{0.5,0,1}		
\definecolor{col4}{rgb}{0.1,.75,0}		

\hypersetup{pdftex=true, colorlinks=true, breaklinks=true, linkcolor=darkblue, menucolor=darkblue, pagecolor=darkblue, citecolor=darkblue, urlcolor=darkblue}

\newtheoremstyle{rem}
{6pt}
{6pt}
{\small}
{}
{\bf}
{:}
{.5em}
{}

\theoremstyle{rem}

\input{neco.tex}

\begin{document}

\begin{center}
	
\Large{\bf{
Nonlinear elastodynamic material identification of heterogeneous isogeometric Bernoulli--Euler beams
}}\\

\end{center}

\renewcommand{\thefootnote}{\fnsymbol{footnote}}

\begin{center}
	\large{Bart\l{}omiej \L{}azorczyk$^\mra$ and Roger A.~Sauer$^{\mra,\mrb,\mrc}$\footnote[1]{corresponding author, email: roger.sauer@rub.de}}
	\vspace{3mm}
	
	\small{\textit{
			$^\mra$Department of Structural Mechanics, Gda\'{n}sk University of Technology, Gda\'{n}sk, Poland \\[1mm]
			$^\mrb$Institute for Structural Mechanics, Ruhr University Bochum, Bochum, Germany \\[1mm]
			$^\mrc$Department of Mechanical Engineering, Indian Institute of Technology Guwahati, Assam, India}}
		
	\vspace{3mm}
	
	\small{Published\footnote[2]{This pdf is the personal version of an article whose journal version is available at \href{https://doi.org/10.1016/j.cma.2025.118415}{www.sciencedirect.com}} 
		in \textit{Comput.~Methods Appl.~Mech.~Eng.}, \href{https://doi.org/10.1016/j.cma.2025.118415}{DOI: 10.1016/j.cma.2025.118415} \\
		Submitted on 9 June 2025; Revised on 13 August 2025; Accepted on 15 September 2025} 

\end{center}

\renewcommand{\thefootnote}{\arabic{footnote}}

\vspace{-4mm}

\rule{\linewidth}{.15mm}
{\bf Abstract}

This paper presents a Finite Element Model Updating framework for identifying heterogeneous material distributions in planar Bernoulli--Euler beams based on a rotation-free isogeometric formulation. The procedure follows two steps: First, the elastic properties are identified from quasi-static displacements; then, the density is determined from modal data (low frequencies and mode shapes), given the previously obtained elastic properties. The identification relies on three independent discretizations: the isogeometric finite element mesh, a high-resolution grid of experimental measurements, and a material mesh composed of low-order Lagrange elements. The material mesh approximates the unknown material distributions, with its nodal values serving as design variables. The error between experiments and numerical model is expressed in a least-squares manner. The objective is minimized using local optimization with the trust-region method, providing analytical derivatives to accelerate computations. Several numerical examples exhibiting large displacements are provided to test the proposed approach. To alleviate membrane locking, a hybrid discretization approach is employed when necessary. Quasi-experimental data are generated using refined finite element models with random noise applied up to 4\%. The method yields satisfactory results as long as a sufficient amount of experimental data is available, even for high measurement noise. Regularization is used to ensure a stable solution for dense material meshes. The density can be accurately reconstructed based on the previously identified elastic properties. The proposed framework can be straightforwardly extended to shells and 3D continua.


\vspace{-5mm}
\rule{\linewidth}{.15mm}

\section{Introduction}\label{s:intro}

Modern design and analysis use high-fidelity numerical simulations, which in turn require advanced knowledge of material parameters. Unfortunately, many materials are heterogeneous, and the validity of treating them as homogeneous depends on the physical scale of the analysis. This applies to materials of natural origin, such as soft tissues, bones, and timber, as well as anthropogenic materials, including concrete, textiles, and composites. In addition, materials exhibit various changes during their lifetime, often leading to nonhomogeneous deterioration of their characteristics. Such problems are typical for structures subjected to environmental conditions and are common in civil and industrial engineering. From another perspective, traditional testing requires collecting samples from the examined structure, which is not always possible and provides only local information about the properties. It occurs in soft biological tissues, where \textit{in vivo} tests are preferred since the samples are fragile, difficult to grip in a testing machine, and it is hard to provide appropriate physiological conditions~\citep{Evans2017,Navindaran2023}. The availability of modern full-field measurement techniques, such as Digital Image Correlation (DIC), opens the door to the full utilization of non-destructive inverse methods for material identification~\citep{Pierron2021}. 

Inverse problems are inherently ill-posed, meaning that there is no assurance of the existence, uniqueness, and stability of solutions~\citep{Turco2017}. For nonlinear inverse problems, the multimodality of the objective function is not the only obstacle, as these functions often exhibit plateaus, i.e., they are insensitive to the changes of parameters in some subspace~\citep{Snieder1998}. If a minimum is in such a plateau, this leads to poor convergence and identifiability~\citep{Zhang2022}. Furthermore, the reconstruction of heterogeneous materials leads to high-dimensional parameter spaces. Hence, these problems are inherently more complex, often multimodal and unstable. The choice of a proper parametrization of the unknown material distribution is always an individual task, typically leading to the so-called \textit{bias/variance trade-off}, i.e., the balance between underfitting (high bias, low variance) and overfitting (low bias, high variance)~\citep{Nelles2020}. 

The two most popular inverse approaches in the identification of mechanical properties are the Virtual Fields Method~\citep{Pierron2012} and the Finite Element Model Updating Method (FEMU)~\citep{Kavanagh1971}. VFM uses the virtual work principle and a set of chosen virtual fields to obtain unknown constitutive parameters. For linear elasticity, this leads to explicit computations. However, VFM needs an appropriate choice of virtual fields and full-field measurement data~\citep{Avril2008}. In FEMU, the deviation between experimental data and finite element simulation is minimized in a global least-squares manner. The main advantages of FEMU are straightforward implementation, the ability to model complex structures, and low vulnerability to noise~\citep{Goenezen2012,Roux2020}. On the contrary, FEMU requires the knowledge of boundary conditions and runs the finite element (FE) model iteratively, which is computationally expensive. The latter can be partially mitigated with continuation strategies, see~\cite{Gokhale2008,Goenezen2011}. Various FEMU strategies for material identification were recently reviewed by~\citet{Chen2024}. Performance of VFM, FEMU, and related methods was compared by~\citet{Avril2007,Avril2008}, and recently by~\citet{Martins2018,Roux2020}. Bayesian inference enables probabilistic inverse modeling and has been applied to the identification of spatial material distributions, e.g., in~\citet{Koutsourelakis2009,Hoppe2023}. Among modern approaches, one notable method is EUCLID (Efficient Unsupervised Constitutive Law Identification and Discovery)~\citep{Flaschel2021}, which uses sparse regression with a library of candidate models to discover interpretable constitutive laws from full-field measurement under physical constraints, with extensions including anisotropic hyperelasticity~\citep{Joshi2022}, plasticity~\citep{Xu2025}, and general material models~\citep{Flaschel2023}. A recent review for data-driven material identification techniques can be found in~\citet{Fuhg2025}. Unlike VFM and EUCLID, FEMU can operate on partial or sparse data and calibrates a full numerical model that can serve, for example, as part of a digital twin. However, FEMU requires \textit{a prior} assumption of the constitutive law. EUCLID formulations for heterogeneous material are still at an early stage, while FEMU is a well-established technique. For completeness in the beam identification context, see also exact inversion~\citep{Eberle2022} and Physics-Informed Neural Networks~\citep{Teloli2025}.

Concerning homogeneous bodies, FEMU is commonly applied to the identification of constitutive laws parameters in metal plasticity~\citep{Prates2016}, elastic composites~\citep{Gras2013}, and hyperelastic biological tissues~\citep{Murdock2018}. Among recent, less conventional FEMU applications, it is worth mentioning the work of~\citet{Liu2018}, who identified damage parameters of graphite using a single four-point bending test and a double iterative optimization technique.~\citet{Hachem2019} employed a coupled isotropic hygro-mechanical model and Digital Volume Correlation to assess the Poisson ratio and swelling coefficient of spruce wood cell walls. Finally,~\citet{Shekarchizadeh2021} homogenized a micro-scale model of the pantographic structure with a second-gradient macro-scale model using an energy-based inverse approach.

The identification of heterogeneous material distributions with FEMU has been addressed less commonly, with most studies focusing on soft tissues.~\cite{Goenezen2012} reconstructed parameter maps for the modified Veronda--Westmann law for a 2D continuum, demonstrating their potential in breast cancer diagnosis.~\citet{Affagard2014,Affagard2015} proposed and experimentally validated a displacement-based FEMU framework for \textit{in vivo} identification of compressible neo-Hookean parameters for thigh muscles in plane strain.~\citet{Kroon2008,Kroon2009} and~\citet{Kroon2010a} applied FEMU to identify element-wise constant material distributions of anisotropic nonlinear membranes, which was further extended to more general material distributions by~\citet{Kroon2010b}. Recently,~\citet{Borzeszkowski2022} developed an isogeometric shell FEMU framework enabling the reconstruction of heterogeneous material distributions.~\citet{Lavigne2023} proposed an inverse framework for hyperelastic bodies capable of identifying material parameters and the frictionless contact traction field based only on two known deformed configurations. Beyond biomechanics,~\citet{Liu2019} identified the damage properties of graphite, preceded by the reconstruction of Young's modulus distribution.~\citet{Campos2020} used FEMU to identify piecewise-linear parameters of Swift's hardening model across a friction stir weld.~\citet{Wu2022} applied global-optimized FEMU to identify spatially varying linear elastic properties of a sandstone rock. The examples presented in the previous two paragraphs show that FEMU based on quasi-static experiments has attracted growing interest across various fields of material identification. 

Dynamic data such as natural frequencies, mode shapes, and frequency response functions (FRFs) are widely used for model updating in structural engineering, particularly in model calibration, structural health monitoring, and damage detection~\citep{Mottershead2011,Simoen2015,Ereiz2022}. Dynamic data typically serve to identify the stiffness distribution under the assumption of known mass. For example,~\citet{Liu2002} used harmonic response to identify distributed bending stiffness.~\citet{Michele2010} proposed a damage detection method relying on shifts in natural and antiresonant frequencies.~\citet{Saada2013} combined frequency-based FEMU with global optimization to detect damage in linear elastic beams. In practice, model updating often relies solely on frequencies or point-wise data, although examples for full-field measurements can also be found, see e.g.~\citet{Wang2011}. Mass and stiffness parameters are frequently updated simultaneously, as demonstrated by~\citet{Girardi2020} and~\citet{Pradhan2012}, who used frequencies and FRFs, respectively. While modal-based FEMU is well-established and widely adopted, sequential identification of elastic and mass parameters from large-displacements quasi-static and modal data, considered here, remains uncommon.

Slender structures, such as beams, cables, or rods, can be accurately modeled with Bernoulli--Euler~(BE) beam theory, assuming that transverse shear deformations are insignificant. However, the $C^1$-continuity required by this theory makes the standard displacement-based Finite Element Method (FEM) with Lagrange polynomial shape functions unsuitable. BE beams are therefore commonly discretized using $C^1$-continuous cubic Hermite interpolation. Since FEM is the core of any FEMU framework, the choice of discretization is critical. Isogeometric Analysis (IGA) was introduced by~\citet{Hughes2005} primarily to provide exact geometric representation regardless of discretization, and facilitate the transition between Computer Aided Design (CAD) and FEM. Over the years, IGA gained interest not only due to this but also because of arbitrary smoothness across elements boundaries, elimination of Gibbs phenomena, high accuracy, and robustness per degree of freedom~\citep{Nguyen2015,Schillinger2018}. Importantly, IGA enables \textit{rotation-free} BE formulations, which are particularly suitable for geometrically nonlinear problems, since in such cases rotations have a nonlinear group structure~\citep{Engel2002}. For the same reason, IGA attracted attention in the modeling of plates and shells, especially of Kirchhoff--Love type~\citep{Kiendl2009,Benson2011,Nguyenthanh2011,Tepole2015,Kiendl2015,Duong2017}, where it naturally provides required smoothness without rotational degrees of freedom~(DOFs). The above characteristics make IGA a well-suited choice for the considered inverse problem.

In the past two decades, a variety of isogeometric formulations have been proposed for one-dimensional structures, including Timoshenko beams~\citep{Echter2010,Cazzani2014}, beams with deformable cross-sections~\citep{Choi2021a,Choi2023}, collocation methods for Cosserat rods~\citep{Weeger2017}, and geometrically exact collocation for shear-deformable static~\citep{Marino2016,Marino2017} and dynamic beams~\citep{Marino2019}. Early work on IGA structural vibration was conducted by~\citet{Cottrell2006}. One of the earliest studies for nonlinear rotation-free IGA is~\citet{Raknes2013}, which examined bending-stabilized torsion-free cables for linear and nonlinear statics as well as nonlinear dynamics. Nonlinear dynamics of straight IGA BE beams were studied in~\citet{Weeger2013}. The first study to treat a nonlinear spatial IGA BE beam including torsion appears to be~\citet{Bauer2016}, with other seminal contributions on spatial IGA BE beams in~\citet{Greco2013,Greco2014}. IGA collocation methods for Bernoulli--Euler beams and Kirchhoff plates were proposed, for example, in~\citet{Reali2015}. Arbitrary curved geometrically exact BE beams were studied, for example, in~\citet{Borkovic2018,Borkovic2019,Borkovic2022}. IGA beam formulations have been applied to various inverse problems, including shape optimization~\citep{Weeger2019,Choi2019,Choi2021b}, simultaneous shape and sizing optimization~\citep{Nagy2010,Nagy2011,Weeger2022}, shape sensing~\citep{Zhao2020,Chen2021}, and geometry reconstruction of beams from image data~\citep{Passieux2023}.

In this work, we propose a FEMU framework for identifying heterogeneous elastic properties of planar isogeometric BE beams, followed by the reconstruction of their density distribution. The beams are assumed to be composed of an isotropic linear elastic material. The elastic properties are identified independently using quasi-static experiments that exhibit large deformations. Subsequently, the density distribution is identified from modal data (low frequencies and modes), using the previously identified elastic parameters. For quasi-static problems, large displacements are considered to keep the formulation general and since soft structures are expected to be in that regime. In real experiments, large displacements tend to improve measurements signal-to-noise ratio. The FE mesh-independent low-order discretization of the unknown material distributions facilitates capturing material discontinuities and adapting the inverse problem size; thus, reducing the risk of overfitting. Our approach is built upon~\citet{Borzeszkowski2022} and extended to density reconstruction. To the best of our knowledge, this is the first time quasi-static and dynamic measurements have been combined for material identification in IGA. The approach can be outlined as follows:
\begin{itemize}[noitemsep,topsep=0pt]
	\item Rotation-free isogeometric FE formulation for nonlinear planar BE beams.
	\item FE-mesh-independent discretization of unknown material parameter distributions.
    \item Least-squares FEMU approach with optional regularization.
	\item Elastic properties are identified from quasi-static measurements and used to estimate the density from modal data.
	\item Gradient-based optimization, accelerated by analytical derivatives.
	\item A study of several numerical examples using synthetic experimental data to analyze the effect of various error sources.
	\item A hybrid discretization approach is used if notable membrane locking occurs in the FE solution.
\end{itemize}

The remainder of this paper is organized as follows: Sec.~\ref{s:theo} describes the governing equations of planar BE beams. The finite element formulation and discretization of the unknown material fields are presented in Sec.~\ref{s:FE}. The proposed inverse framework with derivation of analytical sensitivities is shown in Sec.~\ref{s:Invese}, which is followed by several numerical examples in Sec.~\ref{s:Nex}. The article concludes with Sec.~\ref{s:concl}.  

\section{Planar Bernoulli--Euler beam theory}\label{s:theo}

This section briefly describes BE theory for planar beams under finite deformations and linear elastic material behavior. The formulation is derived directly from a 3D curve. It can also be degenerated from nonlinear Kirchhoff--Love shell theory~\citep{Naghdi1973} with the Koiter shell model~\citep{Ciarlet2005} by taking $\ba_2$ normal to the plane of the beam and assuming zero Poisson's ratio.
\subsection{Kinematics}\label{s:kine}

The deformed configuration of a beam axis $\sL$ embedded in 2D space can be parametrized by 
\eqb{l}
\bx = \bx(\xi)\,,
\eqe
where $\bx$ is the beam axis position and $\xi$ is its parametric coordinate. A basis at $\bx\in\sL$ can be defined with an orthogonal triad: tangent vector $\ba_1 := \bx_{,1}$, out-of-plane unit vector $\ba_2$, and unit normal vector $\bn := \ba_1\times\ba_2/\norm{\ba_1\times\ba_2}$. Here, a comma denotes the parametric derivative $\dots_{,1} = \partial\dots/\partial\xi$. Owing to the above assumptions, the basis is characterized by the single covariant and contravariant metric components
\eqb{l}
a_{11}:= \ba_1\cdot\ba_1 \,,\quad a^{11} := 1/a_{11} \,,
\eqe
respectively. Since the basis is orthogonal but not necessarily orthonormal, contravariant vectors are introduced by a scaling, $\ba^1 := \ba_1/a_{11}$ and $\ba^2 := \ba_2$. The curvature of the beam is given by 
\eqb{l}
b_{11} := \bn\cdot\ba_{1,1} = -\bn_{,1}\cdot\ba_1 \,.
\eqe
Further, $\ba_{1;1} := \ba_{1,1} - \Gamma^1_{11}\ba_1$ denotes the covariant derivative of $\ba_1$, where $\Gamma^1_{11} = \ba_{1,1}\cdot\ba^1$ is the Christoffel symbol of the second kind. All quantities mentioned up to this point can be defined on the reference curve $\sL_0$ analogously, as $\bX \,, \bA_1 \,, \bA_2 \,, \bN \,, A_{11} \,, B_{11}$. The \textit{Jacobian} of the deformation, i.e., the stretch of the curve, is given by $\lambda = \sqrt{a_{11}/A_{11}}$. The Green--Lagrange and Almansi strain tensors for the beam are
\eqb{l}
\bE = \eps_{11} \bA^1 \otimes \bA^1 \,, \quad	\be = \eps_{11} \ba^1 \otimes \ba^1 \,,
\label{e:epsTen}\eqe
and the material and spatial relative curvature tensors are
\eqb{l}
\bK := \kappa_{11} \bA^1 \otimes \bA^1 \,, \quad	\bk := \kappa_{11} \ba^1 \otimes \ba^1 \,.
\label{e:kapTen}\eqe
They are defined by their covariant components
\eqb{l}
\eps_{11} := \ds\frac{1}{2}\big(a_{11} - A_{11}\big) \,, \qquad \kappa_{11} := b_{11} - B_{11} \,.
\label{e:epskap}\eqe
Introducing the unit vector $\bnu = \ba_1/\sqrt{a_{11}}$, the Almansi strain and spatial relative curvature tensors can also be expressed as
\eqb{l}
\be := \eps\, \bnu \otimes \bnu \,, \qquad \bk := \kappa\, \bnu \otimes \bnu \,,
\label{e:epskapPhs1}\eqe
where $\eps := \eps_{11}/a_{11}$ and $\kappa := \kappa_{11}/a_{11}$ are the physical strain and curvature components. Likewise, introducing the unit vector $\bnu_0 = \bA_1/\sqrt{A_{11}}$, the Green--Lagrange and material relative curvature tensors become
\eqb{l}
\bE := \eps_0\, \bnu_0 \otimes \bnu_0 \,, \qquad \bK := \kappa_0\, \bnu_0 \otimes \bnu_0 \,,
\label{e:epskapPhs2}\eqe
where $\eps_0 := \eps_{11}/A_{11}$ and $\kappa_0 := \kappa_{11}/A_{11}$. The components of \eqref{e:epskapPhs1} and \eqref{e:epskapPhs2} are nonlinear; thus, linearization is still necessary to obtain infinitesimal strains. The variations of \eqref{e:epskap} are given by 
\eqb{l}
\delta \eps_{11} = \ds\frac{1}{2}\delta a_{11} = \ba_1\cdot\delta\ba_1 \,, \quad	\delta \kappa_{11} = \delta b_{11} = \big(\delta\ba_{1,1} - \Gamma^1_{11}\delta\ba_1\big)\cdot\bn \,,
\label{e:dab1}\eqe
see, e.g.,~\citet{Sauer2015} for more details.

\subsection{Constitution}\label{s:const}

The constitutive law can be formulated directly on the beam axis. The normal force, $N_0^{11}$, and bending moment $M_0^{11}$~w.r.t.~basis $\bA_1$ of the reference configuration are defined as
\eqb{l}
N_0^{11} := \Astf\,\eps^{11}\,, \quad
M_0^{11} := \Cstf\,\kappa^{11}\,,
\label{e:forceAC}\eqe
where $\eps^{11} = \eps_{11}/(A_{11})^2$, $\kappa^{11} = \kappa_{11}/(A_{11})^2$; $\Astf$ and $\Cstf$ represent the axial and bending stiffness, respectively. In analogy to Eq.~\eqref{e:epskapPhs2}, the corresponding forces w.r.t.~basis $\bnu_0$ are given by 
\eqb{l}
N_0 := \Astf\,\eps_0\,, \quad
M_0 := \Cstf\,\kappa_0\,.
\label{e:forcePhys}\eqe
Further, the forces~w.r.t.~the current basis $\ba_1$ are defined as $N^{11} := N_0^{11}/\lambda$ and $M^{11} := M_0^{11}/\lambda$, with their physical counterparts (w.r.t.~basis $\bnu$) given by $N = \Astf\,\lambda^3\eps$ and $M = \Cstf\,\lambda^3\kappa$. Assuming a rectangular cross-section, the axial and bending stiffness of the beam are given by
\eqb{l}
\Astf = EBT \,,\quad \Cstf = EBT^3/12 \,,
\label{e:AI}\eqe
where $E$ is Young's modulus, $B$ is the beam width, and $T$ is its thickness. It is assumed here that $B$ and $T$ remain unchanged during deformation. In the inverse analysis, $\Astf$ and $\Cstf$ are identified, and the values of $E$ and $T$ can be determined for known $B$.

\subsection{Weak form}\label{s:WF}
The weak form (or principle of virtual work) can be written as
\eqb{l}
G(\bx,\delta\bx) = G_\mrin + G_\mrint - G_\mrext= 0 \quad \forall\,\delta\bx\in\sV\,,
\label{e:WF}\eqe
where $\delta\bx\in\sV$ is a kinematically admissible variation. The inertial virtual work is expressed by
\eqb{l}
G_\mrin = \ds\int_{\sL_0}\delta\bx\cdot\rho_0\ddot{\bu}\,\dif L\,,
\label{e:Gin}\eqe
in which $\rho_0$ denotes the density of the material per beam length in the reference configuration. For quasi-static conditions, the inertial term vanishes. It is discussed in further detail for dynamic eigenvalue problem in Sec.~\ref{s:MD}. The internal virtual work is given by
\eqb{l}
G_\mrint = \ds\ds\int_{\sL_0}\delta \eps_{11}\,N_0^{11}\,\dif L + \int_{\sL_0}\delta \kappa_{11}\,M_0^{11}\,\dif L \,,
\label{e:Gint}\eqe
where the material model presented in Sec.~\ref{s:const} is applied. For a planar beam, the external virtual work is given by
\eqb{l}
G_\mrext = \ds\int_{\sL}\delta\bx\cdot\bff\,\dif \ell + \big[\delta\bx\cdot\bt\:\big] + \big[\delta\bn\cdot \bar M\,\bnu\big]\,,
\label{e:Gext}\eqe
where $\bff = \bff_0/\lambda + p\,\bn$ is the body force, consisting of dead load $\bff_0$ and live pressure $p$, both per length of the beam; $\bt = \bar N\bnu + \bar S \bn$, where $\bar N$, $\bar S$ and $\bar M$ denote prescribed end forces and end moments. Distributed moments are not considered here.

The Newton--Raphson method for solving the weak form \eqref{e:WF} requires the linearization of Eqs.~\eqref{e:Gint}~and~\eqref{e:Gext}. This can be found, e.g., in~\citet{Duong2017}. 

\section{Finite element discretization}\label{s:FE}

Two different discretizations are discussed in this section. Firstly, the isogeometric FE formulation is introduced and used to approximate weak form \eqref{e:WF}, and its corresponding dynamic eigenvalue problem. Secondly, the independent discretization of the material fields with Lagrange interpolation is defined. The mapping between the FE analysis mesh and material mesh is also provided. 

\subsection{Isogeometric curve discretization}\label{s:surfFE}

Since the BE beam formulation contains second derivatives, at least $C^1$-continuous discretization is necessary to solve the weak form in Eq.~\eqref{e:WF} with FE.
To satisfy this, the curve $\sL$ is discretized with NURBS following the concept of \textit{isogeometric analysis}, introduced by~\citet{Hughes2005}. In order to recover the standard structure of FEM, the B\'ezier extraction operator $\mC_e$ proposed by~\citet{Borden2011} is used. Each element $\Omega^e$ contains $n_e$ NURBS basis functions $\{N_I\}_{I=1}^{n_e}$, where $n_e$ is the number of control points of the element. The NURBS basis functions are defined by 
\eqb{l}
N_I(\xi) = \ds\frac{w_I\hat N_I^e(\xi)}{\sum_{I=1}^{n_e} w_I\hat N_I^e(\xi)} \,,
\label{e:shpfcn}\eqe
where $\{\hat N_I^e\}_{I=1}^{n_e}$ are the B-spline basis functions. The geometry of the reference and current curve $\sL$, the displacements, and accelerations are approximated from the corresponding quantities at control points, respectively, as
\eqb{l}
\bX = \mN_e\,\mX_e\,, \quad \bx = \mN_e\,\mx_e\,, \quad \bu = \mN_e\,\muu_e \,, \quad \ddot{\bu} = \mN_e\,\ddot{\muu}_e \,,
\label{e:Xxuu}\eqe
where $\mN_e := [N_1\bone,N_2\bone,\dots,N_{n_e}\bone]$ is a matrix of the nodal shape functions defined in Eq.~\eqref{e:shpfcn}, and $\bone$ is the identity tensor in $d$-dimensional space. With \eqref{e:Xxuu}, the covariant tangent vectors become
\eqb{l}
\ba_1 = \bx_{,1} \approx \mN_{e,1}\,\mx_e \,,\quad  \,
\bA_1 = \bX_{,1} \approx \mN_{e,1}\,\mX_e \,,
\eqe
while the variations of $\bx$, $\ba_1$, and $\bn$ are
\eqb{l}
\delta\bx \approx \mN_e\,\delta\mx_e \,,\quad  
\delta\ba_1 \approx \mN_{e,1}\,\delta\mx_e \,, \quad 
\delta\bn = -\left(\ba^1\otimes\bn\right)\delta\ba_1 \,,
\eqe
see~\citet{Sauer2015} for more details.

\subsection{FE approximation}\label{s:WFFE}

With the discretization scheme from the previous section, one obtains
\eqb{l}
G(\mx,\delta\mx) \approx \ds\sum_{e=1}^{n_\mrel}\left(G_\mrin^e + G_\mrint^e - G_\mrext^e\right)= 0 \quad \forall\,\delta\mx\in\sV\,,
\label{e:dWF}\eqe 
where $n_\mrel$ is the number of finite elements. The elemental inertial contribution to the weak form \eqref{e:dWF} is given by
\eqb{l}
G_\mrin^e = \delta\mx_e\T\mf^e_\mrin \,, 
\label{e:Gine}\eqe
where the inertial FE force vector is defined as
\eqb{l}
\mf^e_\mrin := \mm_e\ddot{\muu}_e\,,
\label{e:fine1}\eqe
and 
\eqb{l}
\mm_e := \ds\intooe \rho_0\,\mN_e\T\mN_e\,\dif L \,.
\label{e:fine2}\eqe
is the elemental mass matrix. In the same manner, 
\eqb{l}
G_\mrint^e = \delta\mx_e\T \mf^e_\mrint =\delta\mx_e\T \left(\mf^e_{\mrint N} + \mf^e_{\mrint M}\right) \,,
\label{e:Ginte}\eqe
in which the internal FE force vectors from $N_0^{11}$ and $M_0^{11}$ are 
\eqb{l}
\mf^e_{\mrint N} := \ds\intooe N_0^{11}\,\mN_{e,1}\T\,\ba_1 \,\dif L \,,\quad \mf^e_{\mrint M} := \ds\intooe M_0^{11}\,\mN_{e;11}\T\,\bn \,\dif L \,,
\label{e:finte}\eqe
and $\mN_{e;11} := \mN_{e,11} - \Gamma^1_{11}\mN_{e,1}$. The elemental external virtual work follows as
\eqb{l}
G_\mrext^e = \delta\mx_e\T \mf^e_\mrext = \delta\mx_e\T \left(\mf^e_{\mrext0} + \mf^e_{\mrext p} + \mf^e_{\mrext t} + \mf^e_{\mrext m}\right) \,,
\label{e:Gexte}\eqe 
with the external FE force vectors
\eqb{l}
\mf^e_{\mrext0} :=\ds\intooe \mN_e\T\,\bff_0 \,\dif L \,, \quad \mf^e_{\mrext p} :=\ds\intoe \mN_e\T\,p\bn \,\dif \ell \,,
\label{e:fext0p}\eqe
and
\eqb{l}
\mf^e_{\mrext t} := \mN_e\T\bt \,, \quad \mf^e_{\mrext M} := -\mN_{e,1}\T\,\nu^1\bar M\,\bn \,. 
\label{e:fexttm}\eqe

With Eqs.~\eqref{e:fine1}, \eqref{e:finte}, \eqref{e:fext0p}, and \eqref{e:fexttm}, the weak form in Eq.~\eqref{e:dWF} yields
\eqb{l}
\delta\mx\T \left(\mf_\mrin + \mf_\mrint - \mf_\mrext\right) = 0 \quad \forall\,\delta\mx\in\sV\,,
\label{e:eqFEdelta}\eqe
where
\eqb{l}
\mf_\mrin = \ds\sum_{e=1}^{n_\mrel} \mf^e_\mrin = \mM\ddot{\muu} \,, \quad \mf_\mrint = \ds\sum_{e=1}^{n_\mrel} \mf^e_\mrint \,, \quad \mf_\mrext = \ds\sum_{e=1}^{n_\mrel} \mf^e_\mrext \,,
\eqe
are obtained from the usual assembly of the corresponding elemental contributions. The nodal variations $\delta\mx$ equal zero at the nodes on the Dirichlet boundary. For the remaining part of the body, Eq.~\eqref{e:eqFEdelta} implies
\eqb{l}
\mf(\muu) = \mf_\mrin + \mf_\mrint - \mf_\mrext = \bf0 \,,
\label{e:eqFE}\eqe
which is the discretized global equilibrium equation solved for the unknown nodal displacement vector $\muu$. This vector contains $d\,n_\mrno$ components, where $n_\mrno$ is the number of free control points. Since the considered beam is planar, the out-of-plane DOFs are fixed; thus $d = 2$. For quasi-static conditions, the inertial term in Eq.~\eqref{e:eqFE} vanishes.

It is worth noting that in the presented formulation no mapping of derivatives between reference and deformed configuration is required. No introduction of a local, Cartesian basis is needed either.

\subsection{Modal dynamics}\label{s:MD}

If the deformation of the structure remains small and no external load exists, Eq.~\eqref{e:eqFE} can be further approximated as
\eqb{l}
\mf_\mrin + \mf_\mrint \approx \mM\ddot{\muu} + \mK\muu = \bf0 \,,
\label{e:eqMD}\eqe
where $\mK$ is the tangent stiffness matrix. The general solution of Eq.~\eqref{e:eqMD} is $\muu = \tmuu_i\exp(\mri\,\omega_i t)$, which leads to the linear eigenvalue problem~\citep{Zienkiewicz2000}
\eqb{l}
\left(-\omega_i^2\mM + \mK\right)\tmuu_i = \bf0 \,,
\label{e:eigprobl}\eqe
where $\omega_i$ and $\tmuu_i$ denote the $i^{\text{th}}$ eigenvalue (natural frequency) and the $i^{\text{th}}$ eigenvector (normal mode) of the beam, respectively. The eigenvectors are made unique by normalization, such that
\eqb{l}
\tmuu_i\T\mM\tmuu_i = 1 \,, \quad i = 1,2,3,\dots\,.
\label{e:MDnorm}\eqe
In addition, by the property of modal orthogonality, one obtains 
\eqb{l}
\tmuu_i\T\mK\tmuu_i = \omega_i^2 \,.
\label{e:mdortho}\eqe

\subsection{Discretization of the material parameters}\label{s:matFE}

The unknown material fields are discretized with a \textit{material mesh}, introduced in~\citet{Borzeszkowski2022} and briefly described here. The elastic parameters $\Astf$ and $\Cstf$, or the density $\rho_0$, are defined over the curve $\sL_0$ as a scalar field $q(\xi)$, which is approximated within each material element (ME), $\bar\Omega^{\bre}$, using $\brn_e$ nodal values and interpolation functions $\brN_I$ as
\eqb{l}
q = q(\xi) \approx \ds\sum_{I=1}^{\brn_e} \brN_I(\xi)q_I = \mbN_{\bre}\,\mq_{\bre}  \,,
\label{e:qdiscr}\eqe
where $\mbN_{\bre} := [\brN_1,\brN_2,\dots,\brN_{\brn_e}]$ and $\mq_{\bre} := [q_1,q_2,\dots,q_{\brn_e}]\T$ are matrices containing all $\brN_I$ and $q_I$ of the material element. In this work, except in Sec.~\ref{s:BsplineMatMesh}, the material mesh consists of constant 1-node or linear 2-node Lagrange elements. By means of the material mesh, the field of unknown parameters is represented by the global vector
\eqb{l}
\mq = \begin{bmatrix}
	  \bq_1 \\ \bq_2 \\ \vdots \\ \bq_{\brn_\mrno}
	  \end{bmatrix}
\label{e:qfull}\eqe
in which each nodal entry $\bq_I$ contains $[\Astf_I,\Cstf_I]^T$, or $\rho_{0I}$. The design vector \eqref{e:qfull} consists of $n_\mrvr = \bar d\,\bar n_\mrno$ unknown components, where $\bar n _\mrno$ is the number of material nodes and $\bar d$ is the number of material parameters per material node. Note that the elastic parameters are discretized with a single material mesh, while the density utilizes a separate material mesh. While this approach may not be optimal, it is sufficient for the investigated numerical tests.

\begin{figure}[!ht]
\begin{center}
	\includegraphics{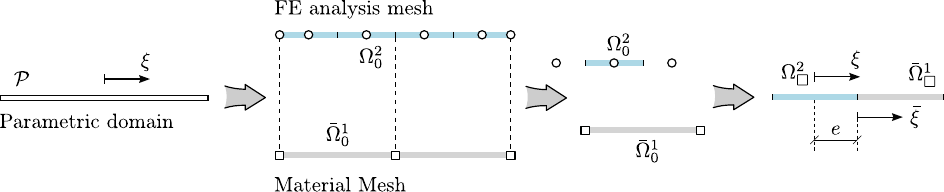}
	\caption{Example of mapping $\xi\mapsto\bar\xi$. Here, $n_\mrel = 4$, $\brn_\mrel = 2$, $m = 2$, and $e = -1$.} 
	\label{fig:matFEmapping}
\end{center}
\end{figure}

To integrate the material mesh into the FE analysis, the mapping between $\muu$ and $\mq$ must be established. Two conforming meshes are defined in the parameter domain $\sP$, as shown in Fig.~\ref{fig:matFEmapping}. Finite elements are considered to satisfy the relation $\Omega_\square^e\subset\bar\Omega _\square^e$, where $\Omega_\square^e$ and $\bar\Omega _\square^e$ denote the element domains in $\sP$ for the FE analysis and the material mesh, respectively. It is noted that this consideration is a present choice, not a necessity. It will be generalized in the example Sec.~\ref{s:Ex3_stat}, see App.~\ref{s:B2M1_map}.

The mapping between domains $\xi\mapsto\bar\xi$ is given by the following linear affine transformation
\eqb{l}
\bar\xi = \ds\frac{1}{m}(\xi + e) \,,
\label{e:matmap}\eqe
where $m$ is the number of $\Omega_0^e\subset\Omega_0^{\bre}$ and $e$ is the offset between the centers of $\bar\Omega_0^{\bre}$ and $\Omega_0^e$. Hence, $\brN_I = \brN\left(\bar\xi(\xi)\right)$ becomes a function of $\xi$, allowing numerical integration in FE domain $\sP$.

Alternatively, one can employ a full isogeometric approach, utilizing the same NURBS basis as in the FE analysis. However, such an approach would quickly lead to a large number of model parameters and overfitting. To mitigate this, a coarser NURBS representation might be selected for the unknown material fields. Both material and FE analysis NURBS representation can be related through a series of knot insertions, resulting in a projection operator similar to B\'ezier extraction operator~\citep{Borden2011}. The performances of the isogeometric and Lagrange material mesh are compared for selected cases in Sec.~\ref{s:BsplineMatMesh}. 

\section{Inverse analysis}\label{s:Invese}

The inverse identification of the unknown material parameter vector $\mq$ is formulated as a constrained nonlinear least-squares problem, which is solved using a gradient-based local optimization algorithm. To speed up calculations and avoid computationally expensive finite differences, the analytical gradient $\mg(\mq)$ and Hessian $\mH(\mq)$ of the objective function $f(\mq)$ are used. 

\subsection{Objective function}\label{s:objfcn}

The inverse problem for the unknown vector $\mq$ is solved by the constrained minimization of the objective function
\eqb{l}
\min\limits_{\mq} f(\mq) \,,
\label{e:minf}\eqe
where the $n_\mrvr$ components of $\mq$ are subject to the bounds $0 < q_\mathrm{min} \leq q_I \leq q_\mathrm{max}$ and satisfy the discrete equilibrium equation \eqref{e:eqFE} for elastic parameter identification or the eigenvalue problem \eqref{e:eigprobl} for density identification. The objective function describes the difference between the FE model response and experimental data. 

In the case of identification of the elastic parameters, the objective function is based on quasi-static experiments and takes the form\footnote{For pure Dirichlet problems, each material parameter is only obtainable with Eq.~\eqref{e:objfnc} up to a constant due to the lack of force data. Therefore, an extra term consisting of the reaction forces $\mR$ is necessary to make the problem determinable. See~\citet{Borzeszkowski2022} for details.}
\eqb{l}
f(\mq) := \ds\sum_{i=1}^{n_\mrlc} \frac{\left\| \mU_{\mrex\,i} - \mU_{\mrfe\,i}(\mq) \right\|^2}{\left\| \mU_{\mrex\,i} \right\|^2} + \alpha^2\norm{\mL\mq}^2 \,,
\label{e:objfnc}\eqe
where $n_\mrlc$ is the number of independent load cases considered, $\alpha$ is the regularization parameter, and $\mL$ is a penalty matrix. The second term in \eqref{e:objfnc} represents Tikhonov regularization (see, e.g.,~\citet{Hansen2013}) and is optional. For each separate load case
\eqb{l}
\mU_\mrex = \begin{bmatrix}
			\bu^\mrex_1 \\[1mm] \bu^\mrex_2 \\ \vdots \\ \bu^\mrex_{n_\mrex}
			\end{bmatrix}
\label{e:Uexp}\eqe
is a vector containing $n_\mrex$ experimental measurements $\bu^\mrex_I$, $I=1,2,\dots,n_\mrex$, at location $\bx^\mrex_I\in\sL$ and 
\eqb{l}
\mU_\mrfe(\mq) = \begin{bmatrix}
				 \bu(\bx_1^\mrex,\mq) \\[1mm] \bu(\bx_2^\mrex,\mq) \\ \vdots \\ \bu(\bx_{n_\mrex}^\mrex,\mq)
				 \end{bmatrix}
\label{e:UFE}\eqe
is a vector containing the corresponding FE displacements at $\bx^\mrex_I$, which is given through \eqref{e:Xxuu} as  
\eqb{l}
\bu_I^\mrex(\mq) = \mN_e(\bx^\mrex_I)\,\muu_e(\mq) \,.
\label{e:uFE}\eqe

In the case of identification of the density, the objective function is based on modal dynamics and defined as
\eqb{l}
	f(\mq) :=
	\ds\sum_{i=1}^{n_\mrmd} \left[  
	w_{Ui}\left\|\hat\mU_{\mrex\,i} - \hat\mU_{\mrfe\,i}(\mq) \right\|^2 + 					
	w_{\omega\,i}\frac{\left(\omega_{\mrex\,i} - \omega_{\mrfe\,i}\right)^2}{\omega_{\mrex\,i}^2} 	
	\right] 
	+ \alpha^2\norm{\mL\mq}^2 \,,
\label{e:objfncMD}\eqe
where $\hat\mU_{\bullet\,i} = \mU_{\bullet\,i} / \left\| \mU_{\bullet\,i} \right \|$ is a unit vector representing the $i^{\text{th}}$ normal mode with $n_\mrex$ measurements at location $\bx^\mrex_I\in\sL$ analogously to Eqs.~\eqref{e:Uexp}, \eqref{e:UFE}, and \eqref{e:uFE}. Following this, $n_\mrmd$ is the number of normal modes, $\omega_\mrex$ and $\omega_\mrfe$ are the experimental and FE natural frequencies, respectively. The weights $w_{Ui}$ and $w_{\omega\,i}$ are set to unity and will be omitted for brevity in the remainder of this paper. Note that the density parameters are only obtainable up to a constant with the modes normalized in such a way. Hence, the term consisting of the frequency differences is added. 

One difference between Eqs.~\eqref{e:objfnc} and \eqref{e:objfncMD} is the different normalization of the quasi-static displacements and normal modes. A natural way to normalize the eigenvectors is to use \eqref{e:MDnorm} or \eqref{e:mdortho}. However, this would require modifying the experimental results depending on the FE mesh, which is not straightforward since the mass matrix $\mM$ is not known \textit{a priori}. In addition, using \eqref{e:mdortho} requires knowledge of the stiffness matrix $\mK$ and frequencies. In contrast, the approach in Eq.~\eqref{e:objfncMD} relies only on the frequencies.

\subsection{Optimization algorithm}\label{s:TIR}

To solve the problem posed in Eq.~\eqref{e:minf}, a trust-region approach is employed. Trust-region methods are a family of iterative algorithms whose main idea is to approximate the minimized function $f(\mq)$ in the neighborhood (\textit{trust-region}) $\sN$ of the current guess of solution $\mq_k$. Typically, they require providing the gradient $\mg(\mq)$ and Hessian $\mH(\mq)$ of the objective function, at least in an approximate form. At each iteration the algorithm minimizes the approximated model $\mh_k(\mq_k+\mss_k)$ over $\sN$. This results in solution $\mss_k$ called the \textit{trial step}. If $f(\mq_k+\mss_k) < f(\mq_k)$, $\mq_k$ is updated. If not, it remains unchanged, $\sN$ is shrunk, and $\mh_k$ is minimized again. The optimization algorithm iterates until $\mq$ and $f$ converge, meaning that the two following stopping criteria are both met,
\begin{align}
	\left \| \mq_{k+1} - \mq_{k} \right\| &\le \epsilon \,, \label{e:qconv}\\[3pt]
	\left| f(\mq_{k+1}) - f(\mq_{k}) \right| &\le \epsilon\left(1 + \left|f(\mq_{k})\right| \right) \,, \label{e:fconv}
\end{align}
where $\epsilon$ is a small tolerance. In general, trust-region methods can handle non-convex approximated models $\mh_k(\mq_k+\mss_k)$, are reliable and robust, and can be applied to ill-conditioned problems~\citep{Yuan1999}. A comprehensive description of trust-region methods can be found in~\citet{Conn2000}.

One often knows a coarse range of sought material parameters, which are typically positive. Since providing nonphysical material and density values to the FE solver can lead to its failure, an approach capable of imposing box constraints is required. An example is the Trust-region Interior Reflective (TIR) method, as described in~\citet{Coleman1996}. Here, the \texttt{lsqnonlin} solver from the MATLAB Optimization Toolbox\textsuperscript{\tiny{TM}}~is used, which employs TIR and allows adding analytical Jacobians, which are specified in Sec.~\ref{s:analytical}.

\subsection{Inverse framework overview}\label{s:inv_overview}

\begin{figure}[h]
	\begin{center}
		\includegraphics[width=0.87\textwidth]{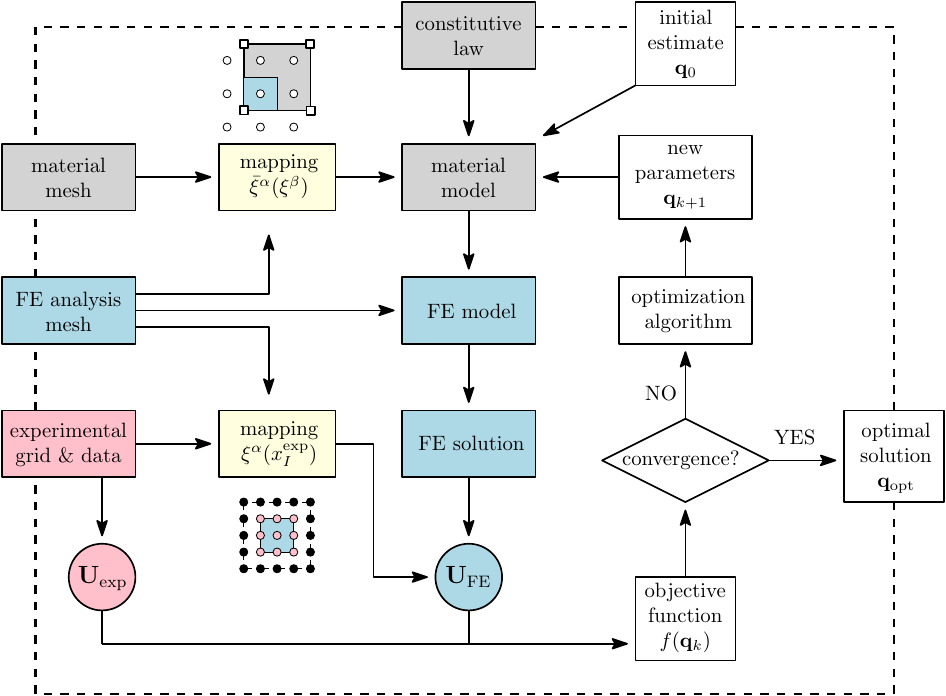}
		\caption{Flow chart of the inverse identification algorithm: Given the experimental data, constitutive law, and the initial guess $\mq_0$, the algorithm calculates the optimal solution of the material parameters $\mq_\mrop$ for chosen FE and material meshes. Source:~\citet{Borzeszkowski2022}.}
		\label{fig:theflowchart}
	\end{center}
\end{figure}

Fig.~\ref{fig:theflowchart} shows a flow chart of the inverse identification algorithm, which consists of four main components: the FE model (blue), the material model with the material mesh (gray), the experimental data (pink), and optimization steps (white). 
The FE mesh is related to the material mesh and the experimental grid through corresponding mappings (see Sec.~\ref{s:matFE} and Eq.~\eqref{e:dUFEdu}). At the input, the algorithm takes the FE discretization, constitutive law, material mesh with the initial guess $\mq_0$, and experimental data. The output is the vector of nodal material values $\mq_\mrop$ that minimizes \eqref{e:minf}. The procedure remains the same for the identification of the elastic and density parameters.

\begin{figure}[h]
	\begin{center}	
		\includegraphics[width=0.65\textwidth]{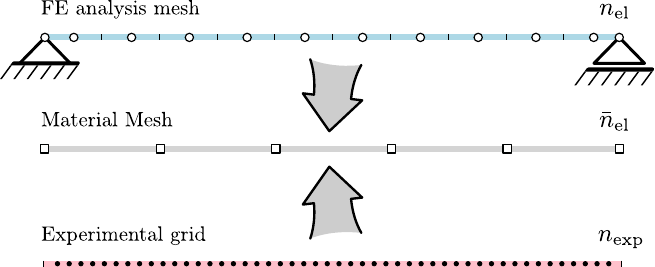}
		\caption{The inverse analysis is based on three separately discretized fields. The resolution of the FE analysis mesh and the experimental grid influences the reconstruction of the unknown material parameters of the material mesh.}
		\label{fig:3meshes}
	\end{center}
\end{figure}

Fig.~\ref{fig:3meshes} illustrates an example of the threefold discretization for a simply supported beam. Each discretized field affects the inverse identification differently. The FE mesh determines the accuracy and computational cost of the forward problems \eqref{e:eqFE} and \eqref{e:eigprobl}, while the material mesh defines the size and computational cost of the inverse problem in \eqref{e:minf}. Both introduce separate sources of error due to the difference between the approximation and the unknown exact field.
Finally, the experimental grid contributes to the computational cost of the inverse problem and introduces errors arising from noise in experimental measurements. The impact of these error sources is analyzed through a convergence study of the forward FE problem, providing known fields of material properties, and simulating the measurement error with random noise. To mitigate analysis bias, known as \textit{inverse crime}, i.e., using the same model to both generate and invert synthetic data~\citep{Wirgin2004}, the FE mesh used for generating synthetic experimental data is significantly denser than the one used in the inverse analysis. 

Prior to density reconstruction from modal data, the elastic parameters are determined from the inverse analysis based on quasi-static measurements. As a result, inaccurate elastic properties affect the density estimates. Since the discretization errors in elastic and density parameters arise from different sources, they are analyzed separately in Sec.~\ref{s:Nex}.

\subsection{Analytical derivatives}\label{s:analytical}

Gradient-based optimization algorithms, such as TIR, rely on the gradient $\mg(\mq)$ and often the Hessian $\mH(\mq)$ of the objective function $f(\mq)$. They can be computed using the finite difference method. However, this approach is time-consuming and inexact. In contrast, the consistent FE formulation enables the derivation of analytical derivatives. The following section provides the derivatives of objectives \eqref{e:objfnc} and \eqref{e:objfncMD}.

\subsubsection{Nonlinear statics}\label{s:analytical_stat}

For nonlinear statics, the analytical gradients are derived in~\citet{Borzeszkowski2022}. Here, they are summarized briefly. A contribution from a single load case to the objective function \eqref{e:objfnc} can be formulated as
\eqb{l}
f(\mq) = \bar\mU_\mrR\T\,\bar{\mU}_\mrR \,,
\label{e:objfncR}\eqe
where the residual is defined as
\eqb{l}
\bar\mU_\mrR := \bar\mU_\mrex - \bar\mU_\mrfe := \ds\frac{\mU_\mrex-\mU_\mrfe(\mq)}{\norm{\mU_\mrex}} \,.
\label{e:res}\eqe
Consequently, the gradient and Hessian are expressed by  
\eqb{l}
\mg(\mq) = \ds2\pa{f(\mq)}{\mq} = 2\mJ(\mq)\T\bar\mU_\mrR(\mq)\,, \quad \mH(\mq) = \ds2\paq{f(\mq)}{\mq} \approx 2\mJ\T\mJ \,,
\label{e:gH}\eqe
where $\mJ$ is the \textit{Jacobian} of the residual 
\eqb{l}
\mJ =  \ds\pa{\bar\mU_\mrR}{\mq} = -\frac{1}{\norm{\mU_\mrex}}\pa{\mU_\mrfe}{\muu}\pa{\muu}{\mq} \,,
\label{e:Jstat}\eqe
in which
\eqb{l}
\ds\pa{\muu}{\mq} = -\mK^{-1}\pa{\mf_\mrint}{\mq} \,,
\label{e:dudq}\eqe
and $\partial\mU_\mrex/\partial\muu$ is a matrix that maps the experimental grid to the FE mesh, which follows from Eqs.~\eqref{e:UFE} and \eqref{e:uFE}, and can be assembled from the $n_\mrel$ elemental contributions
\eqb{l}
\ds\pa{\mU_\mrfe}{\muu_e} = \begin{bmatrix}
							\mN_e(\bx_1^\mrex) \\[1mm] \mN_e(\bx_2^\mrex) \\ \vdots \\ \mN_e\bigl(\bx_{n_\mrex}^\mrex\bigr)
							\end{bmatrix} \,, \quad e=1,2,\dots,n_\mrel \,.
\label{e:dUFEdu}\eqe
Once the contributions from all load cases are summed, the derivatives of the regularization term, $g_\mathrm{reg}(\mq) = 2\alpha^2\mL\T\mL\mq$ and $H_\mathrm{reg}(\mq) = 2\alpha^2\mL\T\mL$, can be directly added to~\eqref{e:gH} or incorporated into the residual and Jacobian by concatenation. The same rule follows for modal dynamics below.

\subsubsection{Modal dynamics}\label{s:analytical_MD}

In analogy to Eqs.~\eqref{e:objfncR} and \eqref{e:res}, and by introducing
\eqb{l}
	\mU_{\mrR\,i} := \hat\mU_{\mrex\,i} - \hat\mU_{\mrfe\,i} \,, \quad
	\omega_{\mrR\,i} := \ds\frac{\omega_{\mrex\,i} - \omega_{\mrfe\,i}}{\omega_{\mrex\,i}} \,,
\eqe
the first, unregularized part of the objective function \eqref{e:objfncMD} can be expressed as
\eqb{l}
	f(\mq) = \ds\sum_{i=1}^{n_\mrmd} \left[\hat\mU_{\mrR\,i}\T \hat\mU_{\mrR\,i} + \omega_{\mrR\,i}^2\right] =  
	\sum_{i=1}^{n_\mrmd} \pmb{\Theta}_{\mrR\,i}\T \pmb{\Theta}_{\mrR\,i} \,,
\label{e:objfncMDR}\eqe
where $\pmb{\Theta}_{\mrR\,i} := \bigr[\hat\mU_{\mrR\,i} \,, \omega_{\mrR\,i}\bigr]\T$.
Moreover, Eq.~\eqref{e:objfncMDR} can be simplified by concatenating all components of the summation in one column vector. By differentiation of \eqref{e:objfncMDR} w.r.t.~the material unknowns vector $\mq$, the gradient is given by
\eqb{l}
	\ds\mg(\mq) = \pa{f(\mq)}{\mq} = 2\sum_{i=1}^{n_\mrmd} \mJ_i\T \pmb{\Theta}_{\mrR\,i} \,,
\eqe
where
\eqb{l}
	\mJ_i = -
	\begin{bmatrix}
		\ds\pa{\hat\mU_{\mrfe\,i}}{\mq} \\[12pt]
		\ds\frac{1}{\omega_{\mrex\,i}} \pa{\omega_{\mrfe\,i}}{\mq} 
	\end{bmatrix}
\label{e:jacobian}\eqe
is the \textit{Jacobian} of $f(\mq)$ in \eqref{e:objfncMDR}. The Hessian $\mH_i(\mq)$ then follows from Eq.~\eqref{e:gH}b. From now on, all derivations are written for a particular dynamic mode; thus, subscript $i$ is dropped. Also, $\omega$ is used instead of $\omega_\mrfe$. The first component of \eqref{e:jacobian} can be expanded as follows
\eqb{l}
	\ds\pa{\hat\mU_\mrfe}{\mq} =
	\pa{}{\mU_\mrfe}\left(\frac{\mU_\mrfe}{\left\|\mU_\mrfe\right\|}\right) \pa{\mU_\mrfe}{\mq} = \frac{1}{\left\|\mU_\mrfe\right\|} \left( \bone - \hat\mU_\mrfe \hat\mU_\mrfe\T\right) \pa{\mU_\mrfe}{\tmuu}\pa{\tmuu}{\mq} \,,
\label{e:L2diff}\eqe
where the identity $\partial\norm{\mx}/\partial\mx = \mx/\norm{\mx}$ is used. Matrix $\partial\mU_\mrfe/\partial\tmuu$ follows directly from Eq.~\eqref{e:dUFEdu}. The second part of Eq.~\eqref{e:jacobian} is obtained by differentiating Eq.~\eqref{e:mdortho} w.r.t.~the unknown variables vector
\eqb{l}
	\ds\pa{\omega^2}{\mq} = \pa{\left( \tmuu\T\mK\tmuu\right)}{\mq} = \tmuu\T\pa{\mK}{\mq}\tmuu + 2\tmuu\T\mK\pa{\tmuu}{\mq}\,.
\label{e:MDdwdq1}\eqe
Since $\mq$ contains only nodal density values in the inverse analysis based on modal dynamics, $\partial\mK / \partial\mq = \bf0$. Finally, by using the chain rule, one obtains
\eqb{l}
    \ds\pa{\omega}{\mq} = \omega^{-1}\,\tmuu\T\mK\pa{\tmuu}{\mq} \,.
\label{e:MDdwdq2}\eqe

Eqs.~\eqref{e:L2diff}, \eqref{e:MDdwdq1}, and \eqref{e:MDdwdq2} require $\partial\tmuu/\partial\mq$. For the nodes at the Dirichlet boundary, $\tmuu$ is prescribed independently of $\mq$ and thus, $\partial\tmuu/\partial\mq$ is zero. For the free nodes, $\partial\tmuu/\partial\mq$ follows from linear eigenvalue problem~\eqref{e:eigprobl}. Therefore, each normal mode must satisfy
\eqb{l}
	\mf(\tmuu(\mq),\omega(\mq),\mq) = 
	\overbrace{-\omega^2\mM\tmuu}^{\mf_\mrin} +
	\overbrace{\vphantom{\omega^2} \mK\tmuu}^{\mf_\mrint}
	= \bf0 \,.
\label{e:eqMD2}\eqe 
Differentiation of Eq.~\eqref{e:eqMD2} w.r.t.~the design vector $\mq$ leads to 
\eqb{l}
	\ds\pad{\mf}{\mq} = 	\pa{\mf_\mrin}{\mq} + 
	\pa{\mf_\mrin}{\tmuu}\pa{\tmuu}{\mq} +
	\pa{\mf_\mrin}{(\omega^2)}\pa{\omega^2}{\mq}+
	\pa{\mf_\mrint}{\mq} +
	\pa{\mf_\mrint}{\tmuu}\pa{\tmuu}{\mq} +
	\pa{\mf_\mrint}{(\omega^2)}\pa{\omega^2}{\mq}
	= \bf0 \, .
\label{e:MDdgdq}\eqe
Since $\mf_\mrint$ does not depend on $\omega$ explicitly, the last term of Eq.~\eqref{e:MDdgdq} is always zero. For modal dynamics, it is more convenient to rewrite \eqref{e:MDdgdq} in terms of $\mM$ and $\mK$. Consequently,  
\eqb{l}
	\ds\pa{\mf_\mrin}{\tmuu} = 
	- \omega^2\mM \; , \qquad
	\ds\pa{\mf_\mrin}{(\omega^2)} = 
	- \mM\tmuu \,.
\label{e:dgdu_dgdw}\eqe
Substituting Eqs.~\eqref{e:dgdu_dgdw} and~\eqref{e:MDdwdq1} to \eqref{e:MDdgdq}, one has 
\eqb{l}
\ds\pa{\mf_\mrin}{\mq} - \omega^2\mM\pa{\tmuu}{\mq} - \mM\tmuu\left(\tmuu\T\pa{\mf_\mrint}{\mq} + 2\tmuu\T\mK\pa{\tmuu}{\mq}\right) +
\pa{\mf_\mrint}{\mq} + \mK\pa{\tmuu}{\mq} = \bf0 \, ,
\eqe
which after rewriting gives
\eqb{l}
\ds\frac{\partial\tmuu}{\partial\mq} = \left(\omega^2\mM + 2\mM\tmuu\,\tmuu\T\mK - \mK\right)^{-1}
\left(\frac{\partial\mf_\mrin}{\partial\mq} - \mM\tmuu\,\tmuu\T\frac{\partial\mf_\mrint}{\partial\mq} + \frac{\partial\mf_\mrint}{\partial\mq}\right) \,,
\eqe
where $\partial\mf_\mrin / \partial\mq$ and $\partial\mf_\mrint / \partial\mq$ are the global, inertial and internal sensitivity matrices $\mS_\bullet$ for an arbitrary normal mode, respectively. If the density is the only parameter to identify, $\partial\mf_\mrint/\partial\mq = \bf0$. Hence
\eqb{l}
\ds\pa{\tmuu}{\mq} = \left(\omega^2\mM + 2\mM\tmuu\,\tmuu\T\mK - \mK\right)^{-1} \pa{\mf_\mrin}{\mq} \,,
\label{e:dutdq}\eqe
which is a formula used later in Sec.~\ref{s:Nex}. It is noted that alternative formulas for the derivatives of eigenvalues and eigenvectors w.r.t.~design variables exists, see e.g.~\citet{Fox1968}. The present approach is adopted for consistency and convenience.

\subsection{Analytical sensitivities}\label{s:sensitivities}

In order to calculate $\partial\muu/\partial\mq$ and $\partial\tmuu_i/\partial\mq$, the derivatives of the FE force vectors w.r.t.~the global design vector $\mq$ are needed. The contribution from an FE to the global \textit{sensitivity matrix} $\mS_\bullet$ is defined as 
\eqb{l}
\mS^{e\bre}_\bullet := \ds\pa{\mf^e_\bullet}{\mq_{\bre}} \,.
\label{e:sens}\eqe
Consequently, the internal force increment due to a change of the nodal values of $\Astf$ and $\Cstf$ is given by
\eqb{l}
\Delta\mf^e_\mrint = \ds\pa{\mf^e_{\mrint N}}{\mE\!\mA_{\bre}}\Delta\mE\!\mA_{\bre} + 
\pa{\mf^e_{\mrint M}}{\mE\mI_{\bre}}\Delta\mE\mI_{\bre} = 
\mS_{\Astf}^{e\bre}\Delta\mE\!\mA_{\bre} + \mS_{\Cstf}^{e\bre}\Delta\mE\mI_{\bre} \,,
\eqe
where 
\eqb{l}
\mS_{\Astf}^{e\bre} := \ds\intooe \mN_{e,1}\T\eps^{11}\ba_1 \, \bar\mN_{\bre} \,\dif L \,,
\label{e:sensA}\eqe
and
\eqb{l}
\mS_{\Cstf}^{e\bre} := \ds\intooe \mN_{e;11}\T\kappa^{11}\bn \, \bar\mN_{\bre} \,\dif L \,,
\label{e:sensC}\eqe
are, respectively, elemental axial and bending stiffness sensitivities, which follow from Eqs.~\eqref{e:forceAC}, \eqref{e:finte}, \eqref{e:qdiscr}, and \eqref{e:sens}. Subsequently, given the change of nodal values of density $\rho_0$\footnote{For convenience, $\rho_0$ will be simply denotes as $\rho$ later.} the inertial force increment associated with the $i^{\text{th}}$ eigenvector is
\eqb{l} 
\Delta\mf_{\mrin\,i} = \ds\pa{\mf_{\mrin\,i}}{\brho}\Delta\brho  = \mS_{\rho\,i}\Delta\brho  \,,
\eqe
where 
\eqb{l}
\mS_{\rho\,i} := -\ds\pa{(\omega^2_i\mM\tmuu_i)}{\brho} =  -\omega^2_i\mZ\,\tmuu_i \,,
\label{e:sensD}\eqe
is the global density sensitivity matrix for the $i^{\text{th}}$ eigenvector, and $\mZ := \partial\mM / \partial\brho$ is a 3-dimensional structure, which can be assembled from the $n_\mrel$ elemental contributions\footnote{Index notation is used here, implying summation over repeated indices.}\addtocounter{footnote}{-1}\addtocounter{Hfootnote}{-1}
\eqb{l}
[\mZ_{e\bre}]_{ijk} =  B\ds\intooe N_{mi}N_{mj}\,\brN_k \,\dif L \,, \quad i,j = 1,2,\dots,d\,n_{\mre} \quad m = 1,\dots,d \quad k = 1,2,\dots,n_{\bre} \,,
\label{e:sensZ}\eqe 
where $N_{mi}$ and $\brN_k$ are the components of the shape function matrices $\mN_e$ and $\mbN_{\bre}$, respectively. Eqs.~\eqref{e:sensD} and \eqref{e:sensZ} results from Eqs.~\eqref{e:fine1}, \eqref{e:eqMD}, \eqref{e:eigprobl}, \eqref{e:qdiscr}, and \eqref{e:sens}. The contraction in Eq.~\eqref{e:sensD} follows the rule $[\mZ\tmuu]_{ik} = Z_{ijk}\tilde u_j$\footnotemark. $\mZ$ depends only on the geometric properties of the body; thus, it can be precalculated once in the reference configuration, saving computational time\footnote{For the implementation of 3-dimensional sparse matrices used here, see \url{https://www.mathworks.com/matlabcentral/fileexchange/29832-n-dimensional-sparse-arrays}, retrieved April 3, 2025.}.

$\mS^{e\bre}_\bullet$ is of size $6\times2$ and $6\times1$ for linear and constant Lagrange material shape function, respectively. Similarly, $\mZ_{e\bre}$ is of size $6\times6\times2$ and $6\times6\times1$. They require numerical integration over element $\Omega_0^e$ and following assembly for all $e=1,2,\dots,n_\mrel$ and $\bre = 1,2,\dots,\brn_\mrel$, the outcome of which is the global sensitivity matrix $\mS_\bullet$ of size $d\,n_\mrno\times\bar d \,\bar n_\mrno$.

\vspace{-\baselineskip}
\section{Numerical examples}\label{s:Nex}

This section presents three independent numerical examples focusing on different aspects of the proposed framework. Each successive example is increasing the complexity of the underlying inverse problem, enabling a step-by-step verification of the proposed approach. The first example, uniaxial tension of a straight bar in Sec.~\ref{s:Ex1}, concentrates on identifying the axial stiffness field, followed by its density reconstruction based on axial vibrations. The second example in Sec.~\ref{s:Ex2} demonstrates the identification of bending stiffness for a beam under gravitational load, followed by the density reconstruction from bending vibrations. The last example, a curved beam in Sec.~\ref{s:Ex3}, involves coupled identification of $\Astf$ and $\Cstf$ with subsequent density reconstruction from bending vibrations. In addition, each example includes a distinct individual analysis: Sec.~\ref{s:Ex1_stat} presents a short sensitivity study; Sec.~\ref{s:Ex2_MD} applies regularization to prevent overfitting in the inverse problem; and Sec.~\ref{s:Ex3} demonstrates the flexibility of the proposed framework by adopting a hybrid approach to alleviate membrane locking. All cases in Secs.~\ref{s:Ex1}--\ref{s:Ex3} use quadratic NURBS for the FE discretization~(except in Sec.~\ref{s:Ex3_stat}) and constant or linear Lagrange polynomials for the material mesh. In Sec.~\ref{s:BsplineMatMesh}, the performance of Lagrange material mesh is compared with that of NURBS material mesh for selected cases. The examples are normalized with $L$, $F$, and $m$ representing an unspecified length, force, and mass scale. For all examples, the out-of-plane DOFs are fixed, as the considered structures are planar. Examined are the errors 
\eqb{l}
e_u := \ds\frac{\norm{\muu_\mrexc - \muu_\mrfe}}{\norm{\muu_\mrexc}} \,,\quad e_\omega := \ds\frac{|\omega_\mrexc - \omega_\mrfe|}{\omega_\mrexc} \,,
\label{e:FEerr}\eqe 
that represent the discrete $L^2$ error of $\muu_\mrfe$ for quasi-static cases, and the relative error of the $i^{\text{th}}$ natural frequency in modal dynamics. In Eq.~\eqref{e:FEerr}, $\muu_\mrect$ and $\omega_\mrect$ are FE reference solutions for a highly refined mesh. Synthetic experimental data for the displacements, normal modes, and frequencies are generated using a very dense FE mesh and reference distributions of $\Astf$, $\Cstf$, and $\rho$. For additional verification, selected cases from Secs.~\ref{s:Ex1_stat}~and~\ref{s:Ex2_stat} are analyzed using quasi-experimental data from 3D bulk FE models, as shown in App.~\ref{s:3D_ver}. Measurements inaccuracies are introduced by the component-wise relative noise:
\eqb{l}
u^\mrex_{Ii} = u_{\mrexc\,i}(\bx^\mrex_I)(1+\gamma_{Ii}) \,,
\label{e:noise}\eqe
where $i=1,2,3$ are the Cartesian components, $u_{\mrect\,i}$ is the reference solution, and $\gamma_{Ii}$ follows a normal distribution with zero mean and standard deviation up to 0.04, which is subsequently referred to as 4\% noise. Frequencies are modified with noise analogously. The relative errors of the identified material parameters are defined as
\eqb{l}
\delta_I := \ds\left| \frac{q_{I,\mrrf} - q_{I,\mrop}}{q_{I,\mrrf}} \right| \,, \quad I=1,\dots,n_\mrvr \,, 
\label{e:error}\eqe
where $q_{I,\mrrf}$ are the reference values of the material parameters and $q_{I,\mrop}$ are the optimal values found from \eqref{e:minf}. For cases with random noise, computations are repeated at least 25 times to analyze the statistics. Hence, the errors in tables report their mean and standard deviation (mean $\pm$ std).

Nonlinear least-squares problems usually have multiple solutions. Since this work is restricted to local optimization, only a local optimum can be found. To mitigate multimodality of \eqref{e:minf}, design variables are bounded, and the initial guess is a random vector between the bounds. In addition, computations are repeated several times. Our preliminary studies indicated that the considered problems are insensitive to the initial guess. Therefore, only results for fixed initial guess are reported\footnote{The only relevant difference between random and fixed initial guess is the number of iterations needed for the optimization algorithm to converge. It is approximately two times larger for a random initial guess.}. A tolerance of $\epsilon = 10^{-6}$ is used for \eqref{e:qconv} and \eqref{e:fconv}. A smaller $\epsilon$ usually leads to longer computations without actual improvement of the solution. In the following examples, the number of load levels $n_\mrll$ typically matches $n_\mrlc$ in \eqref{e:objfnc}; any deviations from this are noted. 

\subsection{Bar under uniaxial deformation}\label{s:Ex1}

In the first example, the axial stiffness $\Astf$ of a straight bar is first reconstructed based on uniaxial tension. Then, the density is identified based on longitudinal vibrations using the previously calculated $\Astf(\xi)$. Only axial deformations are considered.

\subsubsection{Axial stiffness reconstruction from statics}\label{s:Ex1_stat}

\begin{figure}[h]
	\begin{center} \unitlength1cm
		\begin{picture}(0,9.5)
			\put(-7,6.5){\includegraphics[height=30mm]{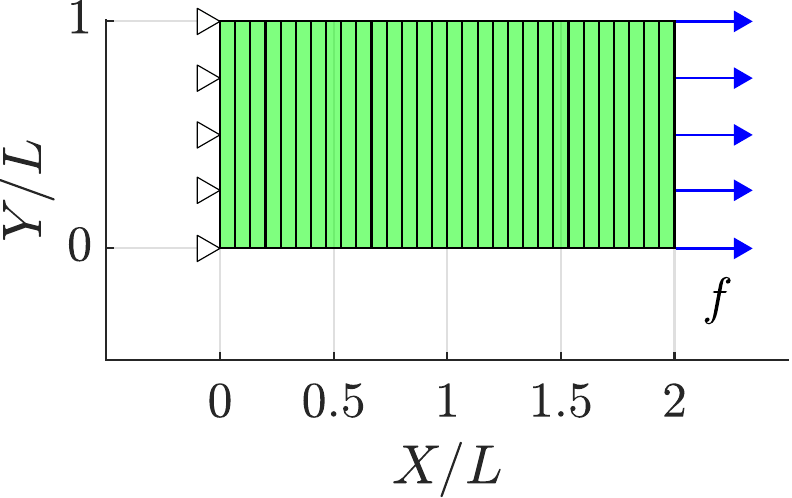}} 						
			\put(-1.5,6.5){\includegraphics[height=30mm]{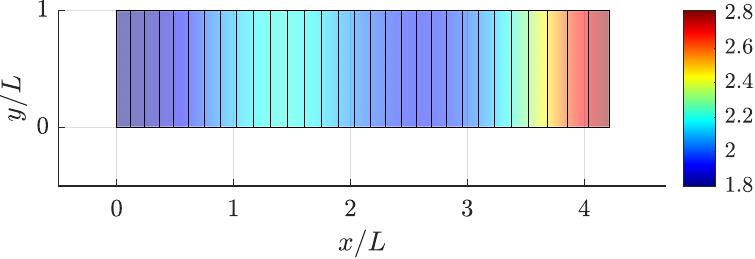}}  						
			\put(-7,0.75){\includegraphics[height=48mm]{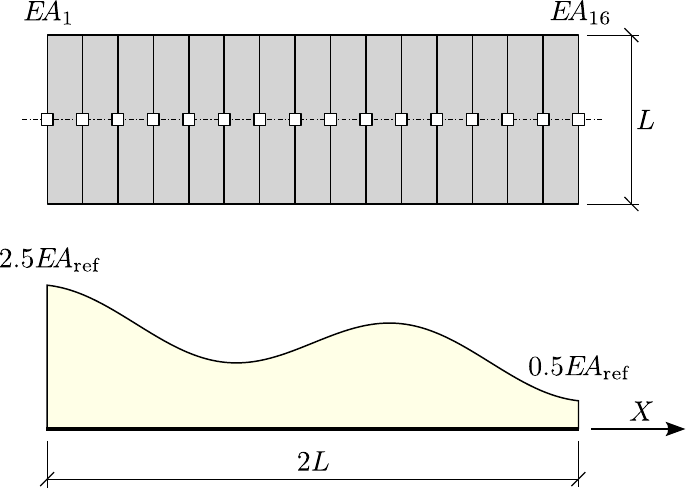}} 	 
			\put(0.75,0){\includegraphics[height=60mm]{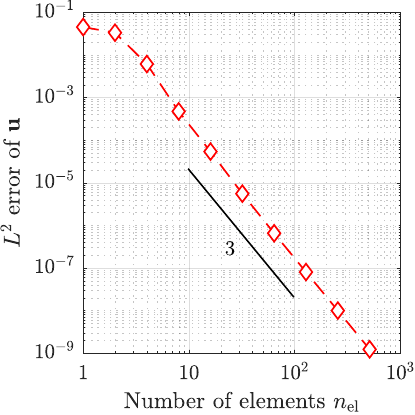}} 					
			\put(-7.35,6.5){\small{a.}}
			\put(-1.5,6.5){\small{b.}}
			\put(-7.35,0.75){\small{c.}}
			\put(0.25,0.75){\small{d.}}
		\end{picture}
		\caption{Uniaxial stretching of a bar: a.~undeformed configuration with boundary conditions; b.~deformed configuration, colored by stretch $\lambda$; c.~material mesh with the reference distribution for $\Astf$; d.~FE convergence of the discrete $L^2$ error w.r.t.~the FE solution for 1024 elements.}
		\label{fig:Ex1_stat}	
	\end{center}
\end{figure}

\paragraph{Problem setup}
The straight bar with length $L_x = 2L$, width $B = L$, and thickness $T = L$, shown in Fig.~\ref{fig:Ex1_stat}a~\&~b is loaded with the point load $P = fBT = 500F$ at the right end ($X=4L$), and fixed at the left end ($X=0$). The chosen reference distribution of the axial stiffness is
\eqb{l}
\Astf(\xi)/\AstfR = \ds 2+0.5\cos(3\pi\xi) - \xi \,,
\label{e:Ex1A}\eqe 
where $\xi = X/2L$ and $\AstfR = 100F$, see Fig.~\ref{fig:Ex1_stat}c. 
In the inverse analysis, between 5 and 30 linear ME are used to approximate this material distribution, leading to $n_\mrvr = \mathinner{\text{6--31}}$ design variables. For the forward FE problem, the optimal convergence ratio, $O(h^3)$, is obtained as Fig.~\ref{fig:Ex1_stat}d shows. One-dimensional synthetic data are generated from 1020 FE, while all inverse analyses are conducted with 30 FE since its $L^2$ error is only $e_u\approx 10^{-5}$. A maximum of four load levels is used in the inverse analysis corresponding to 25\%, 50\%, 75\%, and 100\% of the final load. The lower and upper bounds for $\Astf$ are $0.05\AstfR$ and $5\AstfR$, respectively. The initial guess for $\Astf$ is fixed to $0.545\AstfR$.

Real mechanical tests are conducted on 3D specimens that can have varying material parameters over the cross-section, particularly in the case of thick samples. The suitability of 1D beam models therefore needs to be demonstrated, which is done in App.~\ref{s:3D_verEx1}.

\paragraph{Results for the noise-free cases}
Cases 1.1--1.4 in Tab.~\ref{tab:Ex1_stat} show the convergence of the identification errors w.r.t.~the number of ME for experimental data without noise. The average reconstruction error $\errA$ ranges from $11.06\%$ to $0.22\%$ for 5 and 30 ME, respectively. Since a denser material mesh reduces systematic identification errors but increases sensitivity to random noise, a mesh with $15$ ME is chosen as a trade-off for the remainder of the analysis.
\begin{table}[h]
	\centering
	\begin{tabular}{@{}ccccccccc@{}}
		\toprule
		Case & FE                & mat.                    & exp.                               & load              & noise & ave. iter.& $\errM$               & $\errA$               \\
		     & $n_\mrel$         & $\bar n_\mrel$          & $n_\mrex/n_\mrll$                  & $n_\mrll$         & [\%]  &           & [\%]                  & [\%]                  \\ \midrule
		1.1  & 30                & 5                       & 1000                               & 1                 & 0     & 11        & 31.00                 & 11.06                 \\
		1.2  & 30                & 10                      & 1000                               & 1                 & 0     & 11        & 9.64                  & 2.37                  \\
		1.3  & 30                & 15                      & 1000                               & 1                 & 0     & 11        & 4.39                  & 0.99                  \\
		1.4  & 30                & 30                      & 1000                               & 1                 & 0     & 11        & 1.08                  & 0.22                  \\ \midrule
		1.5  & 30                & 15                      & 1000                               & 1                 & 1     & 12        & $21.34\,\pm\,11.00$   & $4.85\,\pm\,1.49$     \\
		1.6  & 30                & 15                      & 4000                               & 1                 & 1     & 11        & $9.53\,\pm\,4.55$     & $2.30\,\pm\,0.50$     \\
		1.7  & 30                & 15                      & 1000                               & 4                 & 1     & 12        & $9.37\,\pm\,5.41$     & $1.98\,\pm\,0.68$     \\ 
		1.8  & 30                & 15                      & 4000                               & 4                 & 1     & 11        & $5.93\,\pm\,2.12$     & $1.32\,\pm\,0.23$     \\ \midrule
		1.9  & 30                & 15                      & 4000                               & 4                 & 2     & 12        & $8.61\,\pm\,4.04$     & $2.01\,\pm\,0.56$     \\
		1.10 & 30                & 15                      & 4000                               & 4                 & 4     & 13        & $15.04\,\pm\,8.98$    & $3.45\,\pm\,1.04$     \\ 	
		1.11 & 30    		     & 15					   & 2000								& 4                 & 4     & 13 		& $11.79\,\pm\,4.78$    & $4.42\,\pm\,1.05$     \\ 
		\bottomrule
	\end{tabular}
	\caption{Uniaxial stretching of a bar: Studied stiffness reconstruction cases with their FE and material mesh, experimental grid resolution, load levels, noise, average number of iterations, and errors $\errA$, $\errM$. For Case 1.11, $n_\mrlc = 2\times n_\mrll$.}
	\label{tab:Ex1_stat}
\end{table}

\begin{figure}[h]
	\begin{center} \unitlength1cm
		\begin{picture}(0,5)
			\put(-6.85,0.25){\includegraphics[height=45mm]{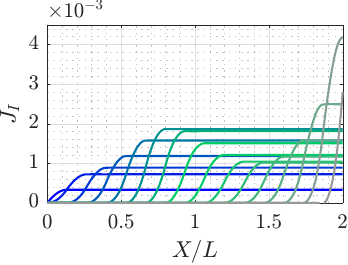}} 					
			\put(1.25,0){\includegraphics[height=50mm]{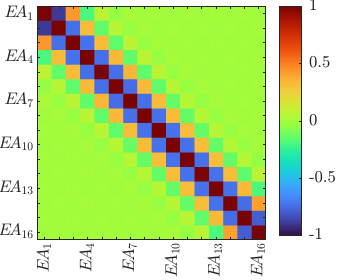}}  						
			\put(-6.85,0){\small{a.}}
			\put(1,0){\small{b.}}
		\end{picture}
		\caption{Uniaxial stretching of a bar, Case 1.3: a.~columns of the Jacobian, $\mJ_{I}=\partial \bar\mU_\mrR / \partial \Astf_{I}$, at the optimal $\mE\!\mA$, plotted on the experimental grid. The values are normalized by $\AstfR$. Colors ranging from blue, through green, to gray correspond to $\Astf_1$--$\Astf_{16}$; b.~correlation matrix for the optimal $\mE\!\mA$, derived from the covariance approximation $(\mJ\T\mJ)^{-1}$, see~\citet{Hansen2013}. For the sake of the correlation matrix, measurement errors are assumed to be uncorrelated, uniform, and Gaussian.}
		\label{fig:Ex1_sens}	
	\end{center}
\end{figure}
One way to assess the sensitivity of the FE solution to the unknown parameters $\mq$ is to analyze the columns of the Jacobian $\mJ_{I}=\partial\bar\mU_\mrR / \partial q_{I}$~\citep{Chen2024}. Generally, a greater sensitivity of $f(\mq)$ to a parameter implies its better identifiability. As Fig.~\ref{fig:Ex1_sens}a shows, the leftmost material nodes affect almost the entire bar response, whereas the rightmost nodes affect only their vicinity. Particular attention should be paid to both ends, where the Jacobian is either small or non-zero only locally. The correlation matrix in Fig.~\ref{fig:Ex1_sens}b shows a banded, oscillatory pattern. Correlations are strongest for physically adjacent material nodes and decay with distance, dropping below $\pm0.06$ beyond four nodes (green color). Slightly larger correlations occur at the bar ends. Denser material meshes exhibit similar behavior~(not shown). 

\paragraph{Influence of noise on results}
Cases 1.5--1.7 in Tab.~\ref{tab:Ex1_stat} examine the influence of the experimental grid density on the reconstruction accuracy in the presence of noise. As expected, a finer grid reduces $\errA$ and $\errM$. Increasing the number of load levels to four while keeping the grid fixed provides also lower errors. This suggests that a dense experimental grid can be substituted with more load levels, which can be helpful when high-resolution measurements are unavailable. 

\begin{figure}[h]
	\begin{center} \unitlength1cm
		\begin{picture}(0,4.5)
			\put(-7,0){\includegraphics[height=45mm]{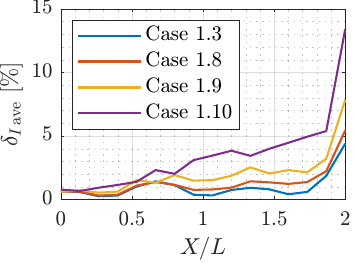}} 					
			\put(0.5,0){\includegraphics[height=45mm]{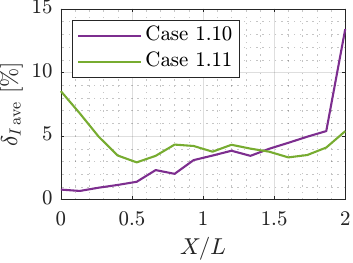}}  						
			\put(-7,0.2){\small{a.}}
			\put(0.5,0.2){\small{b.}}
		\end{picture}
		\caption{Uniaxial stretching of a bar: a.~mean identification error distribution for Cases 1.3~\&~1.8--1.10 (noise 0--4\%). b.~mean identification error distribution for Cases 1.10~\&~1.11 (noise 4\%, various boundary conditions). As seen, the error increases non-uniformly with noise (a.). Combining different boundary conditions reduces the error only on the right side (b.).}
		\label{fig:Ex1_err}	
	\end{center}
\end{figure}

As illustrated in Fig.~\ref{fig:Ex1_err}a, the identification error increases toward the right end of the bar with a peak where the force $P$ is applied. This error peak is primarily induced by the material mesh inexactly capturing $\Astf(\xi)$, as shown in the differences between Cases~1.8--1.10 and Case~1.3 in Fig.~\ref{fig:Ex1_err}a. In contrast, the error at the left end remains unaffected by the noise level, despite the low sensitivity of $f(\mq)$ to $\Astf_1$ (Fig.~\ref{fig:Ex1_sens}a). This discrepancy likely arises from the noise profile relative to the measured displacements (see Eq.~\eqref{e:noise}).

In Case 1.11, data from two different experiments (independent of Case 1.10) are combined to reduce error growth along the bar and its characteristic peak ($n_\mrll = 4$, $n_\mrlc = 2\times4$). These experiments include the one shown in Fig.~\ref{fig:Ex1_stat} and another with the point force and fixation swapped. As shown in Fig.~\ref{fig:Ex1_err}b, this objective is only partially achieved: the maximum error is reduced, but a considerable error is introduced on the left side, likely due to a challenging material distribution. As a result, the average error increases to $\errA=4.42\pm1.05\%$, compared with $\errA=3.45\pm1.04\%$ for Case 1.10. In all cases in Tab.~\ref{tab:Ex1_stat}, the inverse algorithm requires 11--13 iterations, indicating that noise has little effect on optimization convergence.

\subsubsection{Density reconstruction from modal dynamics}\label{s:Ex1_MD}

\begin{figure}[p]
	\begin{center} \unitlength1cm
		\begin{picture}(0,11.2)
			\put(-7.5,5.2){\includegraphics[height=60mm]{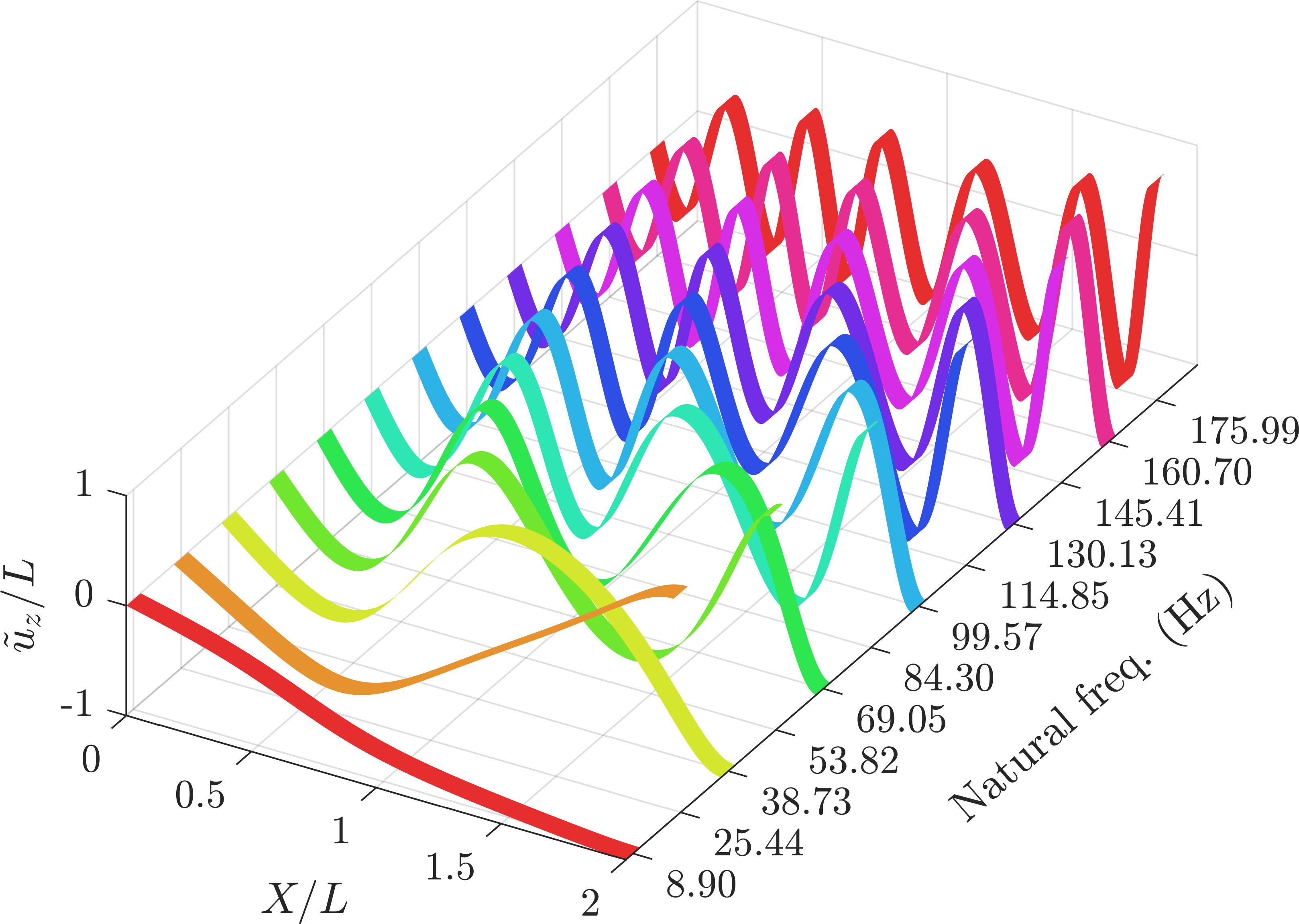}} 						
			\put(2,5.2){\includegraphics[height=60mm]{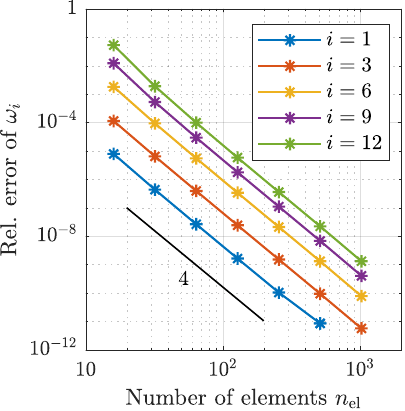}} 						
			\put(-6.35,0){\includegraphics[height=44mm]{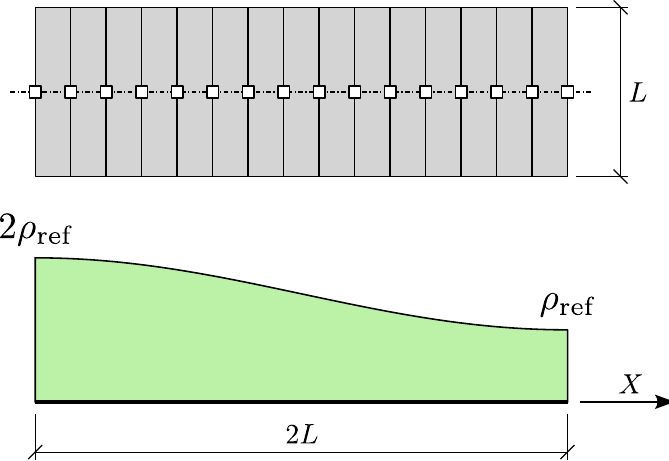}} 
			\put(2.05,0){\includegraphics[height=44mm]{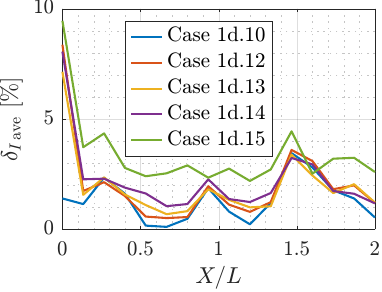}} 	
			\put(-7.5,5.2){\small{a.}}
			\put(2,5.2){\small{b.}}
			\put(-6.75,0){\small{c.}}
			\put(2.05,0){\small{d.}}
		\end{picture}
		\caption{Uniaxial vibrations of a bar: a.~the first 12 axial modes with corresponding $\omega$; the modes are normalized, so that $\max(\mU_\mrfe)=1$ (note that here the $Z$-axis shows the longitudinal displacements); b.~FE convergence of the $i^{\text{th}}$ natural frequency w.r.t.~the FE solution for 2048 elements; c.~material mesh with the reference density distribution. d.~average relative error distribution for Cases 1d.10 and 1d.12--1d.15.}
		\label{fig:Ex1_MD}	
	\end{center}
\end{figure}

\paragraph{Problem setup}
For the same bar, up to the first 12 axial modes (Fig.~\ref{fig:Ex1_MD}a) are used to reconstruct the unknown density field. The bar is assumed to be unloaded and stress-free. 
\begin{table}[p]
	\centering
	\begin{tabular}{@{}cccccccccc@{}}
		\toprule
		Case  & FE   	    & mat.			  & exp.              & modes              & stiffness	  & noise   & iter.     & $\errM$ & $\errA$ \\
		& $n_\mrel$ 	& $\bar n_\mrel$  & $n_\mrex/n_\mrmd$ & $n_\mrmd$          &     		  &         &           & [\%]    & [\%]                  \\ \midrule
		1d.1  & 15   		& 15		      & 100               & 1                  & ref.         & 0       & 9         & 15.81   & 4.33                  \\
		1d.2  & 30   		& 15		      & 100               & 1                  & ref.         & 0     	& 7         & 0.61    & 0.13                  \\
		1d.3  & 15   		& 15			  & 100               & 2                  & ref.         & 0     	& 8         & 8.27    & 3.17                  \\
		1d.4  & 15   		& 15		      & 100               & 3                  & ref.         & 0     	& 9         & 3.39    & 1.62                  \\
		1d.5  & 15   		& 15			  & 100               & 6                  & ref.         & 0     	& 8         & 17.84   & 3.64                  \\ \midrule
		1d.6  & 30   		& 15			  & 100               & 1                  & reconst.     & 0     	& 13        & 169.51  & 35.06                 \\
		1d.7  & 30   		& 15		      & 400               & 1                  & reconst.     & 0     	& 13        & 169.42  & 35.08                 \\
		1d.8  & 60   		& 15			  & 100               & 3                  & reconst.     & 0     	& 12        & 67.31   & 14.81                 \\
		1d.9  & 120  		& 15			  & 100               & 6                  & reconst.     & 0     	& 7         & 4.45    & 1.99                  \\
		1d.10 & 210  		& 15			  & 100               & 9                  & reconst.     & 0     	& 7         & 3.45    & 1.33                  \\ 
		1d.11 & 240  		& 15			  & 100               & 12                 & reconst.     & 0     	& 7         & 6.85    & 1.53                  \\ \bottomrule
	\end{tabular}
	\caption{Uniaxial vibrations of a bar: Cases of density reconstruction with their FE and material mesh, experimental grid resolutions, number of axial modes, type of stiffness distribution (\textit{ref.}~for exact, \textit{reconst.}~for \eqref{e:Ex1Aexp}), noise level, number of iterations, and errors $\errA$, $\errM$.}
	\label{tab:Ex1_MD}
\end{table}
The chosen density distribution is shown in Fig.~\ref{fig:Ex1_MD}c, and defined by
\eqb{l}
\rho(\xi)/\rho_\mrrf = \ds 1.5 + 0.5\cos(\pi\xi) \,,
\label{e:Ex1Rho}\eqe
where $\xi=X/2L$ and $\rhoR = 1 m/L$. In the following cases, the reconstructed stiffness $\Astf(\xi)$ is taken from a sample of Case 1.10 in Tab.~\ref{tab:Ex1_stat}, and is defined by the vector
\eqb{lll}
\mathbf{EA} &= [250.729,\,236.307,\,199.198,\,166.757,\,131.319,\,114.175,\,116.943,\,139.799,\\[2mm]
&\phantom{= [} 160.411,\,180.028,\,188.145,\,164.143,\,143.795,\,97.369,\,64.046,\,50.645] F \,,
\label{e:Ex1Aexp}\eqe
which yields $\errA = 1.75\%$ and $\errM = 6.16\%$ w.r.t.~the exact values from Eq.~\eqref{e:Ex1A}. For the density field, 15 linear ME are chosen, which gives $n_\mrvr = 16$. The convergence study in Fig.~\ref{fig:Ex1_MD}b yields the ideal convergence rate for axial modes, $O(h^4)$~\citep{Cottrell2006}. Based on this, 4090 FE are chosen for the synthetic data generation. Correspondingly, 15--240 FE are used for the inverse analysis, ensuring similar errors for all frequencies. The lower and upper bounds for $\rho$ are $0.1\rhoR$ and $10\rhoR$, respectively. The initial guess is taken as $1.09\rhoR$.

\paragraph{Results for exact and inexact axial stiffness}
Cases 1d.1--1d.5 in Tab.~\ref{tab:Ex1_MD} present results based on the exact axial stiffness $\Astf(\xi)$. In Cases 1d.1 and 1d.3--1d.5, the FE mesh is fixed while the number of modes increases. The errors decreases initially from $\errA = 4.33\%$ (Case 1d.1) to $\errA = 1.62\%$ (Case 1d.4). However, the error rises for 6 modes (Case 1d.5). This occurs because higher modes have larger FE errors (see Fig.~\ref{fig:Ex1_MD}b), increasing discrepancies between the forward and the inverse FE solver. Therefore, the FE mesh resolution should be adjusted to the highest mode used, which is done for the remaining cases. Based on Cases 1d.1 and 1d.2, at least 30 FE are chosen for Cases 1d.6--1d.11. 

Cases 1d.6--1d.11 are based on inexact $\Astf(\xi)$ from \eqref{e:Ex1Aexp}. For Case 1d.7, a denser experimental grid w.r.t.~Case 1d.6 does not improve results, as it cannot compensate for the errors in $\Astf(\xi)$. However, increasing the number of modes reduces $\errA$ from $35.06\%$ (1st mode) to $1.33\%$ (the first 9 modes). Nevertheless, the errors rise beyond this point. A possible explanation for this emerges in Sec.~\ref{s:Ex2_MD}. The error distribution for Case 1d.10 is shown in Fig.~\ref{fig:Ex1_MD}d.  

\paragraph{Influence of noise on results}
Tab.~\ref{tab:Ex1_MD_noise} analyses the impact of noise on Case 1d.10. Up to 4\% noise is added to the natural frequencies, while axial modes are always perturbed with 4\% noise. Random noise in the modes has a moderate effect on the average identification error ($\Delta\errA = 0.68\pm0.46\%$ between Cases 1d.10 and 1d.12, $\approx 50\%$ increase), but a significant effect on the maximum error ($\Delta\errM = 5.48\pm6.30\%$, $\approx 150\%$ increase). Noisy frequencies notably increase average error, particularly at 4\% noise level. As shown in Fig.~\ref{fig:Ex1_MD}d, noise in the modes introduces an error peak at the leftmost material node, where density has the least influence on the forward solution. For 4\% of noise in the frequencies (Case 1d.15), the error distribution is amplified and flattened.
\begin{table}[h]
	\centering
	\begin{tabular}{@{}ccccccccc@{}}
		\toprule
		Case   & FE                & exp.                               & modes               & stiffness     & noise    & ave. iter.& $\errM$              & $\errA$               \\
		& $n_\mrel$         & $n_\mrex/n_\mrmd$                  & $n_\mrmd$           &               & {[}\%{]} &           & {[}\%{]}             & {[}\%{]}              \\ \midrule
		1d.12  & 210               & 100                                & 9                   & reconst.      & [4,0]    & 7         & $8.93\,\pm\,6.30$    & $2.01\,\pm\,0.46$     \\ 
		1d.13  & 210               & 100                                & 9                   & reconst.      & [4,1]    & 7         & $7.76\,\pm\,4.81$    & $1.94\,\pm\,0.43$     \\
		1d.14  & 210               & 100                                & 9                   & reconst.      & [4,2]    & 7         & $8.92\,\pm\,5.56$    & $2.22\,\pm\,0.59$     \\ 
		1d.15  & 210               & 100                                & 9                   & reconst.      & [4,4]    & 8         & $10.62\,\pm\,5.89$   & $3.38\,\pm\,1.05$     \\ \bottomrule
	\end{tabular}
	\caption{Uniaxial vibrations of a bar: Cases of density reconstruction with their FE mesh, experimental grid resolution, number of axial modes, type of axial stiffness distribution (\textit{ref.}~for exact, \textit{reconst.}~for \eqref{e:Ex1Aexp}), noise level, number of iterations, identification errors $\errA$, $\errM$. In the \textit{noise} column, the values in brackets refer to the noise applied to modes and frequencies, in that order. For all cases $\bar n_\mrel = 15$.}
	\label{tab:Ex1_MD_noise}
\end{table}

The first example comprehensively analyzed the inverse problem for axial stiffness and density identification of a 1D bar. In both cases, the proposed framework delivered satisfactory results, even under significant measurement noise. 

\subsection{Bending of an initially straight beam}\label{s:Ex2} 

In the second example, the bending stiffness $\Cstf$ of an initially straight beam subjected to gravitational loading is reconstructed. For this purpose, synthetic experimental data from different boundary conditions are combined. Subsequently, the density is identified using bending vibrations and the previously determined $\Cstf(\xi)$. Furthermore, a study incorporating regularization for density reconstruction is conducted. 

\subsubsection{Bending stiffness reconstruction from statics}\label{s:Ex2_stat}

\begin{figure}[h]
	\begin{center} \unitlength1cm
		\begin{picture}(0,10.5)
			\put(-8,6.7){\includegraphics[height=35mm]{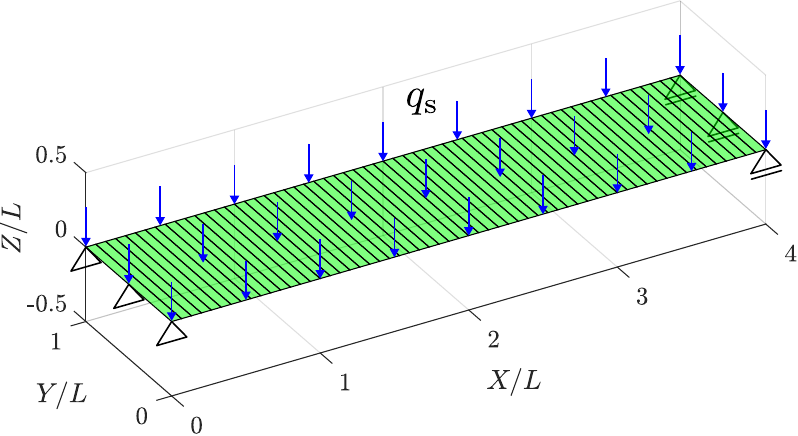}} 						
			\put(0.35,4){\includegraphics[height=60mm]{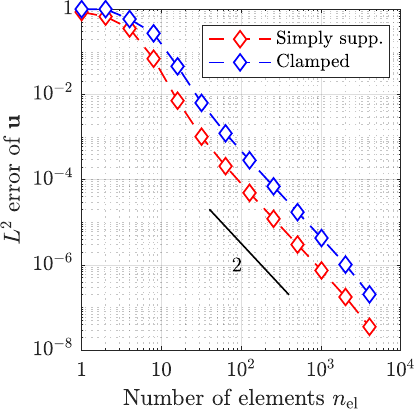}} 					
			\put(-8,0.3){\includegraphics[height=35mm]{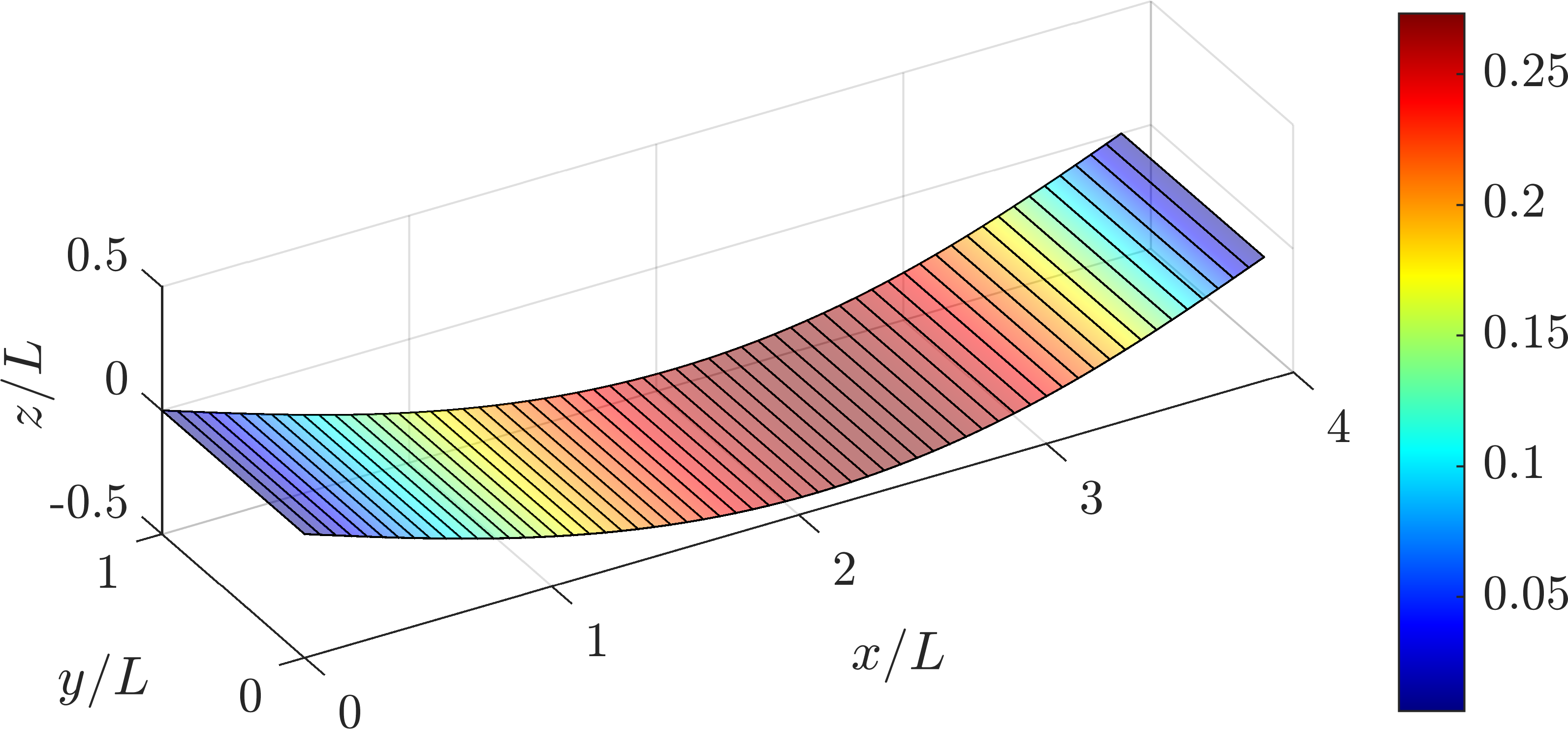}} 						
			\put(-8,3.5){\includegraphics[height=35mm]{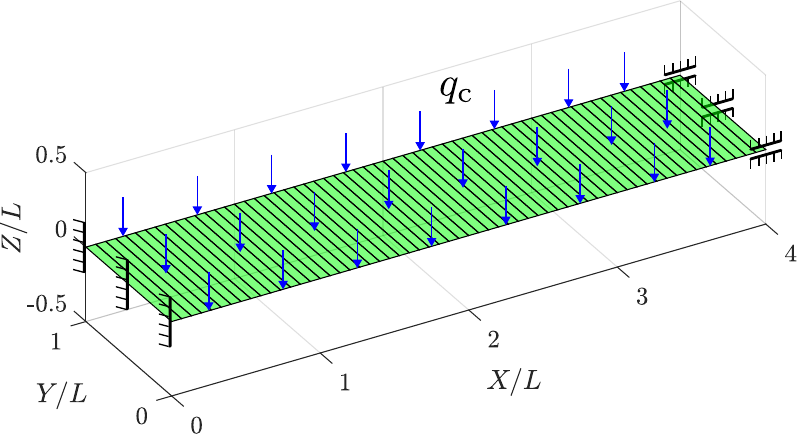}}  						
			\put(0.35,0.3){\includegraphics[height=35mm]{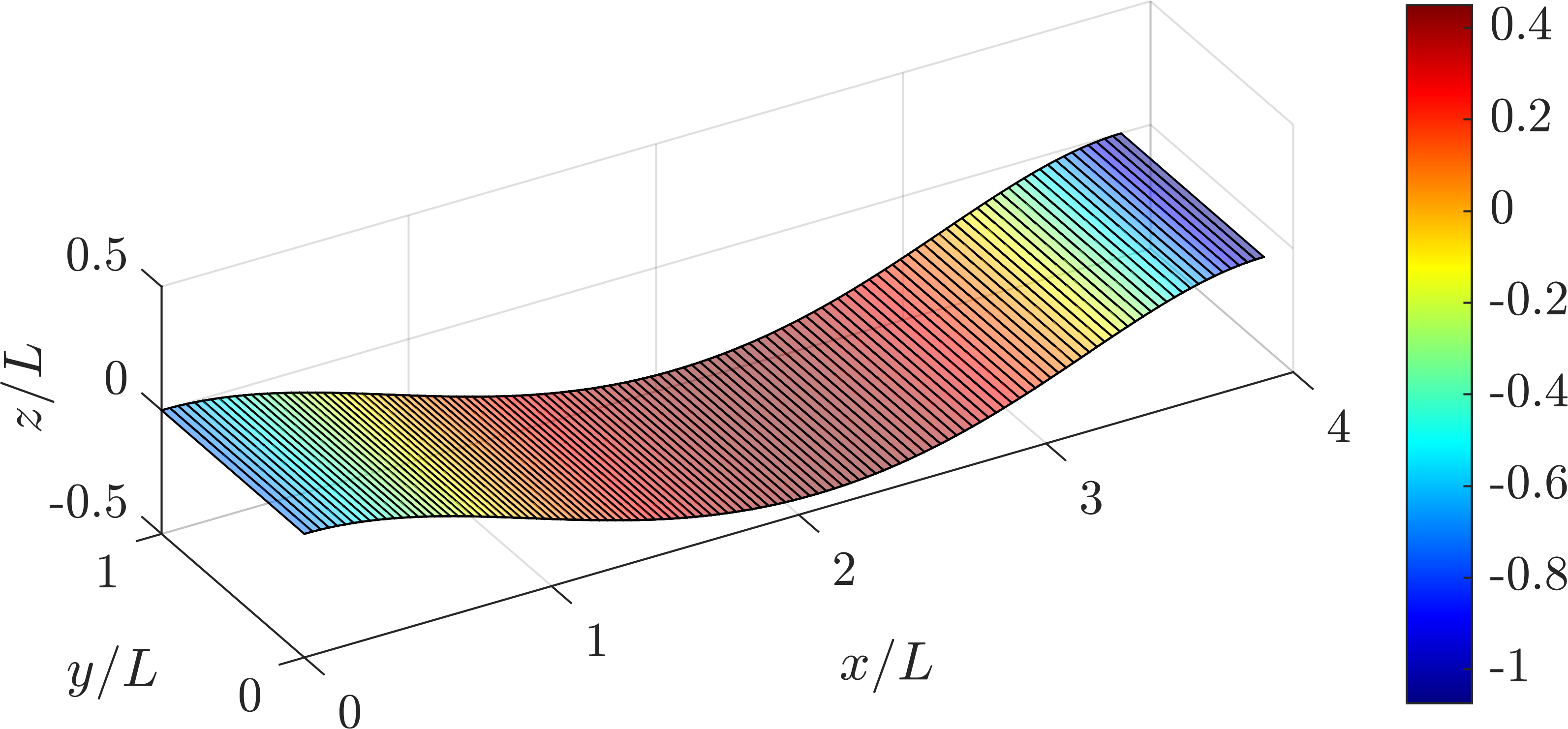}} 					
			\put(-8,6.2){\small{a.}}
			\put(-8,3.3){\small{b.}}
			\put(0.35,4){\small{c.}}
			\put(-8,0){\small{d.}}
			\put(0.35,0){\small{e.}}
			\put(-2.1,3.5){\scriptsize{$[L^{-1}]$}}
			\put(6.3,3.55){\scriptsize{$[L^{-1}]$}}
		\end{picture}
		\caption{Bending of a straight beam: a.~\& b.~undeformed configuration with boundary conditions for the simply supported and clamped beam, respectively; c.~FE convergence of the discrete $L^2$ error w.r.t.~the FE solution for 8192 elements; d.~\&~e.~deformed configuration for the simply supported and clamped beam, respectively, colored by the relative curvature change $\kappa$.}
		\label{fig:Ex2_stat}	
	\end{center}
\end{figure}

\paragraph{Problem setup}
A beam with span $L_x=4L$ and width $B=L$ is loaded with a uniform vertical load on its entire length. Two different boundary conditions are analyzed. For the first one~--~a simply supported beam (Fig.~\ref{fig:Ex2_stat}a~\&~d) under $q_\mrs = 0.002F/L$~--~the left end ($X=0$) is fully fixed, while the right end ($X=4L$) is fixed only in $Z$-~direction. For the second one~--~a clamped beam (Fig.~\ref{fig:Ex2_stat}b~\&~e) under $q_\mrc = 0.01F/L$~--~the rotations at the ends are additionally fixed. The $\Cstf(\xi)$ follows from Eq.~\eqref{e:Ex1Rho} with $\xi = X/4L$ and $\CstfR = 0.01FL^2$ (see Fig.~\ref{fig:Ex1_MD}c for graphical representation), while $\Astf(\xi) = 100F$ is constant. Thickness distribution of the beams can be evaluated from Eqs.~\eqref{e:AI}.
The axial stiffness barely affects the deformation; thus, it is neglected in the identification, and $\Astf$ is treated known. For all cases, 10 linear ME are used, resulting in $n_\mrvr = 11$. The inverse analysis is conducted with a mesh of 60 and 120 FE for the simply supported and clamped beam, respectively, which is accurate up to $e_u\approx3\times10^{-4}$ (see Fig.~\ref{fig:Ex2_stat}c). The synthetic experimental data are generated from 4080 FE with up to four load levels at 25\%, 50\%, 75\%, and 100\% of the final load. The results of the inverse analysis for 3D-~and 1D-based quasi-experimental data are compared in App.~\ref{s:3D_verEx2}, similarly to the first example from Sec.~\ref{s:Ex1_stat}. The lower bound for $\Cstf$ is $0.1\CstfR$, while the upper bound is $10\CstfR$. The initial guess for $\Cstf$ is fixed to $1.09\CstfR$. 

\begin{figure}[h] 
	\begin{center} \unitlength1cm
		\begin{picture}(0,5)
			\put(-7.25,0){\includegraphics[height=45mm]{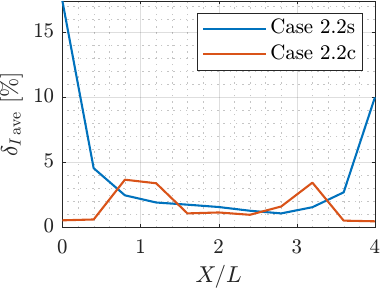}} 					
			\put(0.75,0){\includegraphics[height=45mm]{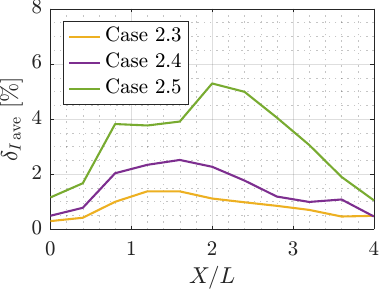}}  						
			\put(-7.25,0.2){\small{a.}}
			\put(0.75,0.2){\small{b.}}
		\end{picture}
		\caption{Bending of a straight beam: a.~average identification error distribution for Cases 2.2s~\&~2.2c; b.~average error distribution for Cases 2.3--2.5.}
		\label{fig:Ex2_Err}	
	\end{center}
\end{figure}

\paragraph{Results}
Cases 2.1s and 2.2s in Tab.~\ref{tab:Ex2_stat} show the results of the inverse analysis for the simply supported beam. Adding 1\% noise to the experimental data leads to $\errA = 4.21\pm1.74\%$ and $\errM = 19.59\pm12.14\%$ in Case 2.2s, even if $4$ load levels are used along with $4000$ measurement points. Cases 2.1c and 2.2c show the analogous analysis for the clamped beam. Even though 1\% noise leads to smaller errors than for the simply supported beam, the errors are still prominent ($\errA = 1.60\pm0.68\%$ and $\errM = 5.29\pm2.82\%$ for Case 2.2c). Fig.~\ref{fig:Ex2_Err}a shows that the error distributions have peaks in characteristic locations, where the curvature of the deformed beam approaches zero\footnote{Or equivalently, the bending moments approach zero.}. This indicates that the deformation is weakly sensitive to bending stiffness in those regions, making $\Cstf(\xi)$ particularly vulnerable to noise.
\begin{table}[h]
	\centering	
	\begin{tabular}{@{}ccccccccc@{}}
		\toprule
		Case & FE                & mat.                    & exp.                               & load              & noise    & ave. iter.& $\errM$               & $\errA$               \\
			 & $n_\mrel$         & $\bar n_\mrel$          & $n_\mrex/n_\mrll$                  & $n_\mrll$         & {[}\%{]} &           & {[}\%{]}              & {[}\%{]}              \\ \midrule
		2.1s & 60                & 10                      & 1000                               & 1                 & 0        & 8         & 0.64                  & 0.22                  \\
  		2.2s & 60                & 10                      & 4000                               & 4                 & 1        & 9         & $19.59\,\pm\,12.14$   & $4.21\,\pm\,1.74$     \\ \midrule
  		2.1c & 120               & 10                      & 1000                               & 1                 & 0        & 10        & 0.53                  & 0.20                  \\
  		2.2c & 120               & 10                      & 4000                               & 4                 & 1        & 9         & $5.29\,\pm\,2.82$     & $1.60\,\pm\,0.68$     \\ \midrule
  		2.3  & [60,120]          & 10                      & 2000                               & 4                 & 1        & 7         & $2.01\,\pm\,0.85$     & $0.82\,\pm\,0.28$     \\
  		2.4  & [60,120]          & 10                      & 2000                               & 4                 & 2        & 7         & $3.81\,\pm\,1.63$     & $1.44\,\pm\,0.68$     \\
  		2.5  & [60,120]          & 10                      & 2000                               & 4                 & 4        & 7         & $7.42\,\pm\,3.59$     & $3.15\,\pm\,1.58$     \\ \bottomrule
	\end{tabular}
	\caption{Bending of a straight beam: Studied stiffness identification cases with their FE and material mesh, experimental grid resolution, load levels, noise, average number of iterations, and errors $\errA$, $\errM$. The double value $[60,120]$ indicates the number of FE used for the simply supported and clamped beams, respectively. For Cases 2.3--2.5, $n_\mrlc = 2\times n_\mrll$.} 
	\label{tab:Ex2_stat}
\end{table}

Combining both boundary conditions in a single inverse analysis ($n_\mrll = 4$, $n_\mrlc = 2\times4$) results in much smaller errors for the same number of experimental points and the same noise (1\% in Case 2.3 in Tab.~\ref{tab:Ex2_stat}). Case 2.4 with 2\% noise yields error levels similar to those of Case 2.2c with 1\% noise. Additionally, the error distributions shown in Fig.~\ref{fig:Ex2_Err}b are more uniform than before, and the peaks are eliminated. 

\subsubsection{Density reconstruction from modal dynamics}\label{s:Ex2_MD}

\begin{figure}[h]
	\begin{center} \unitlength1cm
		\begin{picture}(0,12.8)
			\put(-8,6.8){\includegraphics[height=60mm]{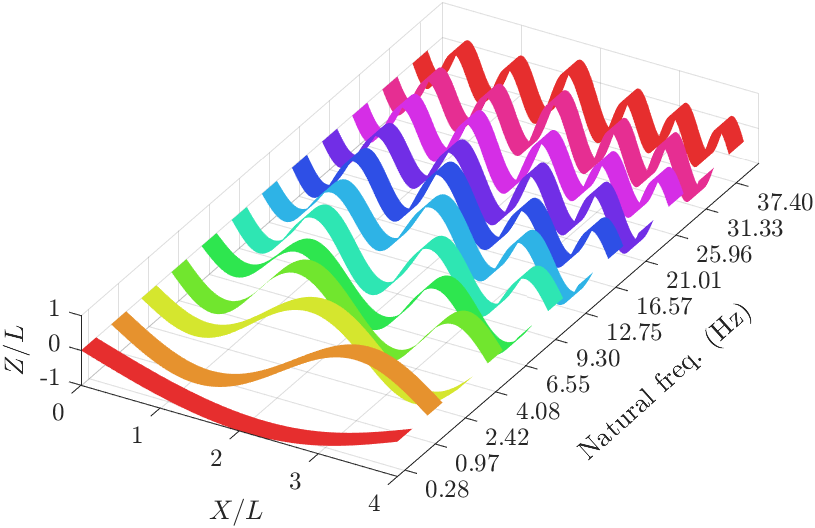}} 								
			\put(1.75,6.8){\includegraphics[height=60mm]{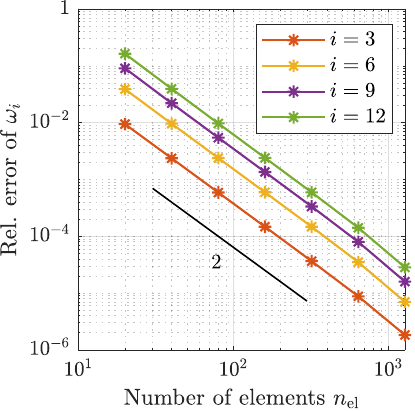}} 							
			\put(-7.75,0){\includegraphics[height=60mm]{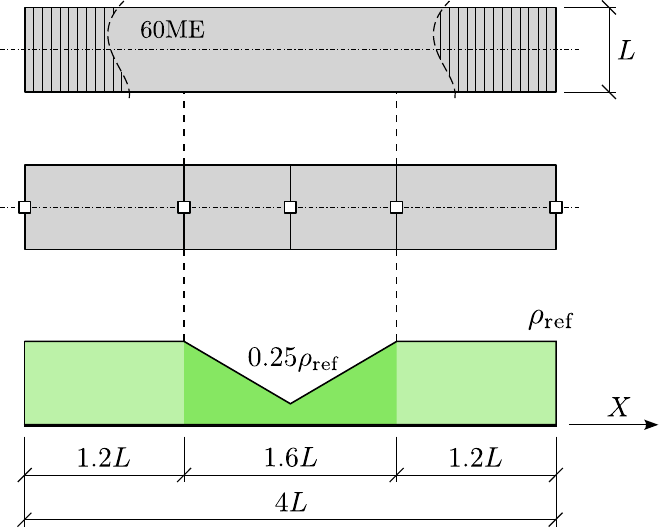}} 	
			\put(0.5,0){\includegraphics[height=55mm]{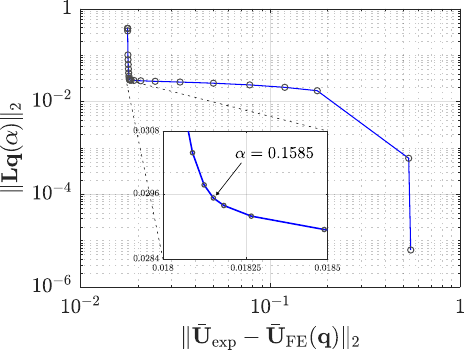}} 
			\put(-8,6.8){\small{a.}}
			\put(1.75,6.8){\small{b.}}
			\put(-8,4.65){\small{c.}}
			\put(-8,2.85){\small{d.}}
			\put(-8,0){\small{e.}}
			\put(0.5,0){\small{f.}}
		\end{picture}
		\caption{Bending of a straight beam: a.~the first 12 bending modes with corresponding frequency $\omega$, the modes are normalized, so that $\max(\mU_\mrfe)=1$; b.~FE convergence of the $i^{\text{th}}$ natural frequency w.r.t.~the FE solution for 2560 elements; c.~uniform material mesh; d.~adapted, nonuniform material mesh; e.~reference density distribution; f.~an example of the L-curve used for the selection of regularization parameter $\alpha$ in Case 2d.3.}
		\label{fig:Ex2_MD}	
	\end{center}
\end{figure}

\paragraph{Problem setup}
For the same beam, the density distribution is reconstructed from modal data of up to the first 12 bending modes. The truncated spectrum of the beam is shown in Fig.~\ref{fig:Ex2_MD}a. The structure is assumed to be unloaded and stress-free. Based on a separate convergence study (see Fig.~\ref{fig:Ex2_MD}b), the synthetic experimental data are generated from 2560 FE, while 60-240 FE are used for the inverse analysis since errors are only $e_\omega\approx10^{-3}$. 
The ideal convergence rate for bending modes, $O(h^2)$, is obtained, even though the material distribution is not smooth. The density distribution (see Fig.~\ref{fig:Ex2_MD}e) is taken as
\eqb{l}
\rho(\xi)/\rho_\mrrf = \left\{ \begin{array}{rcl}
	1     				& \mbox{for} & \xi\in[0,0.3]\cup[0.7,1]\,,\\
	1-3.75(\xi-0.3) 	& \mbox{for} & \xi\in(0.3,0.5]\,,\\
	0.25+3.75(\xi-0.5)  & \mbox{for} & \xi\in(0.5,0.7]\,, \\
\end{array}\right.
\label{e:Ex2Rho}\eqe
where $\xi = X/4L$ and $\rhoR = 0.1m/L$. As this example depends only on bending stiffness, the exact $\Astf(\xi) = 100F$ is used. The inexact distribution $\Cstf(\xi)$ is defined by the vector
\eqb{lll}
\mathbf{EI} & = [2.0689,\,1.9179,\,2.0912,\,1.6498,\,1.6891,\,1.5454,\,1.2641,\,1.2839,\, 1.0536\\[2mm]
&\phantom{= [} 1.0153,\,0.9852] \, 10^{-2}FL^2 \,,
\label{e:Ex2ACexp}\eqe
which corresponds to a sample from Case 2.5 in Tab.~\ref{tab:Ex2_stat} that has $\errA = 4.36\%$ and $\errM = 9.80\%$ error w.r.t.~the exact values from Eq.~\eqref{e:Ex1Rho}. The bounds for $\rho$ are $0.01\rhoR$ and $1\rhoR$, and a fixed initial guess of $0.109\rhoR$ is used. 

\paragraph{Regularized results for exact and inexact bending stiffness}
Two material meshes (Fig.~\ref{fig:Ex2_MD}c~\&~d) are compared: a uniform mesh with 60 linear ME and an adapted mesh with 4 linear ME that ideally capture the unknown $\rho(\xi)$. Since such a dense uniform mesh would inevitably lead to overfitting, regularization is applied. As the penalty matrix $\mL$, a finite-difference approximation of the first derivative operator is chosen~\citep[page 199]{Hansen2013}. The value of $\alpha$ is selected using the L-curve~--~a parametric log-log plot that relates the norms of the regularized solution and the residual~\citep{Hansen1993}. The optimal $\alpha$ corresponds to the leftmost corner of the L-curve, where a balance between solution smoothness and data fit is achieved. An example of the L-curve for Case 2d.3 from Tab.~\ref{tab:Ex2_MD} is shown in Fig.~\ref{fig:Ex2_MD}f.
\begin{table}[h]
	\centering
	\begin{tabular}{@{}cccccccccccc@{}}
		\toprule
		Case  & FE         & exp.                & modes       & stiffness   & $\alpha$ 	     & \multicolumn{2}{c}{iter.}  & \multicolumn{2}{c}{$\errM${[}\%{]}} & \multicolumn{2}{c}{$\errA${[}\%{]}} \\ \cmidrule(l){7-8}\cmidrule(l){9-10}\cmidrule(l){11-12}			
			  & $n_\mrel$  & $n_\mrex/n_\mrmd$ 	 & $n_\mrmd$   &    		 & 				   	 & reg.         & adt.        & reg.           & adt.        		& reg.            & adt.    		  \\ \midrule
		2d.1  & 60         & 100                 & 3           & ref.        & $5\times10^{-4}$  & 24           & 8           & 9.20           & 0.63        		& 0.91            & 0.37    		  \\ \midrule				
		2d.2  & 60         & 100                 & 3           & reconst.    & $0.0282$          & 20           & 8           & 39.54          & 8.98        		& 9.73            & 3.39    		  \\					
		2d.3  & 120        & 100                 & 6           & reconst.    & $0.1585$	         & 18           & 8           & 24.75          & 1.96        		& 2.60            & 0.69              \\					
		2d.4  & 180        & 100                 & 9           & reconst.    & $0.2239$	         & 17           & 7           & 15.00          & 1.24        		& 1.69            & 0.57              \\					
		2d.5  & 240        & 100                 & 12          & reconst.    & $0.3162$	         & 13           & 7           & 11.39          & 1.55        		& 1.37            & 0.61              \\ \bottomrule		
	\end{tabular}
	\caption{Bending of a straight beam: Cases of density reconstruction with their FE mesh, experimental grid resolution, number of bending modes, type of stiffness distribution~(\textit{ref.} for exact, \textit{reconst.} for \eqref{e:Ex2ACexp}), regularization parameter $\alpha$ for the uniform mesh, number of iterations, errors $\errA$, $\errM$. The table compares the results for a uniform material mesh with 60 linear ME and regularization~(\textit{reg.}), and an adapted nonuniform material mesh with 4 linear ME~(\textit{adt.}). No noise is introduced in any of the cases.}
	\label{tab:Ex2_MD}
\end{table} 

\begin{figure}[h] 
	\begin{center} \unitlength1cm
		\begin{picture}(0,5)
			\put(-7.35,0){\includegraphics[height=50mm]{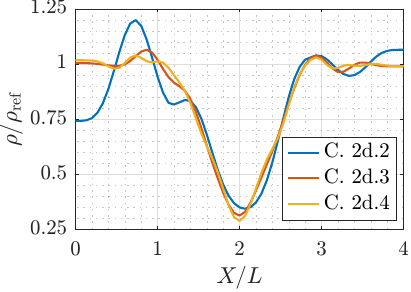}} 					
			\put(0.5,0){\includegraphics[height=50mm]{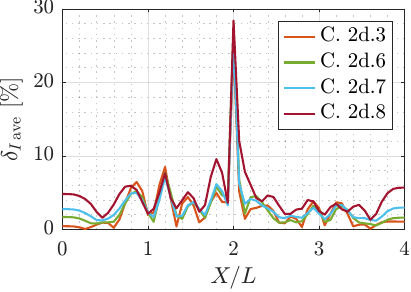}}  						
			\put(-7.35,0.2){\small{a.}}
			\put(0.5,0.2){\small{b.}}
		\end{picture}
		\caption{Bending of a straight beam, using uniform material mesh and regularization for the inverse analysis: a.~normalized, identified density distribution for Cases 2d.2--2d.4. In each case, the sudden decrease of density in the middle is well captured; b.~relative reconstruction error for Cases 2d.3~\&~2d.6--2d.8. The error peak at $X=2L$ corresponds to the minimum of the density distribution, and it is caused by the smoothing effect of regularization.}
		\label{fig:Ex2_MD_err}	
	\end{center}
\end{figure}

Tab.~\ref{tab:Ex2_MD} compares the results obtained with the regularized uniform material mesh and the adapted mesh in the absence of noise. Case 2d.1 uses the reference stiffness, while the others use the inexact distribution from Eq.~\eqref{e:Ex2ACexp}. The average identification error for the uniform mesh, $\errAreg$, is around three times higher than $\errAadt$ for the adapted mesh (e.g., Case 2d.3: $\errAreg/\errAadt\approx3.77$). In contrast, the corresponding ratio for the maximum errors is around 10 (e.g., Case 2d.3: $\errMreg/\errMadt\approx12.63$). Fig.~\ref{fig:Ex2_MD_err}a shows that the uniform mesh qualitatively captures the density drop at the middle of the beam, though Case 2d.2 exhibits notable oscillations. However, the density at the center is overestimated, which is responsible for the high maximum errors reported in Tab.~\ref{tab:Ex2_MD} and shown in Fig.~\ref{fig:Ex2_MD_err}b for Case 2d.3. This stems from the smoothing effect of regularization, which here penalizes the solution slope. 

Alternative regularization techniques based on the $\ell_1$ norm may alleviate this issue~\citep{Tibshirani2011}, while also promoting solution sparsity and facilitate model selection. For example, penalizing the term $\alpha\norm{\mL_2\mq}_1$ in the objective function, where $\mL_2$ is an approximation of the second derivative operator, encourages the clustering of linear ME into larger piecewise-linear segments. For constant ME, a similar effect is obtained using Total Variation regularization~\citep{Vogel2002}.

The adapted mesh achieves the best accuracy for the first 9 modes (Case 2d.4), with errors increasing beyond that, as also seen in Sec.~\ref{s:Ex1_MD}. Interestingly, the same trend appears when the effect of inaccurate $\Cstf(\xi)$ is isolated, i.e., when the same FE mesh is used for both data generation and inverse analysis (inverse crime, not shown in Tab.~\ref{tab:Ex2_MD}). A possible explanation is that higher modes are more sensitive to local stiffness changes and therefore more affected by errors in $\Cstf(\xi)$, whereas lower modes remain insensitive. Although this effect is observed here only for the adapted mesh, it is likely to occur in the regularized case as well, if more modes are included.

\paragraph{Influence of noise on regularized results}
Tab.~\ref{tab:Ex2_MD_noise} shows a study of noise applied to Case 2d.3 from Tab.~\ref{tab:Ex2_MD}, using the uniform mesh. Noise in frequencies ranges from 0\% to 4\%, while noise in modes is always 4\%. For noise applied only to modes, the absolute increases in identification error are similar to those in Sec.~\ref{s:Ex1_MD} ($\Delta\errA = 0.22\pm0.38\%$ and $\Delta\errM = 2.50\pm4.51\%$ between Cases 2d.3~\&~2d.6), but the relative increases are around 10\%, compared to 50\%--150\% in Sec.~\ref{s:Ex1_MD}. When both modes and frequencies are perturbed (Cases 2d.7--2d.9), notable error increments are observed for at least 2\% noise in frequencies. Fig.~\ref{fig:Ex2_MD_err}b shows the average error distributions for Cases 2d.3~\&~2d.6--2d.8. As noise increases, the overall shape of the distribution is preserved. Error grows visibly near the beam ends, but no new peaks emerge, likely due to regularization.
\begin{table}[h]
	\centering
	\begin{tabular}{@{}ccccccccc@{}}
		\toprule
		Case  & FE               & exp.                                & modes               & stiffness & noise    & ave. iter. & $\errM$ 				 & $\errA$ 				 \\
		& $n_\mrel$        & $n_\mrex/n_\mrmd$ 				   & $n_\mrmd$           &           & {[}\%{]} &            & {[}\%{]}              & {[}\%{]}              \\ \midrule
		2d.6  & 120              & 100                                 & 6                   & reconst.  & [4,0]    & 15         & $27.25\,\pm\,4.51$    & $2.82\,\pm\,0.38$     \\ 
		2d.7  & 120              & 100                                 & 6                   & reconst.  & [4,1]    & 16         & $25.67\,\pm\,4.46$    & $3.17\,\pm\,0.53$     \\
		2d.8  & 120              & 100                                 & 6                   & reconst.  & [4,2]    & 15         & $29.51\,\pm\,6.13$    & $4.56\,\pm\,1.60$     \\ 
		2d.9  & 120              & 100                                 & 6                   & reconst.  & [4,4]    & 16         & $29.92\,\pm\,9.77$    & $5.86\,\pm\,2.59$     \\ \bottomrule
	\end{tabular}
	\caption{Bending of a straight beam, using uniform material mesh and regularization for the inverse analysis: Cases of density reconstruction with their FE mesh, experimental grid resolution, number of bending modes, type of stiffness distribution (\textit{ref.} for exact, \textit{reconst.} for \eqref{e:Ex2ACexp}), noise level, number of iterations, errors $\errA$, $\errM$. In the \textit{noise} column, the values in brackets refer to the noise applied to modes and frequencies, in that order. For all cases, $\alpha = 0.631$.}
	\label{tab:Ex2_MD_noise}
\end{table}

In contrast to Sec.~\ref{s:Ex1_stat}, the second example showed that combining different boundary conditions in a single analysis can significantly reduce the identification error. Hence, the FE model should be examined beforehand to avoid parameter indeterminacies, as in Fig.~\ref{fig:Ex2_Err}a. For the density identification, the performance of a dense material mesh with regularization is compared to that of a mesh ideally adapted to the unknown $\rho(\xi)$, yielding results similar to those of Sec.~\ref{s:Ex1_MD}.

\subsection{Curved beam}\label{s:Ex3}

\begin{figure}[h]
	\begin{center} \unitlength1cm
		\begin{picture}(0,5)
			\put(-7.6,0){\includegraphics[height=50mm]{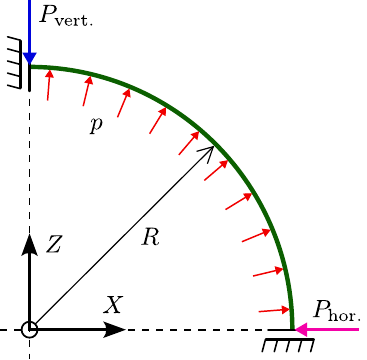}}
			\put(-2.4,0){\includegraphics[height=50mm]{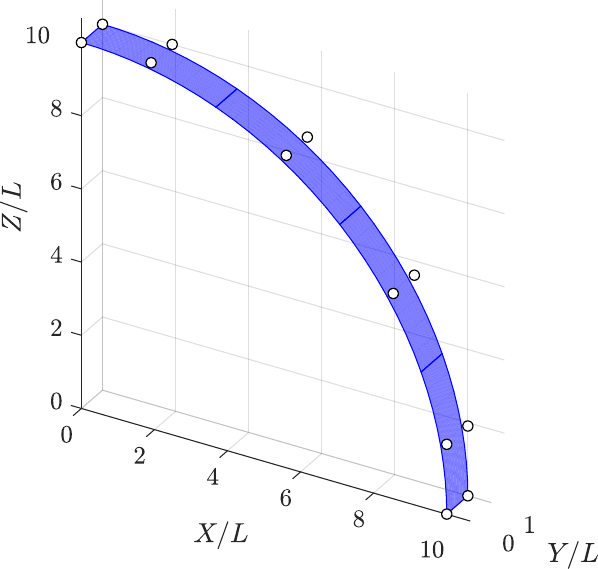}} 	
			\put(2.85,0){\includegraphics[height=50mm]{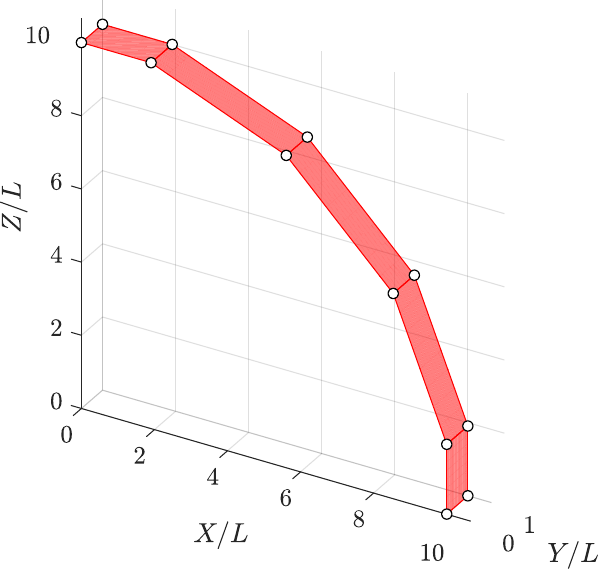}} 				
			\put(-7.9,0){\small{a.}}
			\put(-1.85,0){\small{b.}}
			\put(3.4,0){\small{c.}}
		\end{picture}
		\caption{Curved beam: a.~setup, loading and boundary conditions; An example of IGA mesh (b.)~and Lagrange mesh (c.)~for the B2M1 discretization, consisting of four B2 and five M1 elements.}
		\label{fig:Ex3_Geo}	
	\end{center}
\end{figure}

In the final example, the problem of simultaneous identification of axial and bending stiffness is considered. A $90^\circ$ arc beam is analyzed, as shown in Fig.~\ref{fig:Ex3_Geo}a. The beam has radius $R=10L$ and width $B=L$. An illustrative FE model of the beam consisting of IGA and Lagrange elements is presented in Fig.~\ref{fig:Ex3_Geo}b~\&~c. The role of these two discretizations is clarified later. As shown, the beam is fixed in $X$-~direction at the left end ($X=0$, $Z=10$), and in $Y$-~direction at the right end ($X=10$, $Z=0$). Rotations are fixed at both ends.

\subsubsection{Axial and bending stiffness reconstruction from statics}\label{s:Ex3_stat}

\begin{figure}[p]
	\begin{center} \unitlength1cm
		\begin{picture}(0,11)
			\put(-6.9,6){\includegraphics[height=50mm]{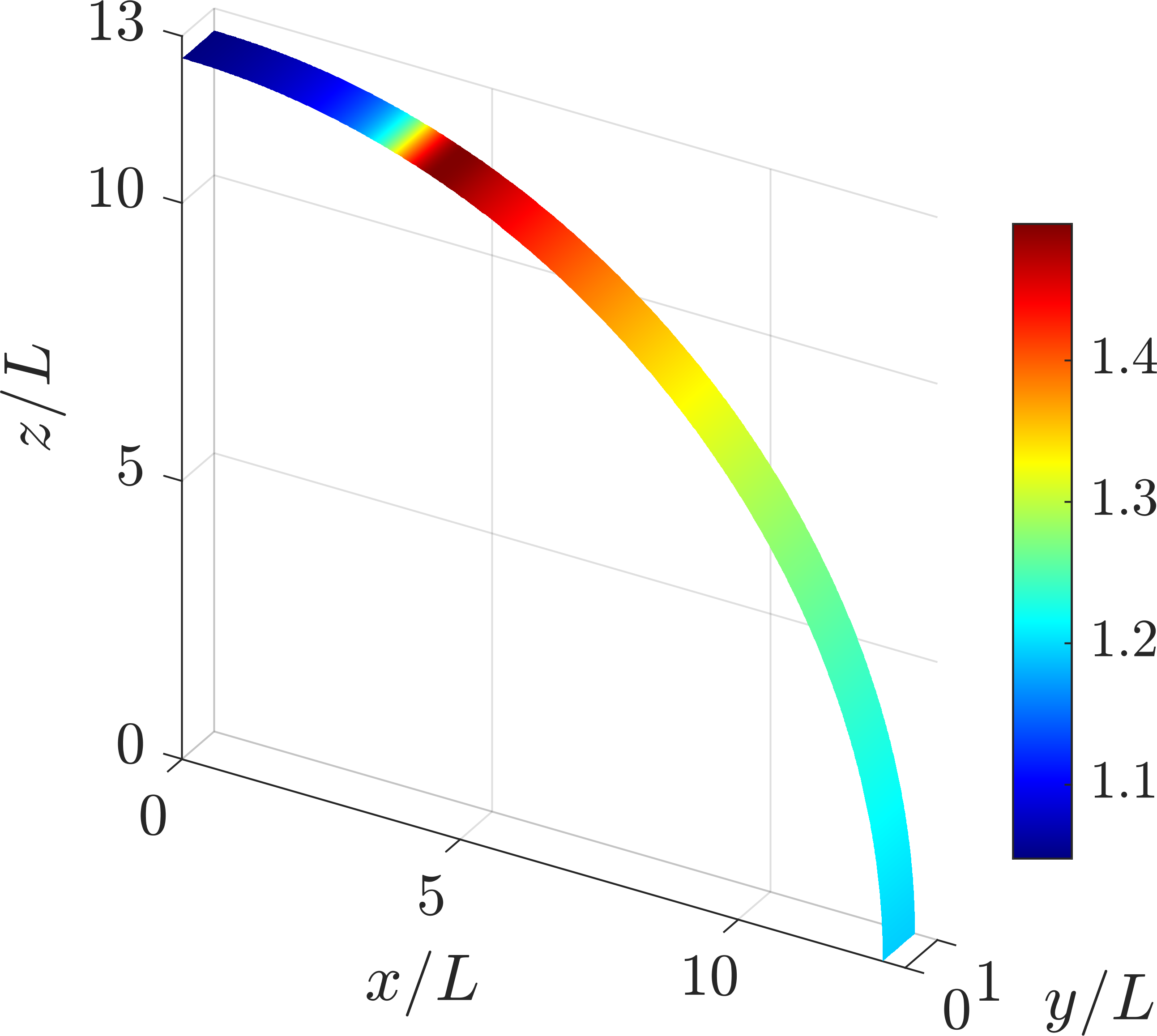}} 			
			\put(1.6,6){\includegraphics[height=50mm]{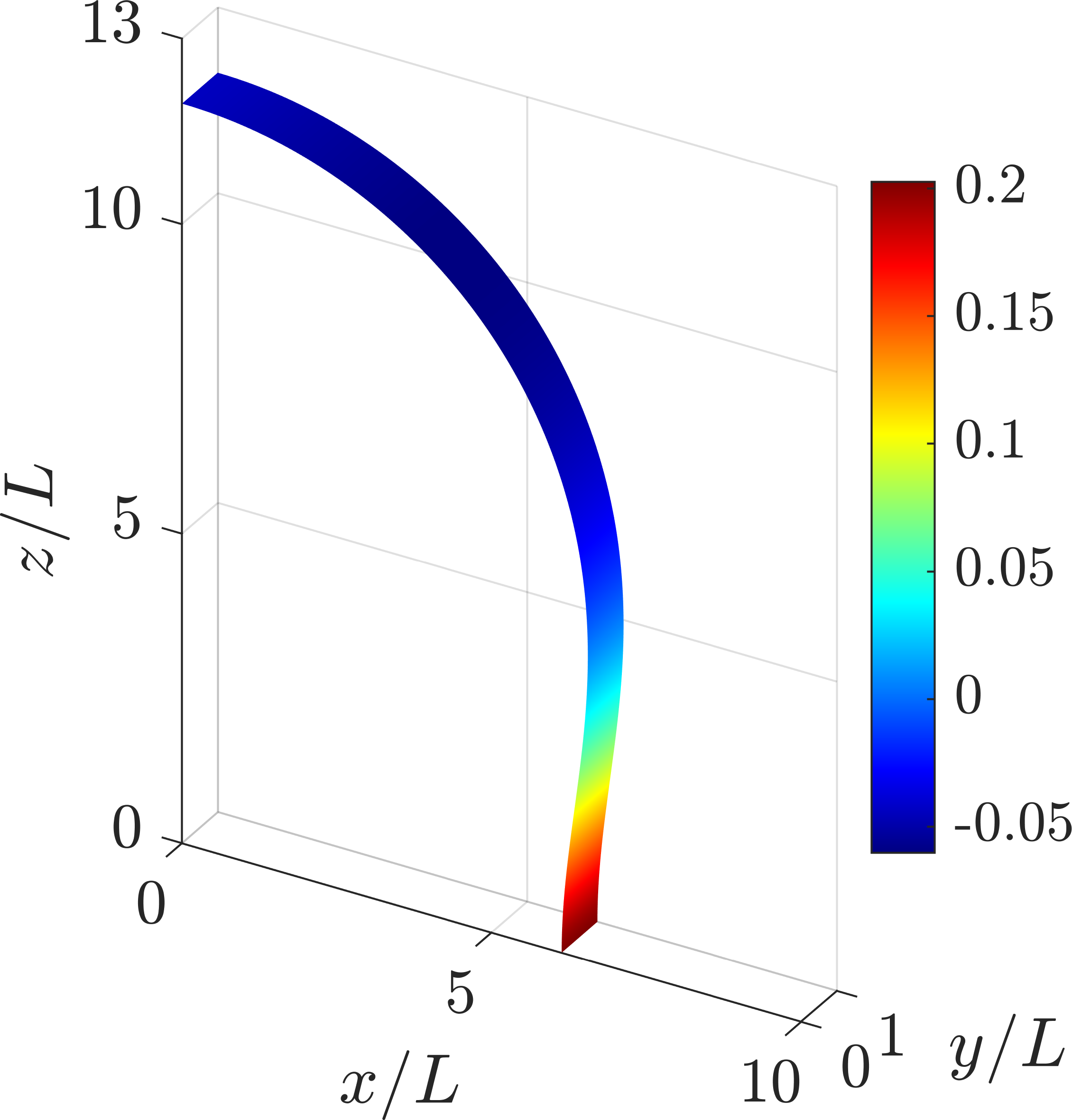}}  			
			\put(-6.9,0.5){\includegraphics[height=50mm]{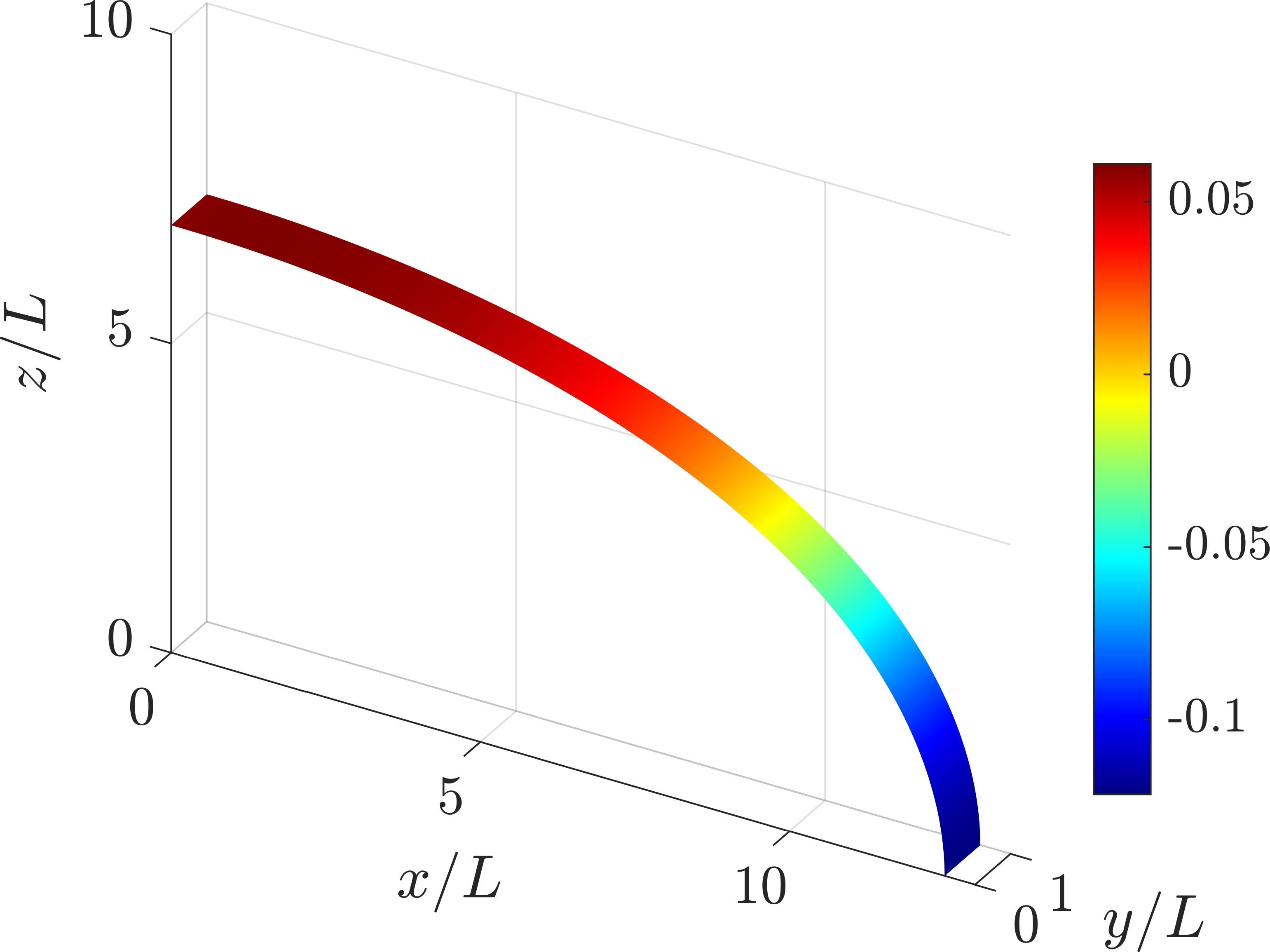}} 								
			\put(1.1,0.35){\includegraphics[height=50mm]{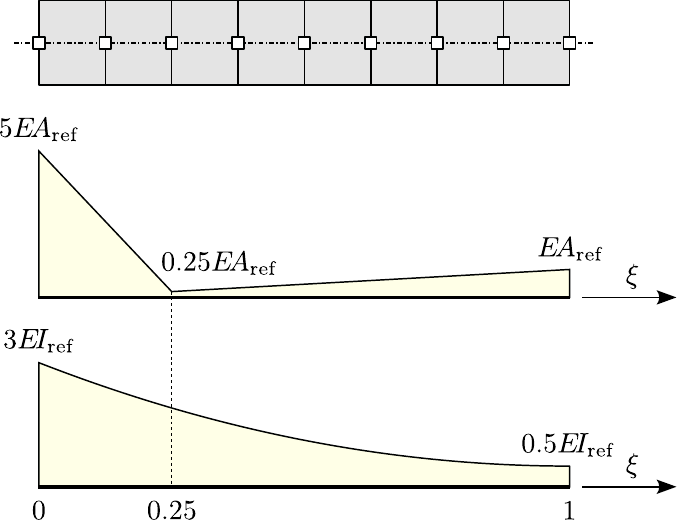}} 			
			\put(-6.9,6){\small{a.}}
			\put(1.6,6){\small{b.}}
			\put(-6.9,0.35){\small{c.}}
			\put(0.8,0.35){\small{d.}}
			\put(-2,4.4){\scriptsize{$[L^{-1}]$}}
			\put(4.7,9.95){\scriptsize{$[L^{-1}]$}}
		\end{picture}
		\caption{Statics of a curved beam: a.~deformed configuration for uniform pressure $p$, colored by stretch $\lambda$; b.~\&~c.~deformed configuration for loading with a horizontal ($P_{\,\mathrm{hor.}}$) and vertical ($P_{\,\mathrm{vert.}}$) force, respectively, colored by relative curvature change $\kappa$; d.~material mesh with the reference distribution for axial and bending stiffness.}
		\label{fig:Ex3_stat1}	
	\end{center}
\end{figure}

\paragraph{Problem setup}
Three independent load cases are analyzed: inflation with uniform internal pressure $p = 2F/L$, horizontal force $P_{\,\mathrm{hor.}} = 2\times10^{-5}F$, and vertical force $P_{\,\mathrm{vert.}} = 2\times10^{-5}F$ (see Fig.~\ref{fig:Ex3_Geo}a). The deformed configurations for these three load cases are shown in Fig.~\ref{fig:Ex3_stat1}a--c. Three load levels for each load case are used (10\%, 50\%, and 100\% of the final load); thus, for all cases $n_\mrlc = 3\times3$. The chosen $\Astf(\xi)$ is defined by
\eqb{l}
\Astf(\xi)/\AstfR = \left\{ \begin{array}{rcl}
	5-19\xi         & \mbox{for} & \xi\in[0,0.25]\,,\\
	0.25+(\xi-0.25) & \mbox{for} & \xi\in(0.25,1]\,, \\
\end{array}\right.
\label{e:Ex3A}\eqe
where $\AstfR = 100F$, while $\Cstf(\xi)$ is specified as
\eqb{l}
\Cstf(\xi)/\CstfR = 2.5\xi^2-5\xi+3 \,,\quad \xi\in[0,1]\,,
\label{e:Ex3C}\eqe
in which $\CstfR = 0.001FL^2$. A uniform mesh of 8 linear ME is chosen to identify the unknown stiffness fields, which gives $n_\mrvr = 18$ (Fig.~\ref{fig:Ex3_stat1}d). The unknowns are bounded between $0.1\AstfR$ and $10\AstfR$, and between $0.1\CstfR$ and $10\CstfR$, respectively. The initial guess for $\Astf$ and $\Cstf$ is fixed to $1.09\AstfR$ and $1.09\CstfR$, respectively. 
\begin{figure}[p]
	\begin{center} \unitlength1cm
		\begin{picture}(0,7)	
			\put(-7.1,0.6){\includegraphics[height=60mm]{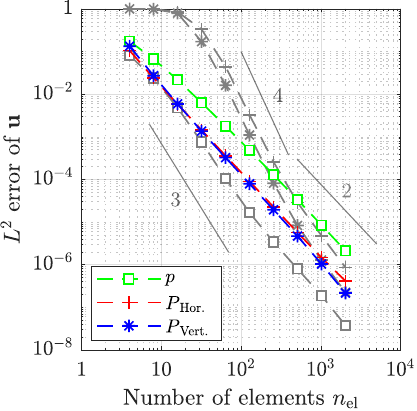}} 		
			\put(1.4,0){\includegraphics[height=70mm]{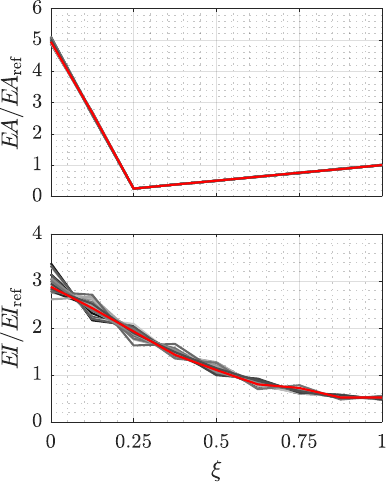}} 						
			\put(-7.4,0.6){\small{a.}}
			\put(1.4,0){\small{b.}}
		\end{picture}
		\caption{Statics of a curved beam: a.~FE convergence of the discrete $L^2$ error for all studied load cases w.r.t.~the FE solution with B2M2 discretization for 4096 elements. Lines in color correspond to locking-free B2M1 discretization, while lines in gray show convergence for B2M2 scheme; b.~distributions of $\Astf$ and $\Cstf$, obtained from the inverse analysis in Case 3.4; each graph contains 25 samples. $\Astf(\xi)$ and $\Cstf(\xi)$ used in Sec.~\ref{s:Ex3_MD} are highlighted in red, see Eq.~\eqref{e:Ex3ACexp}.}
		\label{fig:Ex3_stat2}	
	\end{center}
\end{figure}

Since the inflation depends solely on axial stiffness and point forces induce mostly bending deformation, these load cases can act almost separately in the reconstruction. Combining them in a single inverse analysis enables simultaneous identification of all stiffness parameters while reducing cross-correlation between $\Astf$ and $\Cstf$.

\paragraph{Hybrid B2M1 discretization}
Fig.~\ref{fig:Ex3_stat2}a shows that the load cases with point forces $P_{\,\mathrm{hor.}}$ and $P_{\,\mathrm{vert.}}$ exhibit membrane locking when using a standard discretization with quadratic NURBS, referred to as \textit{B2M2 discretization}. To alleviate locking, the hybrid approach introduced by~\citet{Sauer2024}, known as \textit{B2M1 discretization}, is adopted.  The B2M1 approach uses quadratic NURBS elements for the bending forces in~(\ref{e:finte}.2) and the external forces in Eqs.~\eqref{e:fext0p}~\&~\eqref{e:fexttm}, while linear Lagrange elements are used for the membrane forces in~(\ref{e:finte}.1). This results in two separate discretization, as illustrated in Figs.~\ref{fig:Ex3_Geo}b~\&~c, but a single set of control points/nodes. For more details on the B2M1 discretization, see~\citet{Sauer2024}. 

Color plots in Fig.~\ref{fig:Ex3_stat2}a show that membrane locking is mitigated with the B2M1 approach. After a preliminary analysis, a B2M1 mesh comprised of 64 B2 and 65 M1 elements is selected for the load cases with point forces in the inverse analysis (error $e_u\approx3.5\times10^{-4}$). Membrane locking does not affect the inflation; thus, the standard B2M2 mesh with 32 elements is used in this case (error $e_u\approx7.5\times10^{-4}$). For the generation of the synthetic experimental data, a B2M2 mesh with 4096 elements is employed.

One issue with the B2M1 discretization that requires attention is the non-conforming mapping between the material and M1 elements. Since the B2 and material elements are assumed to be conforming (see Sec.~\ref{s:matFE}), some M1 elements inevitably span across two material elements (compare Figs.~\ref{fig:Ex3_Geo}b~\&~c). This causes discontinuities and kinks in the material distribution within those elements and complicates the assembly of sensitivities. To address this, a dedicated element subroutine divides each affected M1 element into two integration domains, see App.~\ref{s:B2M1_map}. The influence of these elements diminishes with mesh refinement. 

\paragraph{Results}
Case 3.1 in Tab.~\ref{tab:Ex3_stat} shows that the inverse analysis without noise yields higher identification errors for $\Cstf$ than for $\Astf$, despite the use of a denser FE mesh for the load cases with a point force. This behavior is inherent to the convergence of the forward problem. Providing a comparative relative FE error for all load cases does not guarantee similar accuracy in identification. Additionally, the chosen material mesh approximately captures $\Cstf(\xi)$, whereas $\Astf(\xi)$ is represented exactly. The B2M1 discretization allows to obtain comparable results of the inverse analysis using two times fewer DOFs than for the standard B2M2 approach.
\begin{table}[h]
	\centering
	\begin{tabular}{@{}cccccccccc@{}}
		\toprule
		Case & FE                & mat.                    & exp.                               & load              & noise & ave. iter & $q(X)$ & $\errM$ & $\errA$ \\
		& $n_\mrel$ & $\bar n_\mrel$ & $n_\mrex/n_\mrll$ & $n_\mrll$ & [\%]  &           &                 & [\%]                  & [\%]                  \\ \midrule
		3.1  & [32,64]                & 8                       & 4000                               & 3                 & 0     & 18        & $\Astf$         & 1.34                  & 0.31                 \\
		&                   &                         &                                    &                   &       &                     & $\Cstf$         & 3.38                  & 1.44                  \\ \midrule
		3.2  & [32,64]                & 8                       & 4000                               & 3                 & 1     & 18        & $\Astf$         & $1.30\,\pm\,0.32$     & $0.39\,\pm\,0.13$    \\
		&                   &                         &                                    &                   &       &                     & $\Cstf$         & $4.11\,\pm\,1.21$     & $1.75\,\pm\,0.48$     \\ \midrule
		3.3  & [32,64]                & 8                       & 4000                               & 3                 & 2     & 19        & $\Astf$         & $1.25\,\pm\,0.45$     & $0.46\,\pm\,0.17$     \\
		&                   &                         &                                    &                   &       &                     & $\Cstf$         & $6.09\,\pm\,2.49$     & $2.63\,\pm\,1.26$     \\ \midrule
		3.4  & [32,64]                & 8                       & 4000                               & 3                 & 4     & 20        & $\Astf$         & $1.98\,\pm\,0.73$     & $0.83\,\pm\,0.35$     \\
		&                   &                         &                                    &                   &       &                     & $\Cstf$         & $10.08\,\pm\,3.13$    & $4.88\,\pm\,1.64$     \\ \bottomrule
	\end{tabular}
	\caption{Statics of a curved beam: Studied stiffness identification cases with their FE and material mesh, experimental grid resolution, load levels, noise, average number of iterations, and errors $\errA$, $\errM$. A double value of $[32,64]$ indicates the number of FE used for the load cases with pressure and point forces, respectively. For all cases, $n_\mrlc = 3\times n_\mrll$.}
	\label{tab:Ex3_stat}
\end{table}

Cases 3.2--3.4, which include random noise, report greater sensitivity of $\Cstf$ to measurement noise. The increases in $\errAC$ relative to Case 3.1 are typically 4--8 times larger than those in $\errAA$, with even higher ratios observed for the maximum identification errors. This behavior is specific to the chosen set of load cases and does not imply a general relationship. Fig.~\ref{fig:Ex3_stat2}b shows a set of 25 samples of $\Astf(\xi)$ and $\Cstf(\xi)$ for Case 3.4. The reconstructed distributions of $\Cstf$ oscillate evidently, highlighting higher sensitivity of $\Cstf$ to noise. In contrast, no visible oscillations occur for the reconstructed $\Astf(\xi)$. To reduce the oscillations of $\Cstf(\xi)$, one could provide more experimental data, reduce the noise level, or enforce smoothness of the solution with regularization as in Sec.~\ref{s:Ex2_MD}.

\subsubsection{Density reconstruction from modal dynamics}\label{s:Ex3_MD}

\begin{figure}[h]
	\begin{center} \unitlength1cm
		\begin{picture}(0,12.5)
			\put(-7.25,6.5){\includegraphics[height=60mm]{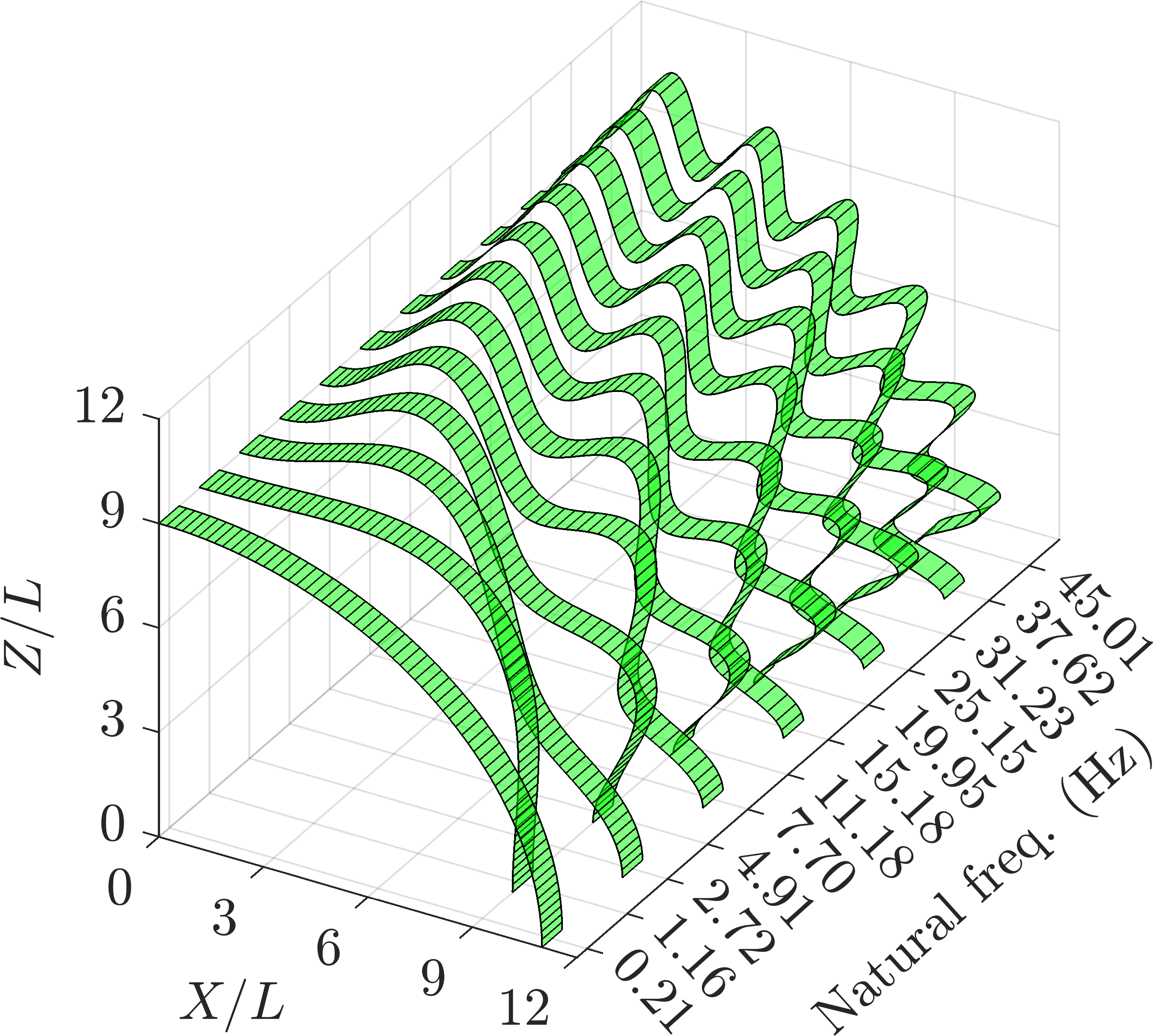}} 					
			\put(1.25,6.5){\includegraphics[height=60mm]{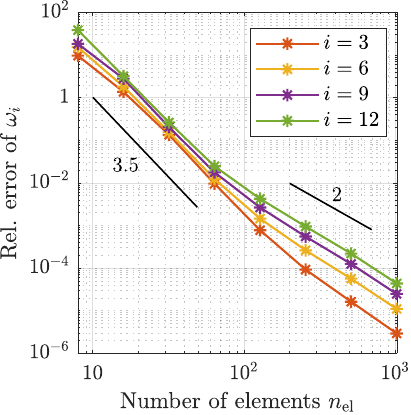}}  					
			\put(-6.75,1.25){\includegraphics[height=35mm]{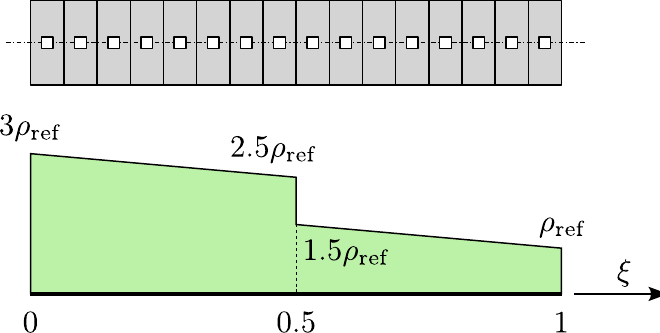}} 	
			\put(1.35,0){\includegraphics[height=60mm]{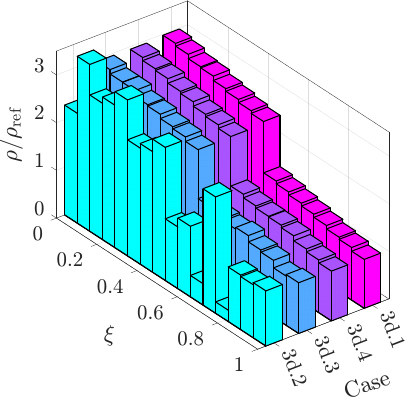}} 									
			\put(-7.25,6.5){\small{a.}}
			\put(1.25,6.5){\small{b.}}
			\put(-7.15,1.25){\small{c.}}
			\put(1.25,0.5){\small{d.}}
		\end{picture}
		\caption{Modal dynamics of a curved beam: a.~the first 12 bending modes with corresponding $\omega$, the modes are normalized, so that $\max(\mU_\mrfe)=1$; b.~FE convergence of the $i^{\text{th}}$ natural frequency w.r.t.~the FE solution for 2048 elements; c.~material mesh with the reference density distribution; d.~normalized results of the inverse analysis for Cases 3d.1--3d.4.}
		\label{fig:Ex3_MD}	
	\end{center}
\end{figure}

\paragraph{Problem setup}
The density distribution is identified using up to the first 12 bending modes (Fig.~\ref{fig:Ex3_MD}a). As before, the structure is assumed to be unloaded and stress-free. 
As locking is not an issue in this case, the standard B2M2 discretization is used. The synthetic experimental data are generated from 2048 FE, while the inverse analysis is conducted with a mesh of 128 FE, with error $e_\omega\approx10^{-3}$ (see Fig.~\ref{fig:Ex3_MD}b). The reference $\rho(\xi)$ is given by 
\eqb{l}
\rho(\xi)/\rhoR = \left\{ \begin{array}{rcl}
	3-\xi & \mbox{for} & \xi\in[0,0.5]\,, \\
	2-\xi & \mbox{for} & \xi\in(0.5,1]\,, \\
\end{array}\right.
\label{e:Ex3Rho}\eqe
where $\rho_\mrrf = 10^{-5}m/L$ (see Fig.~\ref{fig:Ex3_MD}c). The inexact $\Astf(\xi)$ and $\Cstf(\xi)$ are taken from a sample of Case 3.4 in Tab.~\ref{tab:Ex3_stat}, and are given by
\eqb{lll}
\mathbf{EA} \is [493.740,\,266.759,\,24.645,\,37.500,\,50.005,\,62.548,\,75.418,\,87.097,\,99.900] F \,, \\[2mm]
\mathbf{EI} \is [2.873,\,2.419,\,1.921,\,1.423,\,1.099,\,0.795,\,0.719,\,0.515,\,0.515] \, 10^{-3}FL^2 \,,
\label{e:Ex3ACexp}\eqe
which have identification errors $\errAA = 0.61\%$, $\errMA = 1.62\%$, $\errAC = 3.86\%$, and $\errMC = 9.49\%$ w.r.t.~the exact values from Eqs.~\eqref{e:Ex3A}~\&~\eqref{e:Ex3C}. A material mesh consisting of 16 constant ME is used for the density, leading to $n_\mrvr = 16$. The lower and upper bounds for $\rho$ are $0.1\rhoR$ and $10\rhoR$, respectively. The initial guess is taken as $1.09\rhoR$.

\paragraph{Results}
Case 3d.1 in Tab.~\ref{tab:Ex3_MD} shows that the chosen mesh approximates the discontinuous distribution from Fig.~\ref{fig:Ex3_MD}c well when the exact stiffness data are used. Cases 3d.2--3d.5 illustrate the impact of the number of modes on the identification errors for inexact stiffness from \eqref{e:Ex3ACexp}. With the first 9 modes, the algorithm achieves satisfactory results of $\errA=1.36\%$ and $\errM=4.06\%$. Similarly to the previous examples, when the number of modes exceeds a certain threshold (here, the first 9 modes), identification errors begin to stagnate or even grow, which can be observed between Cases 3d.4 and 3d.5 in Tab.~\ref{tab:Ex3_MD}. 
\begin{table}[h]
	\centering
	\begin{tabular}{@{}cccccccccc@{}}
		\toprule
		Case & FE                & mat.                    & exp.                               & modes              & stiff.   & noise & iter. & $\errM$ & $\errA$ \\
		& $n_\mrel$         & $\bar n_\mrel$  		   & $n_\mrex/n_\mrmd$ 					& $n_\mrmd$ 	     &          & 	    &       & [\%]    & [\%]    \\ \midrule
		3d.1 & 128               & 16                      & 100                                & 3                  & ref.     & 0     & 9     & 0.89    & 0.36    \\ \midrule
		3d.2 & 128               & 16                      & 100                                & 3                  & reconst. & 0     & 14    & 90.34   & 22.60   \\
		3d.3 & 128               & 16                      & 100                                & 6                  & reconst. & 0   	& 6     & 6.47    & 2.08    \\
		3d.4 & 128               & 16                      & 100                                & 9                  & reconst. & 0   	& 6     & 4.06    & 1.36    \\
		3d.5 & 128               & 16                      & 100                                & 12                 & reconst. & 0   	& 6     & 3.87    & 1.42    \\ \bottomrule
	\end{tabular}
	\caption{Modal dynamics of a curved beam: Cases of density reconstruction with their FE and material mesh, experimental grid resolutions, number of normal modes, type of stiffness distribution (\textit{ref.} for exact, \textit{reconst.} for \eqref{e:Ex3ACexp}), noise level, number of iterations, and errors $\errA$, $\errM$.}
	\label{tab:Ex3_MD}
\end{table} 

The third example concludes the numerical examples section. It combines several aspects discussed in the previous sections while addressing a more complex structure and simultaneously identifying both $\Astf$ and $\Cstf$. The B2M1 discretization was used to mitigate membrane locking in the quasi-static calculations. The inverse analysis produced results consistent with earlier findings; thus, the study of noise impact on the density reconstruction is omitted. 

\subsection{Comparison between Lagrange and NURBS material meshes}\label{s:BsplineMatMesh}

\begin{figure}[h]
	\begin{center} \unitlength1cm
		\begin{picture}(0,5.5)
			\put(-7.75,0){\includegraphics[height=55mm]{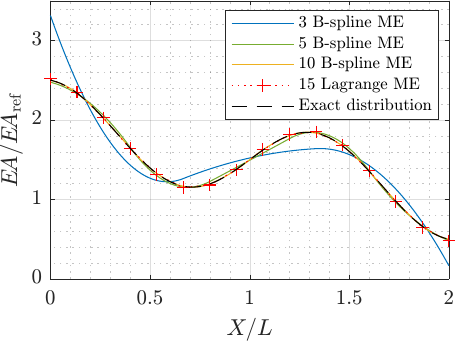}} 					
			\put(0.25,0){\includegraphics[height=55mm]{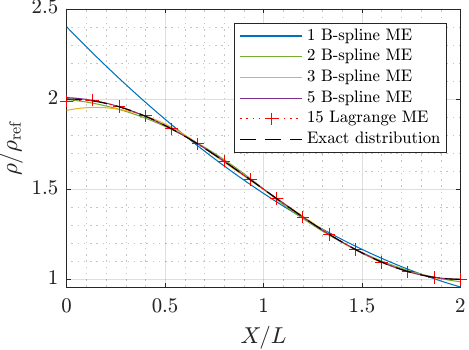}}  						
			\put(-7.75,0){\small{a.}}
			\put(0.25,0){\small{b.}}
		\end{picture}
		\caption{Comparison between Lagrange (red dotted lines) and NURBS material meshes (solid lines) for uniaxial stretching of a bar: a.~reconstructed $\Astf$ distribution for Case~1.3 from Tab.~\ref{tab:Ex1_stat} and the corresponding results for B-spline material meshes consisting of 3--10 quadratic elements; b.~reconstructed density distribution for Case~1d.2 from Tab.~\ref{tab:Ex1_MD} and the corresponding results for B-spline material meshes consisting of 1--5 quadratic elements. The exact distributions are given by Eqs.~\eqref{e:Ex1A}~and~\eqref{e:Ex2Rho}.}
		\label{fig:Ex1_Bspline}	
	\end{center}
\end{figure}

This section compares the performance of Lagrange material mesh introduced in Sec.~\ref{s:matFE} with an alternative NURBS representation of the unknown material distributions. Two smooth and two non-smooth distributions are selected for comparison. Figs.~\ref{fig:Ex1_Bspline}a~and~\ref{fig:Ex1_Bspline}b show the identification results for axial stiffness (Case~1.3 from Sec.~\ref{s:Ex1_stat}) and density (Case~1d.2 from Sec.~\ref{s:Ex1_MD}), respectively, obtained using material meshes composed of linear Lagrange elements and their B-spline counterparts\footnote{In this case, the NURBS representation of the material field reduces to B-splines, as all weights are equal to one.}. The B-spline reconstructions closely match the exact stiffness and density distributions when using 10 and 5 material elements, respectively. As shown, B-spline material meshes outperform their Lagrange alternatives for smooth material distributions.
	
\begin{figure}[h]
	\begin{center} \unitlength1cm
		\begin{picture}(0,5.5)
			\put(-7.75,0){\includegraphics[height=55mm]{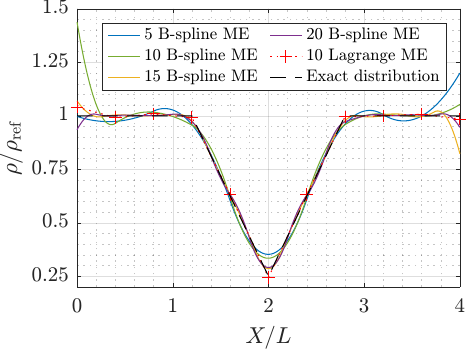}} 					
			\put(0.25,0){\includegraphics[height=55mm]{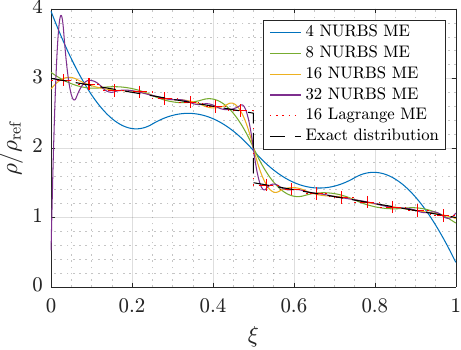}}  						
			\put(-7.75,0){\small{a.}}
			\put(0.25,0){\small{b.}}
		\end{picture}
		\caption{Comparison between Lagrange (red dotted lines) and NURBS material meshes (solid lines): a.~reconstructed density distribution with 10 linear Lagrange ME, analogous to Case~2d.1 in Tab.~\ref{tab:Ex2_MD} and the corresponding results using 5--20 quadratic B-spline ME; b.~reconstructed density distribution for Case~3d.1 in Tab.~\ref{tab:Ex3_MD} and the corresponding results obtained using NURBS material meshes with 4--32 quadratic elements. The exact distributions are given by Eqs.~\eqref{e:Ex2Rho}~and~\eqref{e:Ex3Rho}.}
		\label{fig:Ex23_Bspline}	
	\end{center}
\end{figure}

Fig.~\ref{fig:Ex23_Bspline}a compares the density distribution from Case~2d.1 in Tab.~\ref{tab:Ex2_MD} reconstructed using a uniform mesh of 10 linear Lagrange ME, with results obtained for uniform B-spline material meshes. As shown, B-spline material meshes struggle to capture sharp changes in the material parameters, particularly the middle drop, producing a smoothing effect similar to Tikhonov regularization observed in Fig.~\ref{fig:Ex2_MD_err}a. 
Similarly, Fig.~\ref{fig:Ex23_Bspline}b presents the reconstructed density from Case~3d.1 in Tab.~\ref{tab:Ex3_MD} using a uniform mesh of 16 constant Lagrange ME, and a series of results for the same case with uniform NURBS material meshes. The NURBS meshes fail to capture the density discontinuity in the middle, resulting in oscillations. In both cases, further increasing the number of NURBS material elements leads to overfitting rather than improved results.

\section{Conclusion}\label{s:concl}

This work proposes a FEMU inverse framework for identifying heterogeneous fields of elastic properties and density in nonlinear planar Bernoulli--Euler~(BE) beams. Stiffness distributions, $\Astf(\xi)$ and $\Cstf(\xi)$, are identified from quasi-static displacements under known loads. Then, the density distribution, $\rho(\xi)$, is reconstructed from a finite number of the first modes and frequencies (1 to 12), using the previously identified stiffness. The unknown fields are parameterized using the so-called \textit{material mesh}, introduced by~\cite{Borzeszkowski2022}. Analytical derivatives of the objective function w.r.t.~the discrete parameters of $\Astf$, $\Cstf$, and $\rho$ are derived. Several numerical examples demonstrate the robustness of the framework and highlight key challenges. The results for the identification of elastic parameters align well with those of~\cite{Borzeszkowski2022}, while the density identification gives new insight. The results of the inverse analysis for the quasi-experimental data generated using 1D and full 3D FE models are compared, showing that material heterogeneity over the cross-section does not significantly affect the results for slender beams as long as BE beam theory is valid. A comprehensive study is carried out for the density reconstruction, analyzing the effect of inaccurate stiffness, the number of modes, noise in modal data, and regularization. The framework is modular and extends naturally to shells and bulk structures. Each core component, such as FE formulation, optimization algorithm, or constitutive model, can be easily replaced. This flexibility is demonstrated in Sec.~\ref{s:Ex3_stat}, where the hybrid approach from~\citet{Sauer2024} is used to alleviate membrane locking in selected load cases.

The present inverse analysis confirms that:
\begin{itemize}[noitemsep,topsep=0pt]
	\item Selecting an appropriate set of experiments that are not susceptible to various error sources is crucial. This can be achieved by analyzing sensitivities, avoiding indeterminacies, and applying suitable boundary conditions (Secs.~\ref{s:Ex1_stat}~\&~\ref{s:Ex2_stat}).
	\item Care should be taken when choosing the material mesh. Refined meshes usually lead to oscillations (Sec.~\ref{s:Ex3_stat}) and more pronounced indeterminacies (Sec.~\ref{s:Ex2_stat}). Unless regularization is applied, starting with a coarse material mesh is recommended. 
	\item Accurate FE models are preferred. While the optimal number of load levels and different load cases is case-dependent, more experiments generally improve identification, provided that the FE model captures each case reliably and measurement noise remains consistent.
\end{itemize}
Among all new findings, the most important are:
\begin{itemize}[noitemsep,topsep=0pt]
	\item With inexact stiffness fields, increasing the number of modes reduces density error only up to a certain point, after which the error stagnates or grows (see Secs.~\ref{s:Ex1_MD},~\ref{s:Ex2_MD}~\&~\ref{s:Ex3_MD}). Using the first 6--9 modes appears to be a reasonable choice for coarse material meshes. 
	\item Noise in normal modes moderately affects the density error ($\Delta\errA \le 1\%$ for 4\% noise), unless parameter indeterminacies are present, as in Sec.~\ref{s:Ex1_MD}. Noise in frequencies up to 1\% has little effect, but at higher levels, particularly 4\%, can introduce notable errors. 
	\item Density reconstruction from modal data requires approximately ten times fewer measurement points than stiffness identification from static experiments to achieve similar accuracy.
	\item The material mesh based on linear Lagrange interpolation, proposed in~\citet{Borzeszkowski2022}, captures non-smooth material distributions better than quadratic NURBS. However, it is not as efficient as quadratic NURBS for smooth material fields.
\end{itemize}

Several directions remain for extending the proposed framework. A key challenge is the development of an automatic adaptive material mesh algorithm, with some preliminaries shown in Sec.~\ref{s:Ex2_MD}. Density identification based on modal dynamics should be extended to 3D structures and further investigated, particularly for inaccurate elastic parameters. Since the presented problems only involve up to 61 design variables, computing Eqs.~\eqref{e:dudq}~and~\eqref{e:dutdq} has relatively small cost, but adjoint methods should be considered for large-scale problems. The present framework treats inverse analysis with static and modal dynamics as separate problems; however, integrating them into a unified formulation should be explored. Bayesian approaches will also be considered. Finally, experimental validation remains an important future step.

\bigskip

{\Large{\bf Acknowledgements}}

The authors thank Prof.~Agnieszka Tomaszewska for discussions on the topic and acknowledge the financial support of the Doctoral School at Gda\'{n}sk University of Technology.

\appendix

\section{Inverse analysis with 3D-based synthetic experimental data}\label{s:3D_ver}

This section illustrates the suitability of BE beam theory for slender structures with 3D distributed material properties.

\subsection{Bar under uniaxial deformation}\label{s:3D_verEx1}

\begin{figure}[h]
	\begin{center} \unitlength1cm
		\begin{picture}(0,8.4)
			\put(-8,4){\includegraphics[height=42mm]{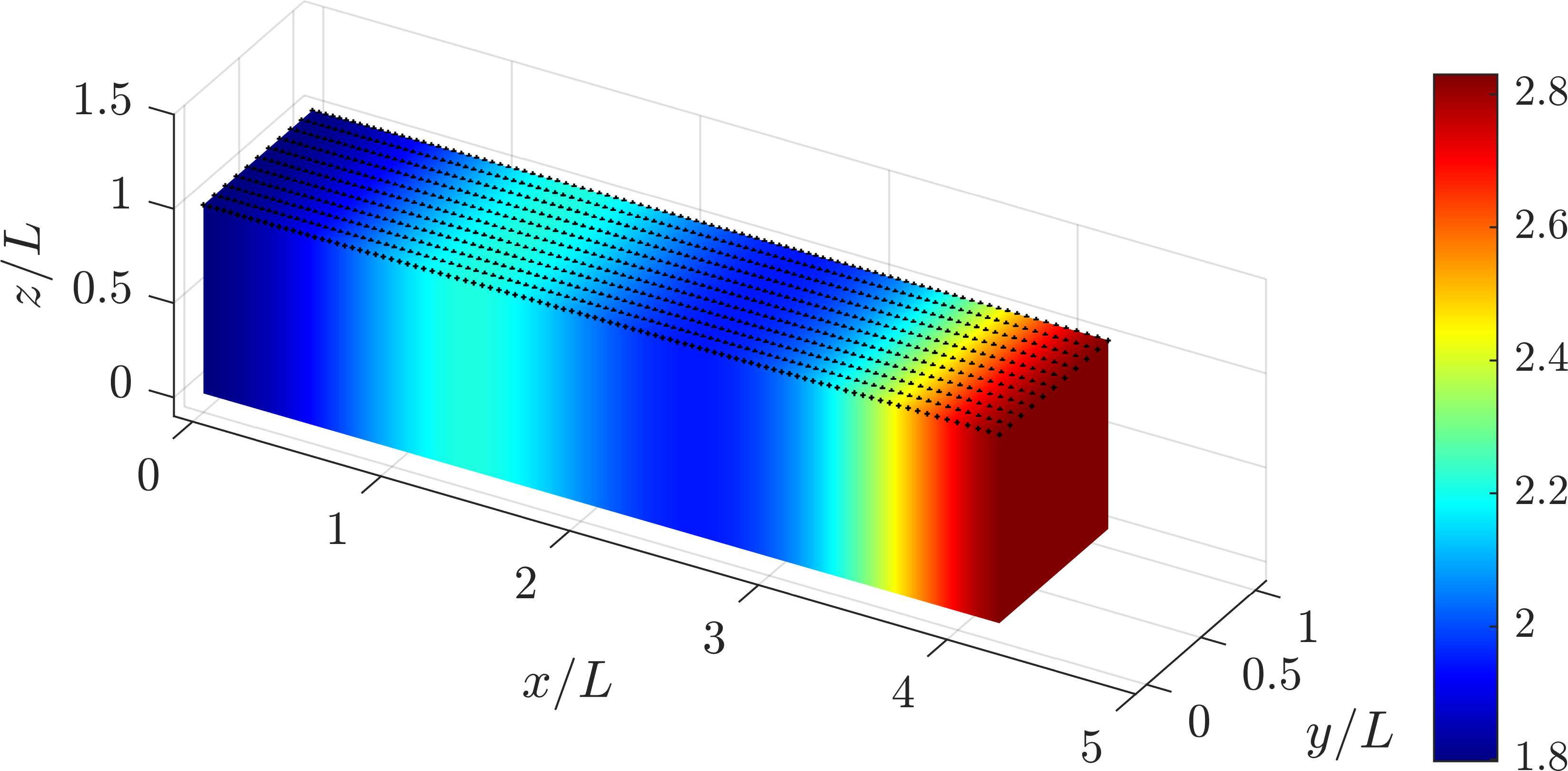}} 	
			\put(-6,0){\includegraphics[height=30mm]{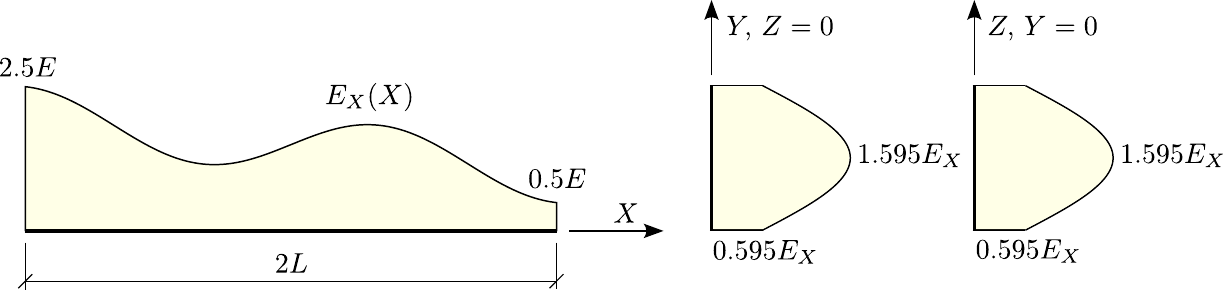}} 						
			\put(1.25,3.5){\includegraphics[height=47mm]{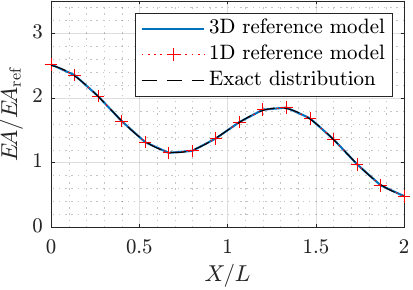}}  						 					
			\put(-8,3.5){\small{a.}}
			\put(-6.4,0){\small{b.}}
			\put(1.25,3.5){\small{c.}}
		\end{picture}	
		\caption{Uniaxial stretching of a bar: a.~deformed 3D reference model colored by the volume change $J := \det \bF$ with black dots representing the experimental grid; b.~reference Young's modulus $E$ spatial distribution from Eq.~\eqref{e:Ex1A_3D}, corresponding to the axial stiffness in Eq.~\eqref{e:Ex1A}; c.~reconstructed axial stiffness using 3D quasi-experimental data and data generated using the 1D model from Fig.~\ref{fig:Ex1_stat}a (Case~1.3 in Tab.~\ref{tab:Ex1_stat}).}
		\label{fig:Ex1_3D_ref}	
	\end{center}
\end{figure}
Fig.~\ref{fig:Ex1_3D_ref}a shows the deformation of a 3D bar consisting of \qty{19652}{} quadratic B-spline FE ($68 \times 17 \times 17$ elements in $X$-, $Y$-, and $Z$-~directions, respectively). The $X = 0$ boundary is fixed along the $X$-~direction, and all DOFs on the $X = 2L$ boundary are constrained to have the same $x$-~displacement. The spatial distribution of Young's modulus shown in Fig.~\ref{fig:Ex1_3D_ref}b, is given by
\eqb{l}
	E(\xi,\eta,\zeta)/E_\mrrf = \ds [2+0.5\cos(3\pi\xi) - \xi]\,[\sin(\pi\eta)\sin(\pi\zeta) + 1 - 4/\pi^2] \,,
\label{e:Ex1A_3D}\eqe
where $E_\mrrf = 100 F/L^2$, $\xi = X/2L$, $\eta = Y/L$, and $\zeta = Z/L$. The distribution is chosen so that the resulting axial stiffness matches that given in Eq.~\eqref{e:Ex1A}. An experimental grid of size $100\times10$ is placed at the top surface of the bar (Fig.~\ref{fig:Ex1_3D_ref}a). To obtain 1D results for the inverse analysis, all measurements are averaged in the $Y$-~direction, resulting in $100$ experimental points. For the inverse analysis, a model with 30 FE and 15 ME is selected.

Fig.~\ref{fig:Ex1_3D_ref}c compares the exact $\Astf(\xi)$ from Eq.~\eqref{e:Ex1A} with the identified $\Astf(\xi)$ based on quasi-experimental data from the full 3D and 1D models. Both 3D and 1D reference models yield similar results with $\errMA = 3.87\%$ and $\errAA = 0.96\%$ for the 3D reference model vs.~$\errMA = 4.39\%$ and $\errAA = 0.99\%$ for the 1D reference model (Case~1.3 in Tab.~\ref{tab:Ex1_stat}). This shows that transverse heterogeneity does not affect the results of the inverse analysis for slender structures. Hence, for the rest of the cases in Secs.~\ref{s:Ex1_stat}~and~\ref{s:Ex1_MD}, quasi-experimental data from 1D models are solely used. Naturally, by the assumption of a beam model, it is not possible to identify heterogeneity in three dimensions. Axial stiffness is a resultant quantity, averaging the behavior over the cross-section. Note also that this study does not take into account inexact boundary conditions, which can be an issue in real experiments and should be examined in future work.

\subsection{Bending of an initially straight beam}\label{s:3D_verEx2}

\begin{figure}[h]
	\begin{center} \unitlength1cm
		\begin{picture}(0,8)
			\put(-8,4.4){\includegraphics[width=75mm]{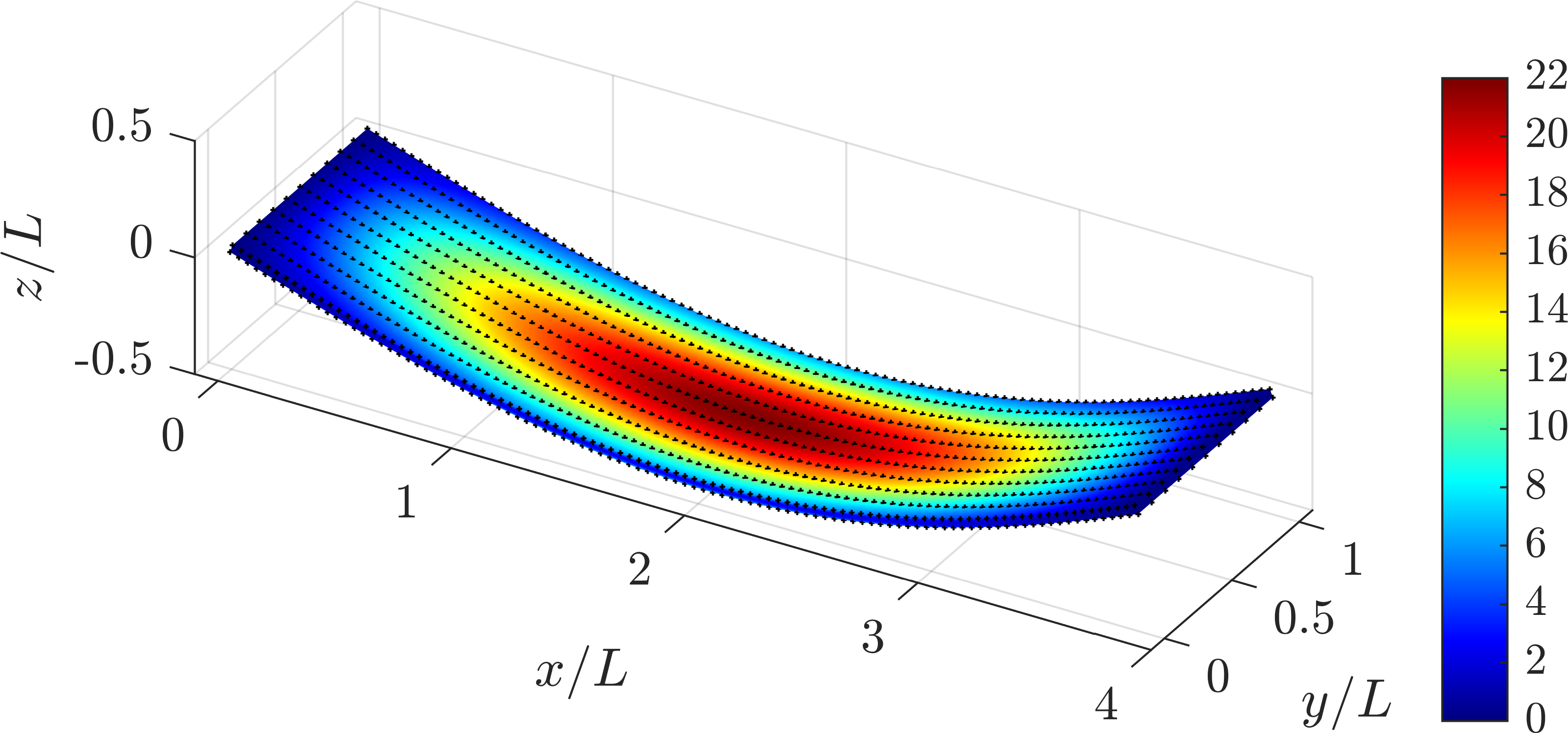}} 	
			\put(0.25,4.2){\includegraphics[width=75mm]{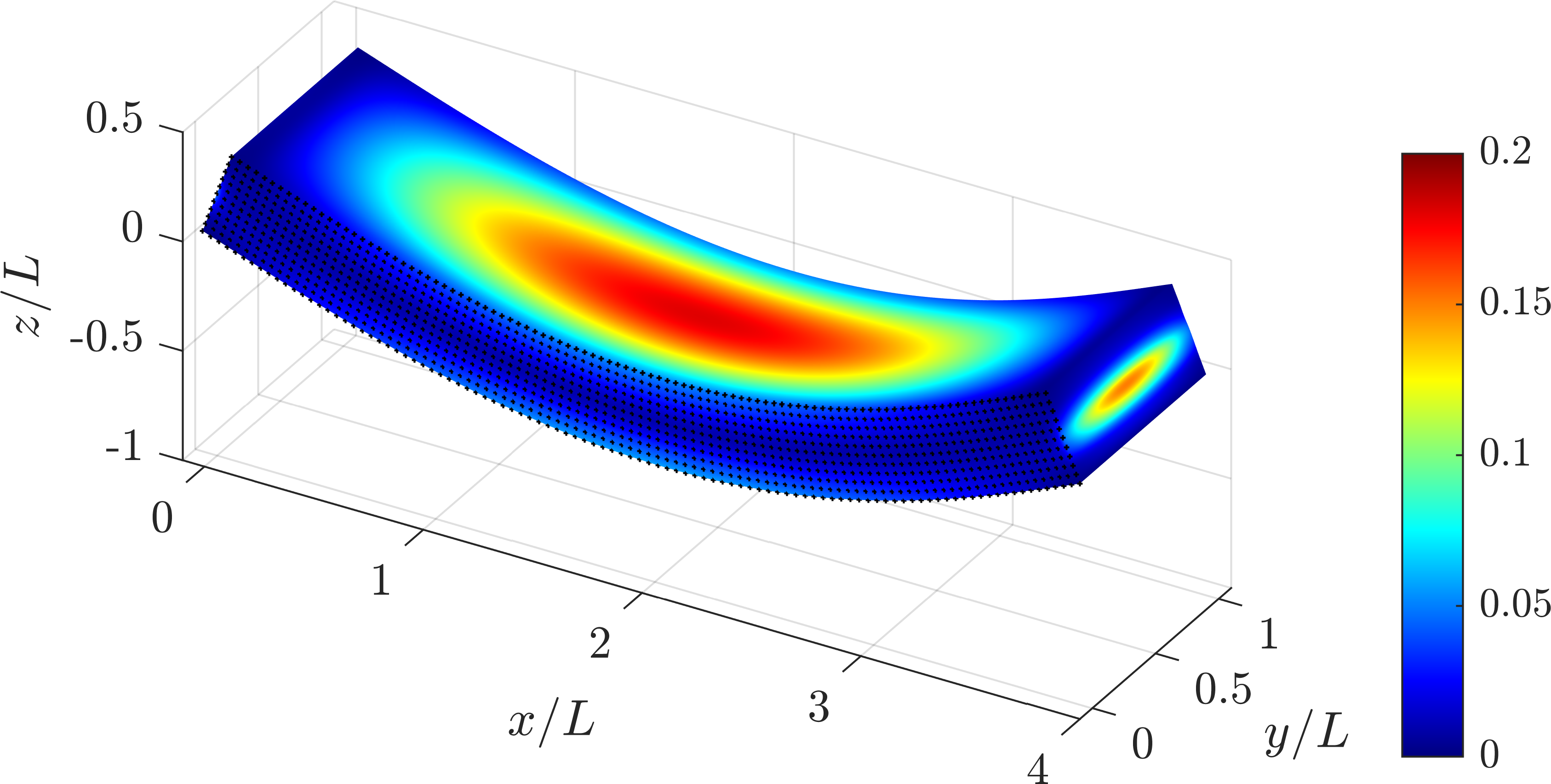}} 						
			\put(-8,0.2){\includegraphics[width=75mm]{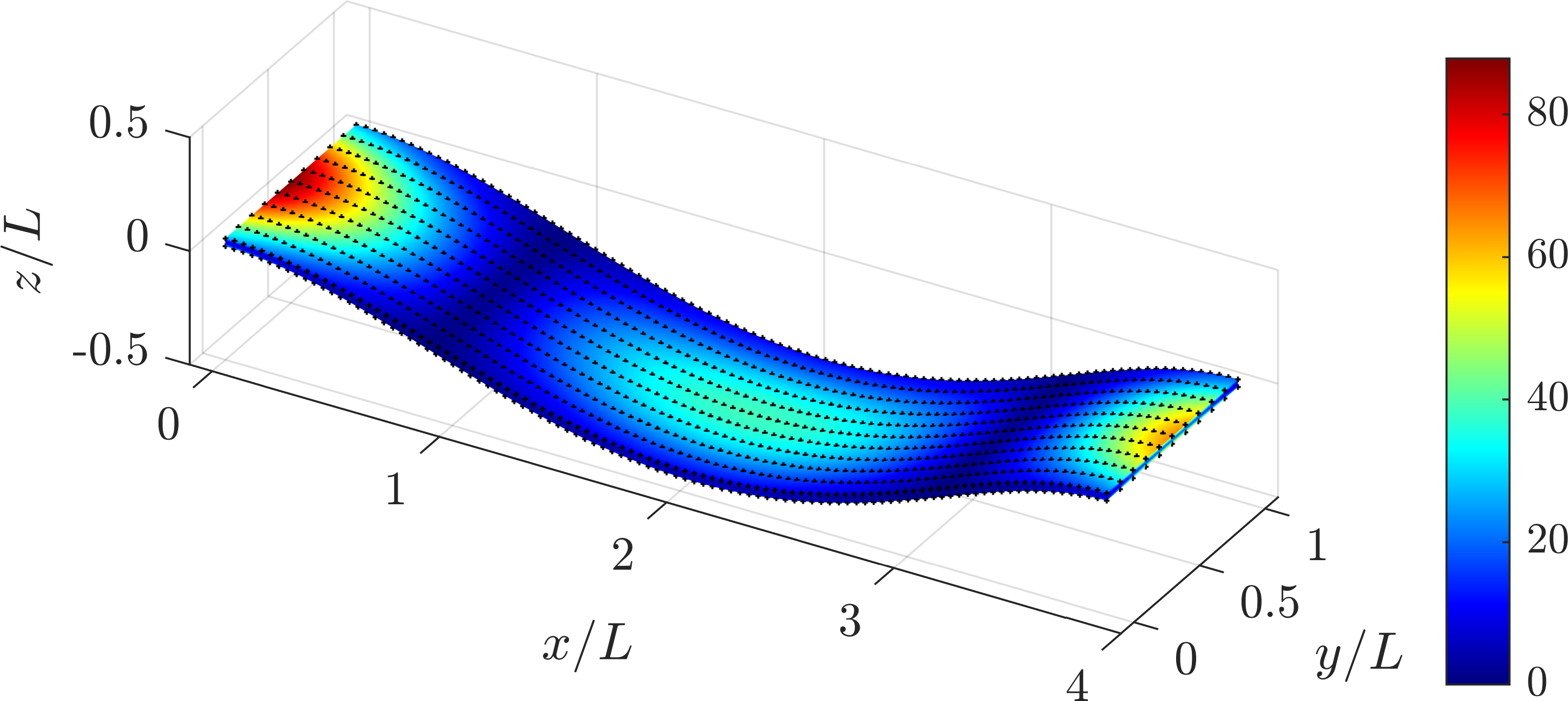}}  	
			\put(0.25,0){\includegraphics[width=75mm]{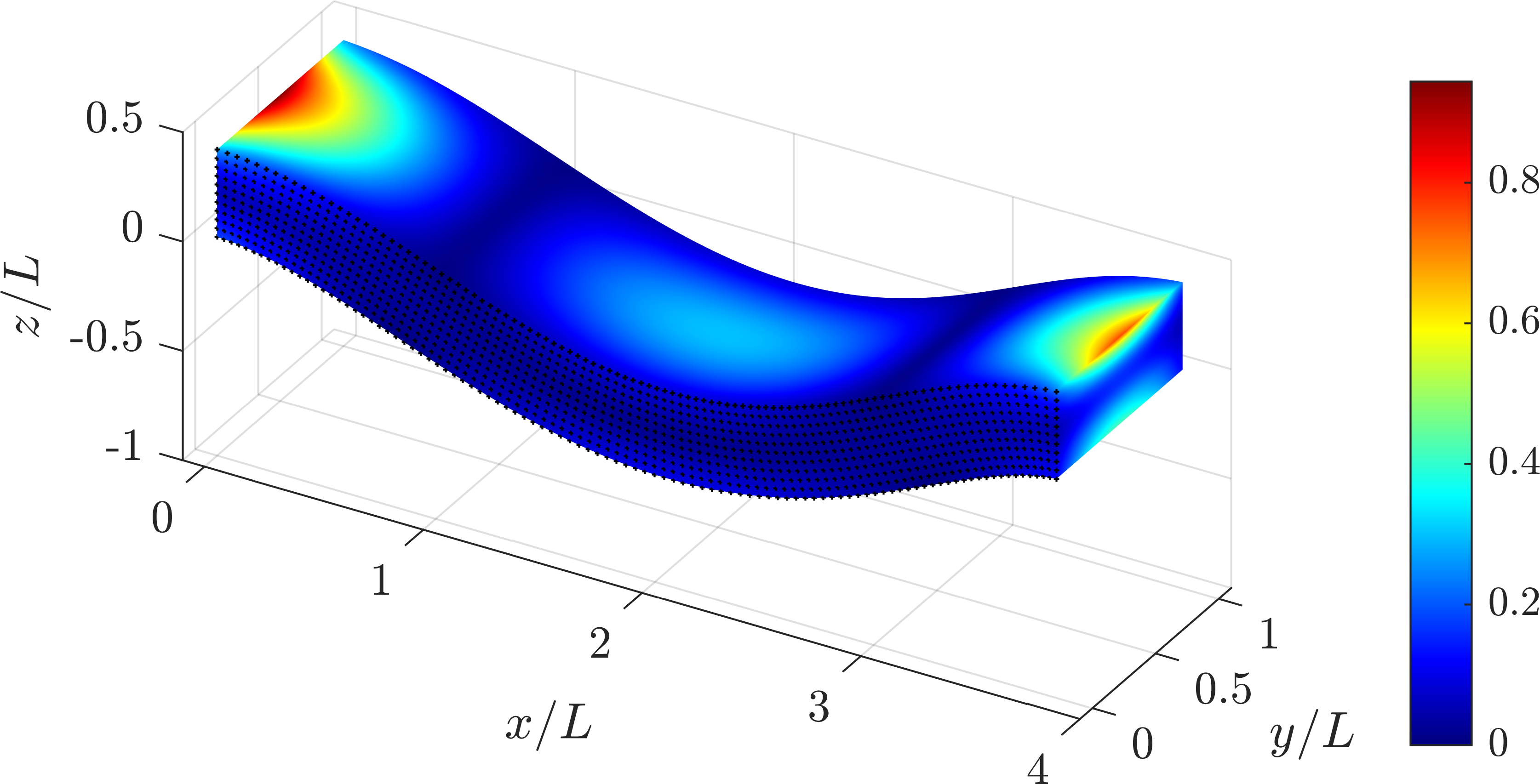}}  					 					
			\put(-8,4.2){\small{a.}}
			\put(0.25,4.2){\small{b.}}
			\put(-8,0){\small{c.}}
			\put(0.25,0){\small{d.}}
			\put(-1.9,7.35){\tiny{$[F/L^2]$}}
			\put(6.25,7.05){\tiny{$[F/L^2]$}}
			\put(-1.9,3.1){\tiny{$[F/L^2]$}}
			\put(6.3,3.25){\tiny{$[F/L^2]$}}
		\end{picture}	
		\caption{Bending of a straight beam: Deformed 3D reference models colored by the von Mises stress: Simply supported beam with $T/L_x\approx0.01$ (a.) and $T/L_x=0.1$ (b.), respectively; clamped beam with $T/L_x\approx0.01$ (c.) and $T/L_x=0.1$ (d.), respectively. Black dots represent the chosen experimental grids.}
		\label{fig:Ex2_3D_ref}	
	\end{center}
\end{figure}
Fig.~\ref{fig:Ex2_3D_ref} shows the deformation of four 3D beams with different slenderness ratios $T/L_x$ and boundary conditions. Thin models in Figs~\ref{fig:Ex2_3D_ref}a~and~\ref{fig:Ex2_3D_ref}c consist of \qty{10800}{} quadratic B-spline FE ($120 \times 30 \times 3$ elements in $X$-, $Y$-, and $Z$-~directions, respectively), while thick beams utilize meshes of $120 \times 30 \times 7 = \qty{25200}{}$~FE. The simply supported beam is pinned at the ends along the neutral axis. The clamped beam is fully fixed at the left end, and at the right end it is fixed in the $Y$- and $Z$-~directions, where zero rotation and plane cross-section are enforced using a projection approach. The spatial distribution of Young's modulus is now given by
\eqb{l}  
	E(\xi,\eta)/E_\mrrf = \ds [\ds 1.5 + 0.5\cos(\pi\xi)]\,[\sin(\pi\eta) + 1 - 2/\pi] \,,
\label{e:Ex2A_3D}\eqe 
where $E_\mrrf \approx 2886.75 F/L^2$ for thin beams and $E_\mrrf = 1.875 F/L^2$ for thick beams, $\xi = X/4L$ and $\eta = Y/L$. The Young's modulus distribution varies only in the $X$- and $Y$-~directions and is chosen so that the resulting bending stiffness matches the distribution given in Eq.~\eqref{e:Ex1Rho}. Since the axial stiffness is constant, Eqs.~\eqref{e:AI} imply a variable beam thickness. This fact is neglected in the following analysis, as it does not introduce significant errors. The considered thin beams use two $100 \times 10$ experimental grids placed on the top and bottom surfaces, while the thick beams use a single $100 \times 10$ grid on the front surface. All measurements are averaged in the $Y$-~direction, resulting in 100 experimental points. The inverse analysis is conducted using a 1D beam model with 60 FE for the simply supported beam and 120 FE for the clamped beam. Ten material elements are used in each case.
\begin{figure}[h]
	\begin{center} \unitlength1cm
		\begin{picture}(0,3)
			\put(-5.8,0){\includegraphics[height=30mm]{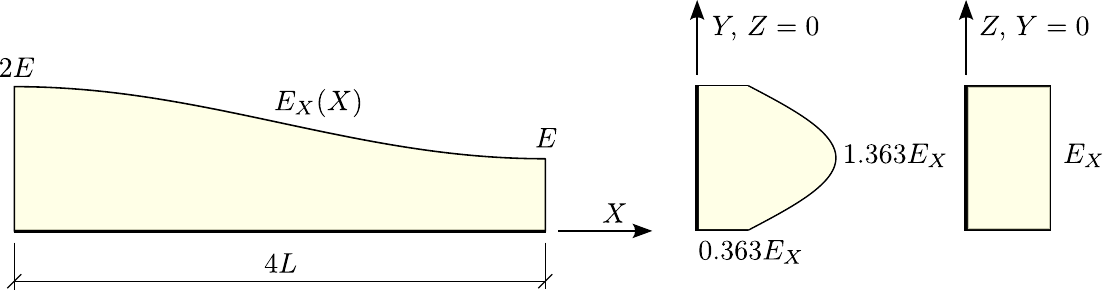}} 	
		\end{picture}
		\caption{Bending of a straight beam: reference Young's modulus $E$ spatial distribution from Eq.~\eqref{e:Ex2A_3D}, producing a bending stiffness with the same distribution as in Eq.~\eqref{e:Ex1Rho}.}
		\label{fig:Ex2_3DmatDisb_mat}	
	\end{center}
\end{figure}

\begin{figure}[h]
	\begin{center} \unitlength1cm
		\begin{picture}(0,5)
			\put(-7.75,0){\includegraphics[height=50mm]{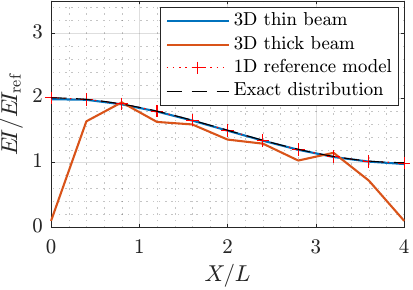}} 	
			\put(0.25,0){\includegraphics[height=50mm]{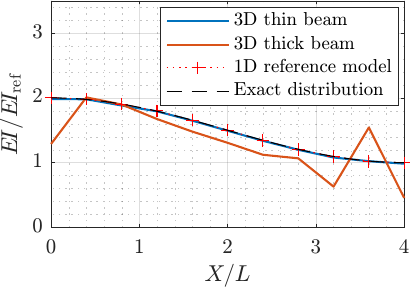}} 											 					
			\put(-7.75,0){\small{a.}}
			\put(0.25,0){\small{b.}}
		\end{picture}
		\caption{Bending of a straight beam: Reconstructed bending stiffness using data generated from 3D and 1D models for a simply supported beam (a.)~and a clamped beam (b.). For beams with $T/L_x\approx0.01$, the stiffness reconstructed using 3D data aligns well with the results for 1D data (Cases~2.1s~and~2.1c from Tab.~\ref{tab:Ex2_stat}) and the exact distribution.}
		\label{fig:Ex2_3D_ref_mat}	
	\end{center}
\end{figure}
Figs.~\ref{fig:Ex2_3D_ref_mat}a~and~b compare the results for the simply supported and clamped beams, respectively. In both cases, the bending stiffness of 3D thin beams is reconstructed accurately, with $\errMC = 2.19\%$ and $\errAC = 0.68\%$ for the simply supported beam, and $\errMC = 1.67\%$ and $\errAC = 0.72\%$ for the clamped beam, as compared with Cases~2.1s~and~2.1c from Tab.~\ref{tab:Ex2_stat}. Conversely, large discrepancies occur at the supports for 3D thick beams: $\errMC = 95.00\%$ and $\errAC = 25.38\%$ for the thick simply supported beam, and $\errMC = 54.58\%$ and $\errAC = 22.17 \%$ for the thick clamped beam. These errors are most likely due to the neglect of shear deformation in BE beams. This shows a limitation of the proposed approach, at least when using BE theory. The remaining findings are consistent with those from the 3D verification in App.~\ref{s:3D_verEx1}; therefore, for the remaining cases in this work, quasi-experimental data from 1D models are used.

\section{Non-conforming mapping between material and M1 elements}\label{s:B2M1_map}

\begin{figure}[h]
	\begin{center} \unitlength1cm
		\begin{picture}(0,7)
			\put(-5.25,0){\includegraphics[height=70mm]{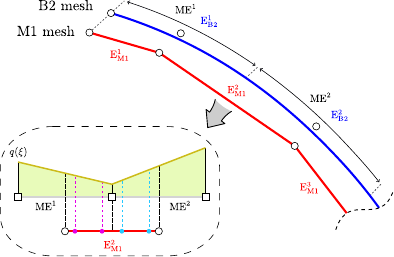}} 					
		\end{picture}
		\caption{An example of mapping between material and M1 elements: Two material elements, such as $\mrM\mrE^1$ and $\mrM\mrE^2$, affect the material distribution $q(\xi)$ within analysis element $\mrE^2_{\mrM1}$. This element is therefore divided into two integration regions, represented by two sets of Gauss points. For the sensitivity evaluation, the first set of Gauss points (magenta) is associated with the material DOFs of $\mrM\mrE^1$, and the second (cyan) with $\mrM\mrE^2$.}
		\label{fig:B2M1map}	
	\end{center}
\end{figure}
Fig.~\ref{fig:B2M1map} illustrates an example of the non-conforming mapping between material and M1 elements based on Figs.~\ref{fig:Ex3_Geo}b~\&~c. The material and B2 meshes conform to each other, which is not the case of the M1 mesh: Interior M1 elements are shifted so that their centers in the parameter domain $\sP$ always correspond to material nodes. The analysis element $\mrE^2_{\mrM1}$ is independently mapped to two material elements, $\mrM\mrE^1$ and $\mrM\mrE^2$, according to Eq.~\eqref{e:matmap}. Then, the numerical integration in $\mrE^2_{\mrM1}$ can be performed in two separated regions, which leads to an element with four Gauss points. Note that the elemental sensitivities have to be split. The contributions from the first set of Gauss points are assigned to the material DOFs of $\mrM\mrE^1$, while the second set belongs to $\mrM\mrE^2$.

\bibliographystyle{apalike}
\bibliography{references}

\end{document}

%% file: neco.tex
\newcommand{\bitm}{\begin{itemize}}
\newcommand{\eitm}{\end{itemize}}
\newcommand{\bnumr}{\begin{enumerate}}
\newcommand{\enumr}{\end{enumerate}}

\newcommand{\Astf}{E\!A}
\newcommand{\AstfR}{E\!A_\mathrm{ref}}
\newcommand{\Cstf}{E\!I}
\newcommand{\CstfR}{E\!I_\mathrm{ref}}
\newcommand{\errA}{\delta_\mathrm{ave}}
\newcommand{\errAA}{\delta_\mathrm{ave}^{\Astf}}
\newcommand{\errAC}{\delta_\mathrm{ave}^{\Cstf}}

\newcommand{\errAreg}{\delta_\mathrm{ave}^\mathrm{reg}}
\newcommand{\errAadt}{\delta_\mathrm{ave}^\mathrm{adt}}
\newcommand{\errM}{\delta_\mathrm{max}}
\newcommand{\errMA}{\delta_\mathrm{max}^{\Astf}}
\newcommand{\errMC}{\delta_\mathrm{max}^{\Cstf}}
\newcommand{\errMreg}{\delta_\mathrm{max}^\mathrm{reg}}
\newcommand{\errMadt}{\delta_\mathrm{max}^\mathrm{adt}}
\newcommand{\rhoR}{\rho_\mathrm{ref}}

\newcommand {\tmuu}{\tilde{\mathbf{u}}}

\newcommand {\T}{^\mathrm{T}}

\newcommand{\mrin}{\mathrm{in}}
\newcommand{\mrel}{\mathrm{el}}
\newcommand{\mrint}{\mathrm{int}}
\newcommand{\mrext}{\mathrm{ext}}
\newcommand{\mrect}{\mathrm{exact}}
\newcommand{\mrll}{\mathrm{ll}}
\newcommand{\mrlc}{\mathrm{lc}}
\newcommand{\mrex}{\mathrm{exp}}
\newcommand{\mrfe}{\mathrm{FE}}
\newcommand{\mrmd}{\mathrm{mode}}
\newcommand{\mrno}{\mathrm{no}}
\newcommand{\mrvr}{\mathrm{var}}
\newcommand{\mrop}{\mathrm{opt}}
\newcommand{\mrexc}{\mathrm{exact}}
\newcommand{\mrrf}{\mathrm{ref}}

\newcommand{\brn}{\bar{n}}
\newcommand{\bre}{\bar{e}}

\newcommand{\brN}{\bar{N}}

\newcommand {\eqb}[1]{\begin{equation}\begin{array}{#1}}
\newcommand {\eqe}{\end{array}\end{equation}}

\newcommand {\esb}[1]{\begin{equation*}\begin{array}{#1}}
\newcommand {\ese}{\end{array}\end{equation*}}
\newcommand {\ds}{\displaystyle}

\newcommand {\pa}[2]{\frac{\partial{#1}}{\partial{#2}}}
\newcommand {\pad}[2]{\frac{\mathrm{d}{#1}}{\mathrm{d}{#2}}}
\newcommand {\paq}[2]{\frac{\partial^2{#1}}{\partial{#2}^2}}

\newcommand {\back}{\! \! \!}
\newcommand {\is}{\back &=& \back}

\newcommand {\norm}[1]{\|#1\|}

\newcommand {\dif}{\mathrm{d}}

\newcommand {\II}{{I\kern-.3em I}}
\newcommand {\III}{{I\kern-.3em I\kern-.3em I}}

\newcommand {\intoe}{\int_{\Omega^e}}

\newcommand {\intooe}{\int_{\Omega_0^e}}

\newcommand {\mra}{\mathrm{a}}
\newcommand {\mrb}{\mathrm{b}}
\newcommand {\mrc}{\mathrm{c}}

\newcommand {\mre}{\mathrm{e}}

\newcommand {\mri}{\mathrm{i}}

\newcommand {\mrs}{\mathrm{s}}

\newcommand {\mrR}{\mathrm{R}}

\newcommand {\mrE}{\mathrm{E}}
\newcommand {\mrM}{\mathrm{M}}

\newcommand {\mf}{\mathbf{f}}
\newcommand {\mg}{\mathbf{g}}
\newcommand {\mh}{\mathbf{h}}

\newcommand {\mm}{\mathbf{m}}

\newcommand {\mq}{\mathbf{q}}

\newcommand {\mss}{\mathbf{s}}

\newcommand {\muu}{\mathbf{u}}

\newcommand {\mx}{\mathbf{x}}

\newcommand {\mbN}{\mathbf{\bar N}}

\newcommand {\ba}{\boldsymbol{a}}

\newcommand {\be}{\boldsymbol{e}}
\newcommand {\bff}{\boldsymbol{f}}

\newcommand {\bk}{\boldsymbol{k}}

\newcommand {\bn}{\boldsymbol{n}}

\newcommand {\bq}{\boldsymbol{q}}

\newcommand {\bt}{\boldsymbol{t}}
\newcommand {\bu}{\boldsymbol{u}}

\newcommand {\bx}{\boldsymbol{x}}

\newcommand {\bnu}{\mbox{\boldmath$\nu$}}

\newcommand {\brho}{\mbox{\boldmath$\rho$}}

\newcommand {\mA}{\mathbf{A}}

\newcommand {\mC}{\mathbf{C}}

\newcommand {\mE}{\mathbf{E}}

\newcommand {\mH}{\mathbf{H}}
\newcommand {\mI}{\mathbf{I}}
\newcommand {\mJ}{\mathbf{J}}
\newcommand {\mK}{\mathbf{K}}
\newcommand {\mL}{\mathbf{L}}
\newcommand {\mM}{\mathbf{M}}
\newcommand {\mN}{\mathbf{N}}

\newcommand {\mR}{\mathbf{R}}
\newcommand {\mS}{\mathbf{S}}

\newcommand {\mU}{\mathbf{U}}

\newcommand {\mX}{\mathbf{X}}

\newcommand {\mZ}{\mathbf{Z}}

\newcommand {\bA}{\boldsymbol{A}}

\newcommand {\bE}{\boldsymbol{E}}
\newcommand {\bF}{\boldsymbol{F}}

\newcommand {\bK}{\boldsymbol{K}}

\newcommand {\bN}{\boldsymbol{N}}

\newcommand {\bX}{\boldsymbol{X}}

\newcommand {\eps}{\varepsilon}

\newcommand {\bone}{\mathbf{1}}

\newcommand {\IR}{{\rm\kern.24em
   \vrule width.02em height1.53ex depth-.05ex
   \kern-.3em R}}
\newcommand {\ic}{{\rm\kern.20em
   \vrule width.02em height1.0ex depth-.05ex
   \kern-.22em c}}
\newcommand {\ia}{{\rm\kern.20em
   \vrule width.02em height1.05ex depth-.0ex
   \kern-.25em a}}
\newcommand {\IC}{{\rm\kern.24em
   \vrule width.02em height1.4ex depth-.05ex
   \kern-.26em C}}
\newcommand {\ID}{{\rm\kern.34em
   \vrule width.02em height1.5ex depth-.05ex
   \kern-.36em D}}
\newcommand {\IS}{{\rm\kern.24em
   \vrule width.02em height1.6ex depth.05ex
   \kern-.26em S}}
\newcommand {\IT}{{\rm\kern.50em
   \vrule width.02em height1.55ex depth-.05ex
   \kern-.52em T}}

\newcommand {\IE}{{\rm\kern.24em
   \vrule width.02em height1.55ex depth-.05ex
   \kern-.33em E}}
\newcommand {\IEa}{{\rm\kern.24em
   \vrule width.02em height1.55ex depth-.05ex
   \kern-.33em E}^{1}_{ijkl}}
\newcommand {\IEb}{{\rm\kern.24em
   \vrule width.02em height1.55ex depth-.05ex
   \kern-.33em E}^{2}_{ijkl}}

\newcommand {\sL}{\mathcal{L}}

\newcommand {\sN}{\mathcal{N}}

\newcommand {\sP}{\mathcal{P}}

\newcommand {\sV}{\mathcal{V}}

\newcommand {\Ass}[2]{\kern 0.9ex \vrule width0.45em height0.2ex depth0ex \kern -2.1ex \bigwedge_{#1}^{#2}}
\newcommand {\ASS}[2]{\kern 1.45ex \vrule width0.5em height0.2ex depth0ex \kern -2.65ex \bigwedge_{#1}^{#2}}

